\documentclass[twocolumn]{aastex631}
\usepackage{apjfonts}
\usepackage[T1]{fontenc}
\DeclareRobustCommand{\VAN}[3]{#2}
\let\VANthebibliography\thebibliography
\def\thebibliography{\DeclareRobustCommand{\VAN}[3]{##3}\VANthebibliography}

\usepackage[utf8]{inputenc}
\usepackage{graphicx}	
\usepackage{amsmath}	
\usepackage{amssymb}    
\usepackage{float}
\usepackage[utf8]{inputenc}
\usepackage{xspace}

\newcommand{\nl}{\newline}

\newcommand{\targa}{TOI-2076\xspace}
\newcommand{\targb}{TOI-1807\xspace}
\newcommand{\tica}{\textit{TIC 27491137}\xspace}
\newcommand{\ticb}{\textit{TIC 180695581}\xspace}
\newcommand{\ch}[1]{\textcolor{red}{#1}\xspace}

\newcommand{\tess}{\emph{TESS}\xspace}

\newcommand{\planb}{TOI-2076b}
\newcommand{\planc}{TOI-2076c}
\newcommand{\pland}{TOI-2076d}
\newcommand{\planbtwo}{TOI-1807b}
\newcommand{\planall}{TOI-2076b/c/d}
\newcommand{\dista}{41.91 pc\xspace}

\newcommand{\agea}{$204 \pm 50$ Myr\xspace}
\newcommand{\ageb}{$180 \pm 40$ Myr\xspace}

\newcommand{\perc}{$17.19342_{-9e-05}^{+9e-05}$ d\xspace}
\newcommand{\perd}{$25.08872_{-2.7e-04}^{+2.7-04}$ d\xspace}

\providecommand{\teff}{\ensuremath{T_{\rm eff}}}

\providecommand{\msun}{\ensuremath{\,M_\Sun}}
\providecommand{\rsun}{\ensuremath{\,R_\Sun}}
\providecommand{\lsun}{\ensuremath{\,L_\Sun}}

\begin{document}

\title{TOI-2076 and TOI-1807: Two young, comoving planetary systems within 50 pc identified by TESS that are ideal candidates for further follow-up.}
\newcommand{\cfa}{Center for Astrophysics \textbar \ Harvard \& Smithsonian, 60 Garden St, Cambridge, MA 02138, USA}
\newcommand{\msu}{Department of Physics and Astronomy, Michigan State University, East Lansing, MI 48824, USA}
\newcommand{\umich}{Astronomy Department, University of Michigan, 1085 S University Avenue, Ann Arbor, MI 48109, USA}
\newcommand{\utaustin}{Department of Astronomy, The University of Texas at Austin, Austin, TX 78712, USA}
\newcommand{\MIT}{Department of Physics and Kavli Institute for Astrophysics and Space Research, Massachusetts Institute of Technology, Cambridge, MA 02139, USA}
\newcommand{\MITEPS}{Department of Earth, Atmospheric and Planetary Sciences, Massachusetts Institute of Technology,  Cambridge,  MA 02139, USA}
\newcommand{\uflorida}{Department of Astronomy, University of Florida, 211 Bryant Space Science Center, Gainesville, FL, 32611, USA}
\newcommand{\riverside}{Department of Earth and Planetary Sciences, University of California,
Riverside, CA 92521, USA}
\newcommand{\usq}{Centre for Astrophysics, University of Southern Queensland, West Street, Toowoomba, QLD 4350, Australia}
\newcommand{\ames}{NASA Ames Research Center, Moffett Field, CA, 94035, USA}
\newcommand{\geneva}{Observatoire de l’Universit\'e de Gen\`eve, 51 chemin des Maillettes,
1290 Versoix, Switzerland}
\newcommand{\uw}{Astronomy Department, University of Washington, Seattle, WA 98195 USA}
\newcommand{\warwick}{Deptartment of Physics, University of Warwick, Gibbet Hill Road, Coventry CV4 7AL, UK}
\newcommand{\warwickceh}{Centre for Exoplanets and Habitability, University of Warwick, Gibbet Hill Road, Coventry CV4 7AL, UK}
\newcommand{\princeton}{Department of Astrophysical Sciences, Princeton University, 4 Ivy Lane, Princeton, NJ, 08544, USA}
\newcommand{\liege}{Space Sciences, Technologies and Astrophysics Research (STAR) Institute, Universit\'e de Li\`ege, 19C All\'ee du 6 Ao\^ut, 4000 Li\`ege, Belgium}
\newcommand{\vanderbilt}{Department of Physics and Astronomy, Vanderbilt University, Nashville, TN 37235, USA}
\newcommand{\fisk}{Department of Physics, Fisk University, 1000 17th Avenue North, Nashville, TN 37208, USA}
\newcommand{\columbia}{Department of Astronomy, Columbia University, 550 West 120th Street, New York, NY 10027, USA}
\newcommand{\toronto}{Dunlap Institute for Astronomy and Astrophysics, University of Toronto, Ontario M5S 3H4, Canada}
\newcommand{\unc}{Department of Physics and Astronomy, University of North Carolina at Chapel Hill, Chapel Hill, NC 27599, USA}
\newcommand{\iac}{Instituto de Astrof\'isica de Canarias (IAC), E-38205 La Laguna, Tenerife, Spain}
\newcommand{\lalaguna}{Departamento de Astrof\'isica, Universidad de La Laguna (ULL), E-38206 La Laguna, Tenerife, Spain}
\newcommand{\louisville}{Department of Physics and Astronomy, University of Louisville, Louisville, KY 40292, USA}
\newcommand{\aavso}{American Association of Variable Star Observers, 49 Bay State Road, Cambridge, MA 02138, USA}
\newcommand{\utokyo}{The University of Tokyo, 7-3-1 Hongo, Bunky\={o}, Tokyo 113-8654, Japan}
\newcommand{\naoj}{National Astronomical Observatory of Japan, 2-21-1 Osawa, Mitaka, Tokyo 181-8588, Japan}
\newcommand{\jstpresto}{JST, PRESTO, 7-3-1 Hongo, Bunkyo-ku, Tokyo 113-0033, Japan}
\newcommand{\astrobiojapan}{Astrobiology Center, 2-21-1 Osawa, Mitaka, Tokyo 181-8588, Japan}
\newcommand{\ctio}{Cerro Tololo Inter-American Observatory, Casilla 603, La Serena, Chile}
\newcommand{\nexsci}{Caltech IPAC -- NASA Exoplanet Science Institute 1200 E. California Ave, Pasadena, CA 91125, USA}
\newcommand{\ucsc}{Department of Astronomy and Astrophysics, University of
California, Santa Cruz, CA 95064, USA}
\newcommand{\gsfc}{Exoplanets and Stellar Astrophysics Laboratory, Code 667, NASA Goddard Space Flight Center, Greenbelt, MD 20771, USA}
\newcommand{\sgtinc}{SGT, Inc./NASA AMES Research Center, Mailstop 269-3, Bldg T35C, P.O. Box 1, Moffett Field, CA 94035, USA}
\newcommand{\chile}{Center of Astro-Engineering UC, Pontificia Universidad Cat\'olica de Chile, Av. Vicu\~{n}a Mackenna 4860, 7820436 Macul, Santiago, Chile}
\newcommand{\Pontificia}{Instituto de Astrof\'isica, Pontificia Universidad Cat\'olica de Chile, Av.\ Vicu\~na Mackenna 4860, Macul, Santiago, Chile}
\newcommand{\Millennium}{Millennium Institute for Astrophysics, Chile}
\newcommand{\maxplank}{Max-Planck-Institut f\"ur Astronomie, K\"onigstuhl 17, Heidelberg 69117, Germany}
\newcommand{\utdallas}{Department of Physics, The University of Texas at Dallas, 800 West
Campbell Road, Richardson, TX 75080-3021 USA}
\newcommand{\MauryLewin}{Maury Lewin Astronomical Observatory, Glendora, CA 91741, USA}
\newcommand{\umbc}{University of Maryland, Baltimore County, 1000 Hilltop Circle, Baltimore, MD 21250, USA}
\newcommand{\osu}{Department of Astronomy, The Ohio State University, 140 West 18th Avenue, Columbus, OH 43210, USA}
\newcommand{\MITAA}{Department of Aeronautics and Astronautics, MIT, 77 Massachusetts Avenue, Cambridge, MA 02139, USA}
\newcommand{\openu}{School of Physical Sciences, The Open University, Milton Keynes MK7 6AA, UK}
\newcommand{\swarthmore}{Department of Physics and Astronomy, Swarthmore College, Swarthmore, PA 19081, USA}
\newcommand{\seti}{SETI Institute, Mountain View, CA 94043, USA}
\newcommand{\lehigh}{Department of Physics, Lehigh University, 16 Memorial Drive East, Bethlehem, PA 18015, USA}
\newcommand{\utah}{Department of Physics and Astronomy, University of Utah, 115 South 1400 East, Salt Lake City, UT 84112, USA}
\newcommand{\USNA}{Department of Physics, United States Naval Academy, 572C Holloway Rd., Annapolis, MD 21402, USA}
\newcommand{\eplcarnegie}{Earth \& Planets Laboratory, Carnegie Institution for Science, 5241 Broad Branch Road, NW, Washington, DC 20015, USA}
\newcommand{\UPenn}{The University of Pennsylvania, Department of Physics and Astronomy, Philadelphia, PA, 19104, USA}
\newcommand{\montana}{Department of Physics and Astronomy, University of Montana, 32 Campus Drive, No. 1080, Missoula, MT 59812 USA}
\newcommand{\psu}{Department of Astronomy \& Astrophysics, The Pennsylvania State University, 525 Davey Lab, University Park, PA 16802, USA}
\newcommand{\psust}{Center for Exoplanets and Habitable Worlds, The Pennsylvania State University, 525 Davey Lab, University Park, PA 16802, USA}
\newcommand{\Kutztown}{Department of Physical Sciences, Kutztown University, Kutztown, PA 19530, USA}
\newcommand{\udel}{Department of Physics \& Astronomy, University of Delaware, Newark, DE 19716, USA}
\newcommand{\Westminster}{Department of Physics, Westminster College, New Wilmington, PA 16172}
\newcommand{\steward}{Department of Astronomy and Steward Observatory, University of Arizona, Tucson, AZ 85721, USA}
\newcommand{\saao}{South African Astronomical Observatory, PO Box 9, Observatory, 7935, Cape Town, South Africa}
\newcommand{\salt}{Southern African Large Telescope, PO Box 9, Observatory, 7935, Cape Town, South Africa}
\newcommand{\ssl}{Societ\`{a} Astronomica Lunae, Italy}
\newcommand{\spot}{Spot Observatory, Nashville, TN 37206, USA}
\newcommand{\txamGP}{George P.\ and Cynthia Woods Mitchell Institute for Fundamental Physics and Astronomy, Texas A\&M University, College Station, TX77843 USA}
\newcommand{\txam}{Department of Physics and Astronomy, Texas A\&M university, College Station, TX 77843 USA}
\newcommand{\wellesley}{Department of Astronomy, Wellesley College, Wellesley, MA 02481, USA}
\newcommand{\byu}{Department of Physics and Astronomy, Brigham Young University, Provo, UT 84602, USA}
\newcommand{\Hazelwood}{Hazelwood Observatory, Churchill, Victoria, Australia}
\newcommand{\pest}{Perth Exoplanet Survey Telescope, Perth, Australia}
\newcommand{\Winer}{Winer Observatory, PO Box 797, Sonoita, AZ 85637, USA}
\newcommand{\icpo}{Ivan Curtis Private Observatory}
\newcommand{\elsauce}{El Sauce Observatory, Chile}
\newcommand{\crow}{Atalaia Group \& CROW Observatory, Portalegre, Portugal}
\newcommand{\dfus}{Dipartimento di Fisica ``E.R.Caianiello'', Universit\`a di Salerno, Via Giovanni Paolo II 132, Fisciano 84084, Italy}
\newcommand{\indfn}{Istituto Nazionale di Fisica Nucleare, Napoli, Italy}
\newcommand{\sotes}{Gabriel Murawski Private Observatory (SOTES)}
\newcommand{\lco}{Las Cumbres Observatory Global Telescope, 6740 Cortona Dr., Suite 102, Goleta, CA 93111, USA}
\newcommand{\ucsb}{Department of Physics, University of California, Santa Barbara, CA 93106-9530, USA}
\newcommand{\carnegie}{The Observatories of the Carnegie Institution for Science, 813 Santa Barbara St., Pasadena, CA 91101, USA}
\newcommand{\wisconsin}{Department of Astronomy, University of Wisconsin-Madison, Madison, WI 53706, USA}
\newcommand{\Patashnick}{Patashnick Voorheesville Observatory, Voorheesville, NY 12186, USA}
\newcommand{\Algonquin}{Algonquin Regional High School, MA, USA}
\newcommand{\crls}{Cambridge Rindge and Latin High School, MA, USA}
\newcommand{\jpl}{Jet Propulsion Laboratory, California Institute of Technology, 4800 Oak Grove Drive, Pasadena, CA 91109, USA}
\newcommand{\stsci}{Space Telescope Science Institute, Baltimore, MD 21218, USA}
\newcommand{\gsfcsellers}{GSFC Sellers Exoplanet Environments Collaboration, NASA Goddard Space Flight Center, Greenbelt, MD 20771 }
\newcommand{\gmu}{George Mason University, 4400 University Drive MS 3F3, Fairfax, VA 22030, USA}
\newcommand{\hawaii}{Institute for Astronomy, University of Hawaii, Maui, HI 96768, USA}
\newcommand{\harvard}{Harvard University, Cambridge, MA 02138, USA}
\newcommand{\ucscchile}{Departamento de Matem\'atica y F\i'sica Aplicadas, Universidad Cat\'olica de la Sant\'isima Concepci\'on, Alonso de Rivera 2850, Concepci\'on, Chile}
\newcommand{\brierfield}{Brierfield Observatory, New South Wales, Australia}
\newcommand{\berkely}{Department of Astronomy, University of California Berkeley, Berkeley, CA 94720-3411, USA}
\newcommand{\flatiron}{Center for Computational Astrophysics, Flatiron Institute, 162 Fifth Ave, New York, NY 10010, USA}
\newcommand{\DGPScaltech}{Division of Geological and Planetary Sciences, Caltech, Pasadena, CA}
\newcommand{\Tsinghua}{Department of Astronomy, Tsinghua University, Beijing 100084, China}
\newcommand{\JSTAKomaba}{Japan Science and Technology Agency, PRESTO, 3-8-1 Komaba, Meguro, Tokyo 153-8902, Japan}
\newcommand{\Komaba}{Komaba Institute for Science, The University of Tokyo, 3-8-1 Komaba, Meguro, Tokyo 153-8902, Japan}
\newcommand{\BayArea}{Bay Area Environmental Research Institute, P.O. Box 25, Moffett Field, CA 94035, USA}
\newcommand{\gemini}{Gemini Observatory/NSF’s NOIRLab, 670 N. A’ohoku Place, Hilo, HI, 96720, USA}
\newcommand{\kansas}{Department of Physics and Astronomy, University of Kansas, 1251 Wescoe Hall Dr., Lawrence, KS 66045, USA}
\newcommand{\amnh}{Department of Astrophysics, American Museum of Natural History, New York, NY 10024, USA}
\newcommand{\dtu}{DTU Space, National Space Institute, Technical University of Denmark, Elektrovej 328, DK-2800 Kgs. Lyngby, Denmark}

\newcommand{\torres}{\altaffiliation{Juan Carlos Torres Fellow}}
\newcommand{\sagan}{\altaffiliation{NASA Sagan Fellow}}
\newcommand{\bernoulli}{\altaffiliation{Bernoulli fellow}}
\newcommand{\gruber}{\altaffiliation{Gruber fellow}}
\newcommand{\kavli}{\altaffiliation{Kavli Fellow}}
\newcommand{\peg}{\altaffiliation{51 Pegasi b Fellow}}
\newcommand{\pappalardo}{\altaffiliation{Pappalardo Fellow}}
\newcommand{\hubble}{\altaffiliation{NASA Hubble Fellow}}
\newcommand{\nsfgf}{\altaffiliation{National Science Foundation Graduate Research Fellow}}
\newcommand{\eberly}{\altaffiliation{Eberly Research Fellow}}

\correspondingauthor{Christina Hedges}
\email{christina.l.hedges@nasa.gov}
\author[0000-0002-3385-8391]{Christina Hedges}
\affiliation{\BayArea}
\affiliation{\ames}

\author{Alex Hughes}
\affiliation{Department of Physics, Loughborough University, Epinal way, Loughborough, Leicestershire, LE11 3TU, UK}

\author[0000-0002-4891-3517]{George Zhou}
\affiliation{\usq}

\author[0000-0001-6534-6246]{Trevor J.\ David}
\affiliation{\flatiron}
\affiliation{\amnh}

\author[0000-0002-7733-4522]{Juliette Becker}
\affiliation{\DGPScaltech}

\author[0000-0002-8965-3969]{Steven Giacalone}
\affil{\berkely}

\author[0000-0001-7246-5438]{Andrew Vanderburg}
\affiliation{\wisconsin}

\author[0000-0001-8812-0565]{Joseph E. Rodriguez} 
\affiliation{\msu}

\author[0000-0001-6637-5401]{Allyson Bieryla} 
\affiliation{\cfa}

\author{Christopher Wirth}
\affiliation{Harvard University, Cambridge, MA 02138, USA.}
\affiliation{\cfa}

\author[0000-0002-6544-1523]{Shaun Atherton}
\affiliation{Department of Physics, Loughborough University, Epinal Way, Loughborough, Leicestershire, UK, LE11 3TU}

\author[0000-0002-3551-279X]{Tara Fetherolf}
\affiliation{\riverside}

\author[0000-0001-6588-9574]{Karen A.\ Collins}
\affiliation{\cfa}

\author[0000-0001-6534-6246]{Adrian M. Price-Whelan}
\affiliation{\flatiron}

\author[0000-0001-9907-7742]{Megan Bedell}
\affiliation{\flatiron}

\author[0000-0002-8964-8377]{Samuel N. Quinn} 
\affiliation{\cfa}

\author[0000-0002-4503-9705]{Tianjun Gan}
\affiliation{\Tsinghua}

\author{George R. Ricker} 
\affiliation{\MIT}

\author[0000-0001-9911-7388]{David W. Latham} 
\affiliation{\cfa}

\author{Roland K. Vanderspek}
\affiliation{\MIT}

\author[0000-0002-6892-6948]{Sara Seager}
\affiliation{\MIT}
\affiliation{\MITEPS}
\affiliation{\MITAA}

\author[0000-0002-4265-047X]{Joshua N. Winn} 
\affiliation{\princeton}

\author[0000-0002-4715-9460]{Jon M. Jenkins}
\affiliation{\ames}

\author[0000-0003-1001-0707]{Ren\'{e} Tronsgaard}
\affiliation{\dtu}

\author[0000-0003-1605-5666]{Lars A. Buchhave}
\affiliation{\dtu}

\author[0000-0003-0497-2651]{John F.\ Kielkopf} 
\affiliation{\louisville}

\author[0000-0001-8227-1020]{Richard P. Schwarz}
\affiliation{\Patashnick}

\author[0000-0001-8189-0233]{Courtney D. Dressing}
\affiliation{\berkely}

\author[0000-0002-9329-2190]{Erica J.\ Gonzales}
\nsfgf
\affiliation{\ucsc}
=
\author{Ian J.\ M.\ Crossfield} 
\affiliation{\kansas}

\author[0000-0003-0593-1560]{Elisabeth C. Matthews}
\affiliation{Observatoire de l’Universit\'e de Gen\`eve, Chemin Pegasi 51, 1290 Versoix, Switzerland}


\author[0000-0002-4625-7333]{Eric L.\ N.\ Jensen}
\affiliation{\swarthmore}

\author[0000-0001-9800-6248]{Elise Furlan}
\affiliation{\nexsci}

\author[0000-0003-2519-6161]{Crystal~L.~Gnilka}
\affil{\ames}

\author[0000-0002-2532-2853]{Steve~B.~Howell}
\affil{\ames}

\author[0000-0002-9903-9911]{Kathryn V. Lester}
\affil{\ames}

\author[0000-0003-1038-9702]{Nicholas~J.~Scott}
\affil{\ames}

\author[0000-0002-2457-7889]{Dax L. Feliz}
\affiliation{\vanderbilt}

\author[0000-0003-2527-1598]{Michael B. Lund}
\affiliation{\nexsci}

\author[0000-0001-5016-3359]{Robert J.\ Siverd}
\affiliation{\gemini}
\affiliation{}

\author[0000-0002-5951-8328]{Daniel J.\ Stevens} 
\eberly
\affiliation{\psu}
\affiliation{\psust}

\author[0000-0001-8511-2981]{N. Narita}
\affiliation{\Komaba}
\affiliation{\JSTAKomaba}
\affiliation{\astrobiojapan}
\affiliation{\iac}

\author[0000-0002-4909-5763]{A. Fukui}
\affiliation{\Komaba}
\affiliation{\iac}

\author[0000-0001-9087-1245]{F. Murgas}
\affiliation{\iac}
\affiliation{\lalaguna}

\author[0000-0003-0987-1593]{Enric Palle}
\affiliation{\iac}
\affiliation{\lalaguna}

\author[0000-0003-3936-4170]{Phil J. Sutton}
\affiliation{School of Mathematics and Physics, University of Lincoln, Brayford Pool Campus, Lincoln, LN6 7TS, UK}

\author[0000-0002-3481-9052]{Keivan G. Stassun} 
\affiliation{\vanderbilt}
\affiliation{\fisk}

\author[0000-0002-0514-5538]{Luke G. Bouma}
\affiliation{Department of Astrophysical Sciences, Princeton University, 4 Ivy Lane, Princeton, NJ 08540, USA}

\author{Michael~Vezie}
\affiliation{\MIT}

\author{Jesus Noel Villase{\~ n}or}
\affiliation{\MIT}

\author[0000-0003-1309-2904]{Elisa~V.~Quintana}
\affiliation{\gsfc}

\author[0000-0002-6148-7903]{Jeffrey~C.~Smith}
\affiliation{\seti}
\affiliation{\ames}

\begin{abstract}
We report the discovery of two planetary systems around comoving stars; \targa (\tica) and \targb (\ticb). \targa is a nearby (41.9 pc) multi-planetary system orbiting a young (\agea), bright (K = 7.115 in TIC v8.1). \targb hosts a single transiting planet, and is similarly nearby (42.58 pc), similarly young (\ageb), and bright. Both targets exhibit significant, periodic variability due to star spots, characteristic of their young ages. Using photometric data collected by \tess we identify three transiting planets around \targa with radii of $R_b$=3.3$\pm$0.04 $R_{\earth}$, $R_c$=4.4$\pm$0.05 $R_{\earth}$, and $R_d$=4.1$\pm$0.07 $R_{\earth}$. Planet \planb{} has a period of $P_b$=10.356 d. For both TOI 2076c and d, \tess observed only two transits, separated by a 2-year interval in which no data were collected, preventing a unique period determination. A range of long periods (>17d) are consistent with the data. We identify a short-period planet around \targb with a radius of $R_b$=1.8$\pm$0.04 $R_{\earth}$ and a period of $P_b$=0.549 d. Their close proximity, and bright, cool host stars, and young ages, make these planets excellent candidates for follow-up. \planbtwo{} is one of the best known small ($R<2$ $R_\Earth$) planets for characterization via eclipse spectroscopy and phase curves with JWST. \planbtwo{} is the youngest ultra-short period planet discovered to date, providing valuable constraints on formation time-scales of short period planets. Given the rarity of young planets, particularly in multiple planet systems, these planets present an unprecedented opportunity to study and compare exoplanet formation, and young planet atmospheres, at a crucial transition age for formation theory. 

\end{abstract}

\section{Introduction}

A primary aim of exoplanetary science is to use the observed properties of planetary systems to constrain theoretical models of planet formation (that which occurs in the protoplanetary disk) and evolution (what occurs after disk dispersal). This problem is approached in a number of ways: by forward modeling of the formation and evolution processes and comparison between simulated and observed exoplanet populations \citep[``planet population synthesis,'' e.g.][]{Mordasini2009}, through measuring the dependence of planet occurrence rates on fundamental stellar properties such as mass \citep[e.g.][]{Howard2012, Yang2020}, metallicity \citep[e.g.][]{Fischer2005, Petigura2018}, or multiplicity \citep[e.g.][]{Wang2014a,Wang2014b}, and via case studies of individual systems that challenge conventional wisdom about the planet formation process \citep[e.g.][]{Carter2012, LopezFortney2013}.

Young exoplanets ($<$1~Gyr) are particularly useful for case studies, as they have had less time to evolve and may therefore have properties that more closely resemble their initial conditions. In older planetary systems, disentangling the effects of planet formation from those of subsequent evolution becomes a more challenging task. However, of the more than 3,300 transiting exoplanets confirmed to date, fewer than 60 (2\%) have securely determined ages $<1$~Gyr.\footnote{NASA Exoplanet Archive \citep{Akeson2013}, accessed in March 2021.} Thus, there is value in identifying and characterizing young planets. This can be done through careful characterization of previously known exoplanet hosts or through targeted planet searches in samples of known young stars \citep[e.g.][]{Nardiello2020, Battley_2020}.

For transiting exoplanets, which are the focus of this work, examining the time-dependence of the planet radius distribution can yield insights into evolutionary processes and timescales. For example, the discovery of a gap in the radius distribution of close-in ($P<$100~d), low-mass ($M_P<$100~$M_\oplus$) exoplanets \citep{CKS} has fueled speculation about its origins. Theoretical studies have demonstrated that a radius gap may result from (1) late-time formation in a gas-poor disk \citep{Lopez2018, Lee2021} or (2) post-formation atmospheric loss via stellar high-energy radiation \citep[``photoevaporation''][]{OwenWu2013, LopezFortney2013}, the luminosity of the planet's cooling core \citep[``core-powered mass loss,''][]{Ginzburg2018, Gupta19, Gupta20}, or impacts \citep{Inamdar2016, Wyatt2020}. In each of these theories the radius gap emerges and evolves on different timescales. Notably, the theoretical models mentioned also predict larger sizes for sub-Neptunes at earlier times (particularly in the first 100~Myr). Efforts to age-date known exoplanet host stars are providing emerging evidence that the size distribution of small planets continues to evolve over billions of years \citep{Berger2020, Sandoval2020} and that the precise location of the radius gap evolves on similar timescales \citep{David2020}. These results are broadly consistent with expectations from the photoevaporation and core-powered mass-loss models. However, the age of any individual field star typically carries large uncertainties. 

The TESS mission \citep{TESS} provides a new opportunity for targeted searches of young exoplanets from precise time-series photometry for millions of targets across most of the sky. For example, the THYME survey has identified several planets in known young associations spanning a diversity of Galactic environments, such as the Tucana-Horologium and Ursa Major moving groups \citep{Newton2019, thyme2}, the Scorpius-Centaurus OB association \citep{thyme1}, the Pisces-Eridanus stream \citep{thyme3}, and even a previously unknown association \citep{Tofflemire2021}. Other searches of TESS data have revealed planets orbiting young stars in the IC 2602 cluster \citep{Bouma2020} and in the field \citep{Zhou2021}.

Here we present the discovery of two young planetary systems; first, a system of three exoplanets orbiting a bright ($K = 7.115$), K-type variable star \targa (\tica), and second a single short period exoplanet orbiting its similarly bright comoving companion \targb (\ticb).  Stellar parameters for these targets are given in Tables~\ref{tbl:LitProps} and ~\ref{tab:seds}, and planet parameters derived for each planet are given in Table~\ref{tab:TOI-2076} and ~\ref{tab:TOI-1807}. We derive ages of \agea and \ageb for \targa and \targb respectively. With its bright magnitude and close proximity of \dista in \emph{Gaia} DR2 \citep{Gaia:2018}, \targa presents an rare opportunity to characterise a range of small radius planets orbiting a young, active star. Owing to their young ages, both \targa and \targb are excellent candidates for studying the atmospheres of close in planets existing around the transition age where photo-evaporation is theorised to cease. Among short-period, small planets, \targb is one of the most amenable to phase curve and eclipse spectroscopy.

Section \ref{sec:obs} of this paper discusses the TESS observations of \targa and \targb. Section \ref{sec:correction} discusses our corrections to the TESS light curves to obtain more precise photometry, and the model fit of the stellar SEDs and planet transits. Sections~\ref{sec:validation} and ~\ref{sec:ground} discusses the statistical validation of all planets in these two systems and ground based follow-up. In Section~\ref{sec:age_estimate} we discuss our age estimates for both targets. We conclude in Section~\ref{sec:discussion} with a discussion of the importance of \targa and \targb to the community, and demonstrate that each of these planets is an excellent candidate for further atmospheric follow-up.

\begin{table*}
\small
\setlength{\tabcolsep}{2pt}
\centering
\caption{Literature and Measured Properties for TOI-1807 and TOI-2076}
\begin{tabular}{llccc}
  \hline
  \hline
Other identifiers\dotfill \\
& & TOI-1807 & TOI-2076 \\
& & TIC 180695581& TIC 27491137 \\
& & HIP 65469& ---\\
& & TYC 3025-00731-1& TYC 3036-00481-1\\
\hline
\hline
Parameter & Description & Value & Value & Source\\
\hline 
$\alpha_{J2000}$\dotfill	&Right Ascension (RA)\dotfill &13:25:07.9959&14:29:34.2428&1\\
$\delta_{J2000}$\dotfill	&Declination (Dec)\dotfill &+38:55:20.9460&+39:47:25.5450&1\\
&                \\
${\rm G}$\dotfill     & Gaia $G$ mag.\dotfill     &9.68$\pm$0.02&8.91$\pm$0.02&1\\
B$_{\rm P}$\dotfill			&Gaia B$_{\rm P}$ mag.\dotfill & 10.26$\pm$0.02&9.37$\pm$0.02&1\\
R$_{\rm P}$\dotfill			&Gaia R$_{\rm P}$ mag.\dotfill & 8.99$\pm$0.02&8.33$\pm$0.02&1\\
${\rm T}$\dotfill     & TESS mag.\dotfill     & 9.036$\pm$0.006& 8.375$\pm$0.006& 2 \\
&                \\
J\dotfill			& 2MASS J mag.\dotfill & 8.103$\pm$ 0.023&7.613$\pm$0.020& 3	\\
H\dotfill			& 2MASS H mag.\dotfill & 7.605$\pm$ 0.020&7.188$\pm$0.027& 3	\\
K$_S^{\ddagger}$\dotfill			& 2MASS ${\rm K_S}$ mag.\dotfill &---&7.115$\pm$0.020& 3	\\
&                \\
\textit{WISE1}\dotfill		& \textit{WISE1} mag.\dotfill & 7.395 $\pm$ 0.03 &7.01$\pm$0.05& 4,5	\\
\textit{WISE2}\dotfill		& \textit{WISE2} mag.\dotfill & 7.508 $\pm$ 0.03 &7.13$\pm$0.03& 4,5	\\
\textit{WISE3}\dotfill		& \textit{WISE3} mag.\dotfill & 7.445 $\pm$ 0.051 &7.09$\pm$0.03& 4,5	\\
\textit{WISE4}\dotfill		& \textit{WISE4} mag.\dotfill & 7.368 $\pm$ 0.115 &7.0$\pm$0.1& 4,5	\\
&                \\
B$_T^\star$\dotfill			&Tycho B$_T$ mag.\dotfill & 11.385$\pm$0.053&10.258$\pm$0.025& 6	\\
V$_T^\star$\dotfill			&Tycho V$_T$ mag.\dotfill & 10.110$\pm$0.027&9.238$\pm$0.015& 6	\\
&                \\
$\mu_{\alpha}$\dotfill		& Gaia DR2 proper motion\dotfill & -124.713 $\pm$ 0.027 &-118.228$\pm$0.036& 1 \\
                    & \hspace{3pt} in RA (mas yr$^{-1}$)	&&                \\
$\mu_{\delta}$\dotfill		& Gaia DR2 proper motion\dotfill 	&  -27.377 $\pm$ 0.039 &-6.973$\pm$0.048& 1 \\
                    & \hspace{3pt} in DEC (mas yr$^{-1}$) &  &                \\
$\pi^\dagger$\dotfill & Gaia Parallax (mas) \dotfill & 23.4877 $\pm$ 0.0423$^{\dagger}$ &23.86220$\pm$0.0384&  1 \\
&                \\
$P_{\rm Rot}$\dotfill & Rotation Period (days)\dotfill &$7.27\pm0.23$&$8.670 \pm 0.048$& \S\ref{sec:gyro} \\
\hline
\end{tabular}
\begin{flushleft}
 \footnotesize{ \textbf{\textsc{NOTES:}}
 The uncertainties of the photometry have a systematic error floor applied. \\
 $\dagger$Values have been corrected for the 30 $\mu$as offset as reported by \citet{Lindegren:2018}. \\
 $\ddagger$The 2MASS K-band value for TOI-1807 had a reported value and uncertainty of 7.56 +/-9.995 mag. Given questions about its reliability, we exclude it from our analysis and this table.\\
 $\star$Not included in the SED analysis.\\
Sources are: $^1$\citet{Gaia:2018},$^2$\citet{TIC},$^3$\citet{Cutri:2003}, $^4$\citet{Cutri2012}, $^5$\citet{Zacharias:2017},$^{6}$]\citet{Hog:2000}\\
}
\end{flushleft}
\label{tbl:LitProps}
\end{table*}

\section{Observations}
\label{sec:obs}
\subsection{TESS Photometry}
\label{sec:photometry}
\targa was observed twice by TESS, once by camera 4 during Sector 16 (11$^{\text{th}}$ September - 07$^{\text{th}}$ October 2019) and then again by camera 2 during Sector 23 (18$^{\text{th}}$ March - 16$^{\text{th}}$ April 2020). \targb was observed in Sector 22 (19$^{\text{th}}$ February 2020 to 17$^{\text{th}}$ March 2020) and Sector 23. Both targets were observed in two minute cadence mode. The literature properties of both targets are shown in Table~\ref{tbl:LitProps}.


\subsubsection{By Eye Search}

\targa was first identified by a student-led, by-eye search. Our by-eye search method was as follows; we downloaded two minute cadence TESS Target Pixel Files (TPFs) for Sector 16 that had been calibrated by the TESS Science Processing Operations Center (SPOC) pipeline, and summed pixels within the pipeline provided aperture to create Simple Aperture Photometry light curves. Outliers were then rejected using a standard deviation of $10\sigma$. Stellar variability was subtracted using the flatten tool from the Python package \texttt{lightkurve}\footnote{\url{https://github.com/keplerGO/lightkurve}}, which applied a Savitsky Golay filter over a 1001 cadence window to remove long term trends on time-scales of 1.5 days. The resulting light curve was plotted and visually inspected. Over 500 targets were processed before \targa was identified as an interesting candidate using Sector 16 data on March 8th 2020.\par

\begin{figure}
    \centering
    \includegraphics[width=0.5\textwidth]{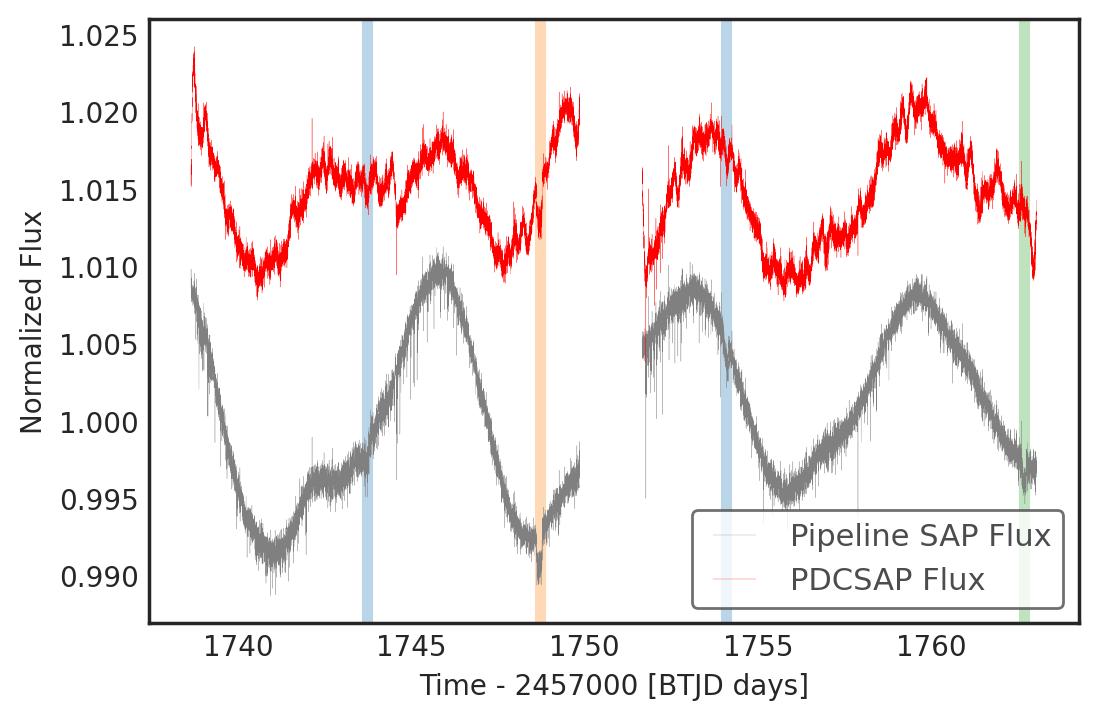}
    \caption{Example of the \tess pipeline PDCSAP Flux from the for Sector 16 of \targa{}. Simple Aperture Photometery (SAP) provided by the pipeline is shown in grey, PDCSAP flux is shown in red. The transits of \planb{}, \planc{}, and \pland{} are highlighted in blue, orange and green respectively. The PDCSAP flux introduces spurious noise that is easily confused with true transiting signals, which hampers detection and modeling efforts when using the PDCSAP processing. \targb{} does not suffer from this issue. Since this work, a reprocessing has become available for PDCSAP flux which remedies this spurious noise.}
    \label{fig:pdcsap}
\end{figure}

\begin{figure*}
    \centering
    \includegraphics[width=\textwidth]{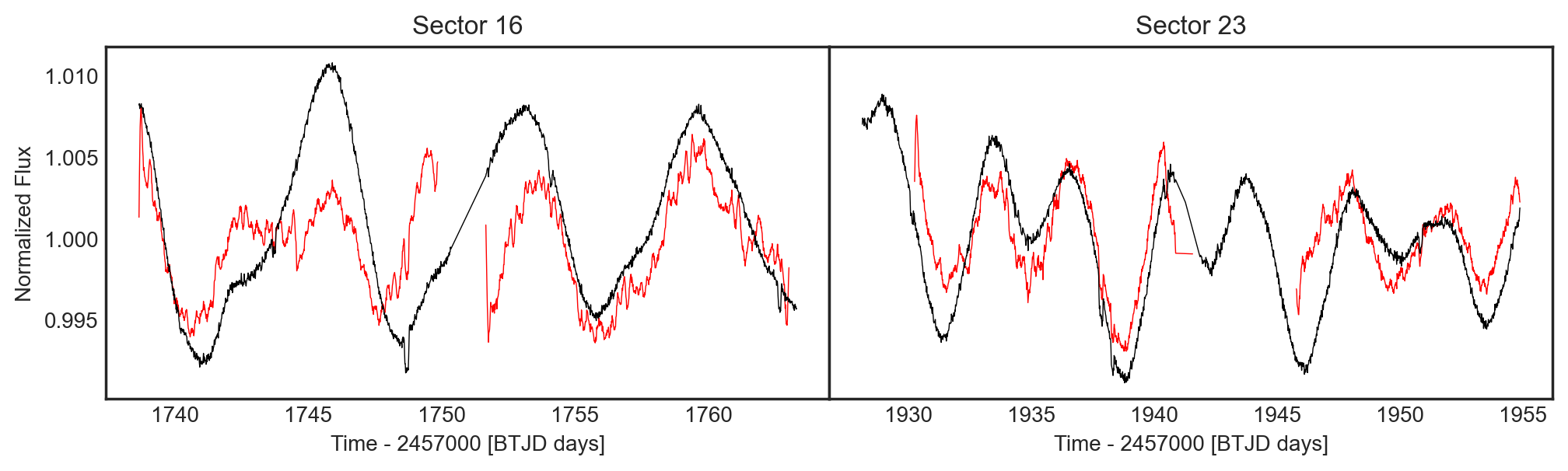}
    \includegraphics[width=\textwidth]{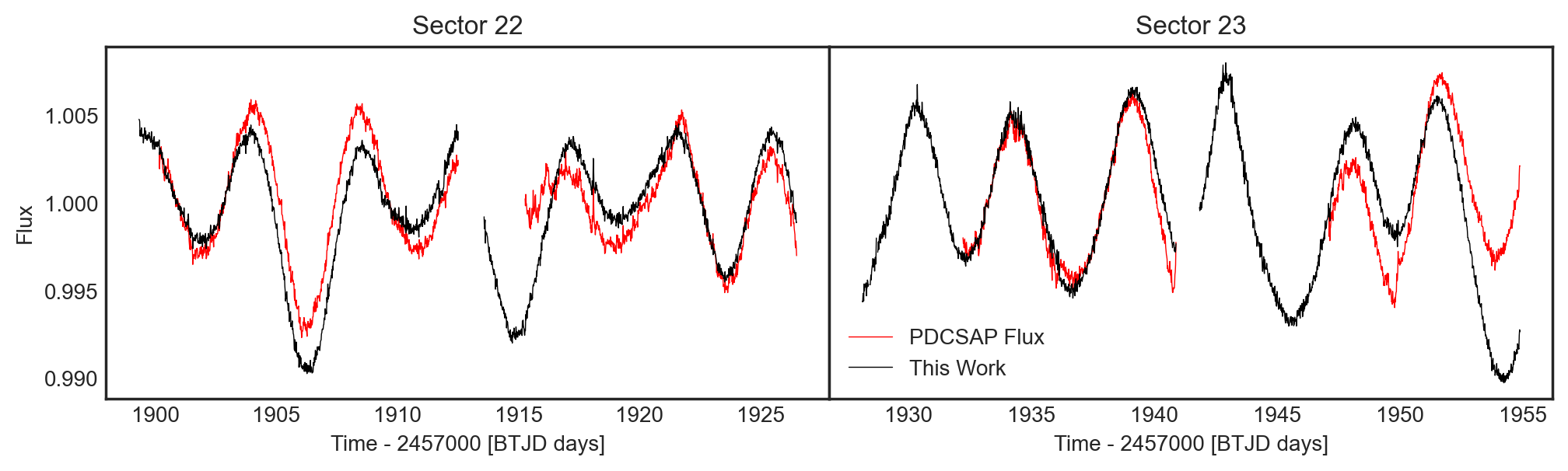}
    \caption{Light curves of TESS photometry for \targa and \targb. Top: TESS Photometry for \targa. Bottom: TESS Photometry for \targb. Black points show data corrected by the method discussed in Section~\ref{sec:correction}, red points show TESS pipeline PDCSAP flux, available during February 2020. In the case of \targa the pipeline correction suffers from spurious signals, reducing the signal to noise ratio of the data. A new reprocessing of the data has since been made available, which greatly improves the detrending of the light curve, and preserves the transits and stellar activity. We include the original data here, as an illustration of why our by-eye search was successful in extracting \pland{}. The light curves we describe in Section~\ref{sec:correction} also preserve more data close to the data downlink, owing to our bespoke background light correction. Data has been binned to a cadence of 20 minutes for clarity. }
    \label{fig:lightcurves}
\end{figure*}

The TESS Pipeline-processed image data for \targa was accessed by our team in February 2020. The pipeline processed Pre Data-search Conditioned Simple Aperture Photometry (PDCSAP) photometry available at that time for \targa suffered from spurious, semi-periodic signals with durations on the order of ~0.59 days, which is at timescales and amplitudes comparable to the planet transits. This ultimately adversely affected the planet transit search and planet modeling efforts (see Figure~\ref{fig:pdcsap}). By performing a by-eye search of Simple Aperture Photometry (SAP) flux generated from the TPFs, with no systematics corrections applied, our team was able to identify three, high signal to noise transiting objects in the Sector 16 data. We use the techniques described in Section~\ref{sec:correction} to detrend the SAP flux derived from the TESS products and improve precision before fitting the transits in the data. We later identified \targb{} as a comoving target also in the TOI list (see Section~\ref{sec:comoving}. Our processed light curves for \targa and \targb are shown in Figure ~\ref{fig:lightcurves}, alongside the PDCSAP flux that was originally obtained. Figure ~\ref{fig:lightcurves} shows that, particularly in the case of \targa, there is an increase in spurious noise, which hampered pipeline detection efforts. Since our access of the data, the TESS pipeline data has been reprocessed, and the new publicly available PDCSAP light curves show greatly improved the correction. We include the original PDCSAP light curves for illustration in Figure ~\ref{fig:pdcsap}, compared to the pipeline provided SAP flux.

\subsubsection{TESS Mission Transit Detections}

The SPOC transit search pipeline \citep{jenk1, jenk2} detected the  transit of \planbtwo{} in March 2020 with a period of 0.55 d and a radius of 1.52 ± 0.94 R$_{\earth}$. A limb-darkened transit model was fitted to the light curve \citep{Li:DVmodelFit2019} and a suite of diagnostic tests were performed to establish the planetary nature of the signal \citep{Twicken:DVdiagnostics2018}. \planbtwo{} passed all the tests, including the odd/even depth test, the ghost diagnostic test, and the difference image centroiding test, which located the source of the transit signature to within 4.9+/-2.7 arcseconds of the target star. This was further reduced to 3.1+\-2.7 arc sec in the multi-sector search of the combined light curves from sectors 16 and 23 conducted in May  2021. The TESS Science Office vetted the data validation results and issued an alert for TOI-1807b on 15 April 2020.
 
\planb{} was detected by the SPOC pipeline in a search of the sector 16 data on 16 March 2020 at a period of 10.357 d and a radius of 2.57 $R_\oplus$. It passed all the data validation tests and the difference image centroid test located the transit source to within 2.44+/-2.6 arcsec.  The transit signature of \planc{} was detected at a period of 33.69 d with a radius of 4.3 R$_{\earth}$, but was clearly a single transit detection according to the DV report. The difference imaging centroid test indicated that the transit source for \planc{} was located within 1.634+/-2.7 arcsec. Alerts for \planb{} and \planc{} were issued on 15 July 2020.

\subsection{Comoving Targets}
\label{sec:comoving}
\targa and \targb were identified as a comoving pair of stars by \cite{comovedisc} because, after accounting for geometric projection, their proper motions are consistent with having the same three-dimensional velocity. Using updated astrometry and radial velocity data from Gaia EDR3, the stars have a mean heliocentric distance of 42.3 pc, a physical separation of 9.2 pc, and an angular separation of $\sim 12.5^\circ$. While the stars have a proper motion difference of $\sim 21.5~{\rm mas}~{\rm yr}^{-1}$, this is largely due to their large angular separation: the 3D velocity difference between the stars is only $\sim 0.6~{\rm km}~{\rm s}^{-1}$ (5th and 95th percentile of 0.39 and 1.58${\rm km}~{\rm s}^{-1}$ respectively). Even though recent Gaia data confirm that these stars are comoving, their large physical separation suggests that these objects are not a bound wide binary, but could instead be part of a small moving group. The shared formation history, (indicated by their three-dimensional velocity and similar ages), and similar stellar parameters of \targa and \targb make them a further interesting laboratory for testing planet formation theory.


\section{Data Analysis}
\label{sec:correction}
After identifying \targa as a planet host by eye, and \targb as a comoving planet host among the public TOI list, we perform the following analysis to extract the planet parameters. In this analysis we use the \texttt{lightkurve} Python package to create Simple Aperture Photometry (SAP) light curves of \targa from the TESS SPOC pipeline \citep{tesspipeline} Target Pixel Files (TPFs). Sky background light from Earth is a significant systematic in \tess, which the pipeline corrects in TPF products. In this work, we use TPFs without background subtraction, since the SPOC pipeline masks cadences where the background is estimated to be severe, leading to data loss. Instead, we perform a bespoke background correction that includes these cadences, in order to preserve the most time-series data. This correction is discussed in Section ~\ref{sec:tpfs}.





\subsection{Light curve creation}
\label{sec:tpfs}

\begin{figure*}
    \centering
    \includegraphics[width=0.49\textwidth]{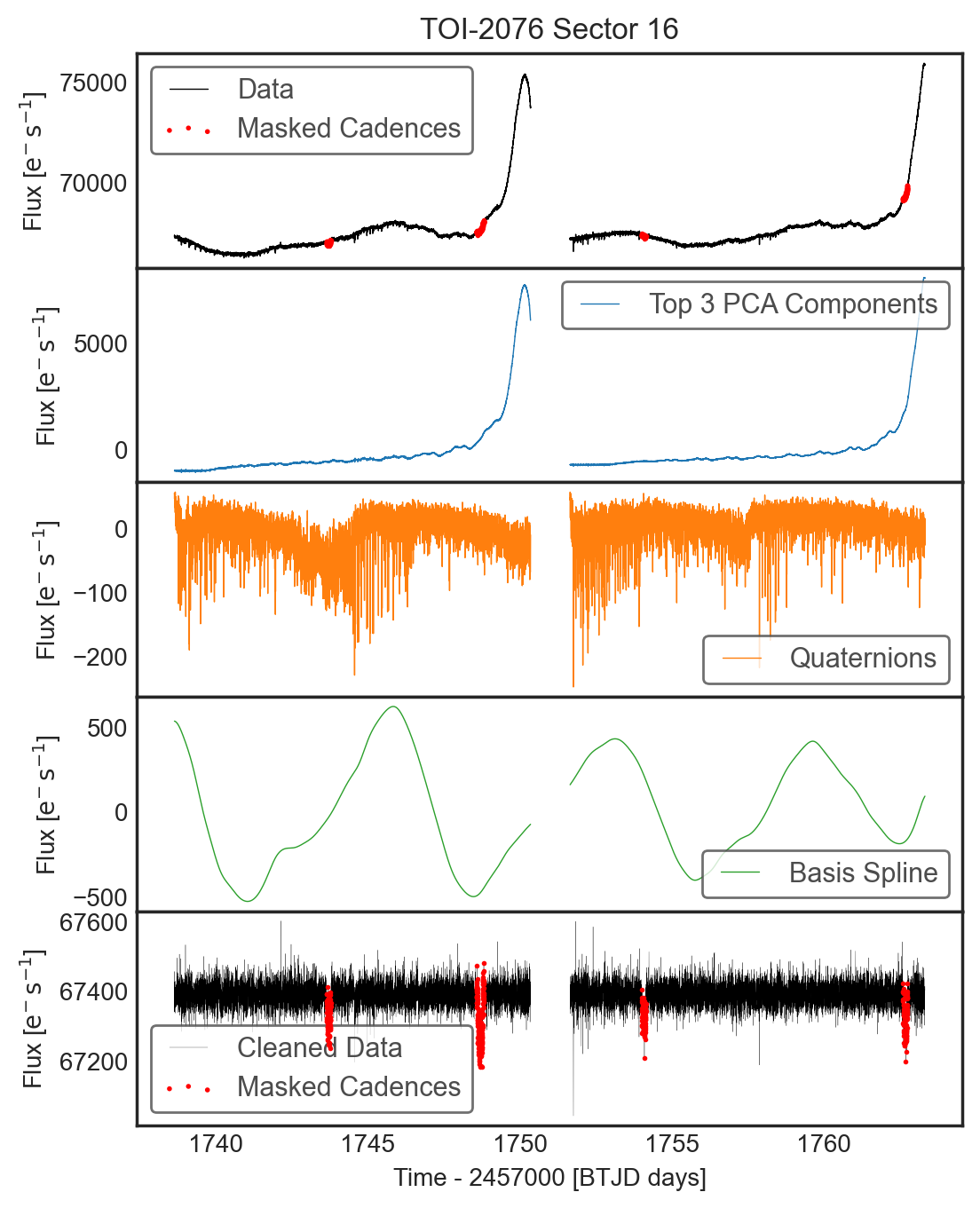}
    \includegraphics[width=0.49\textwidth]{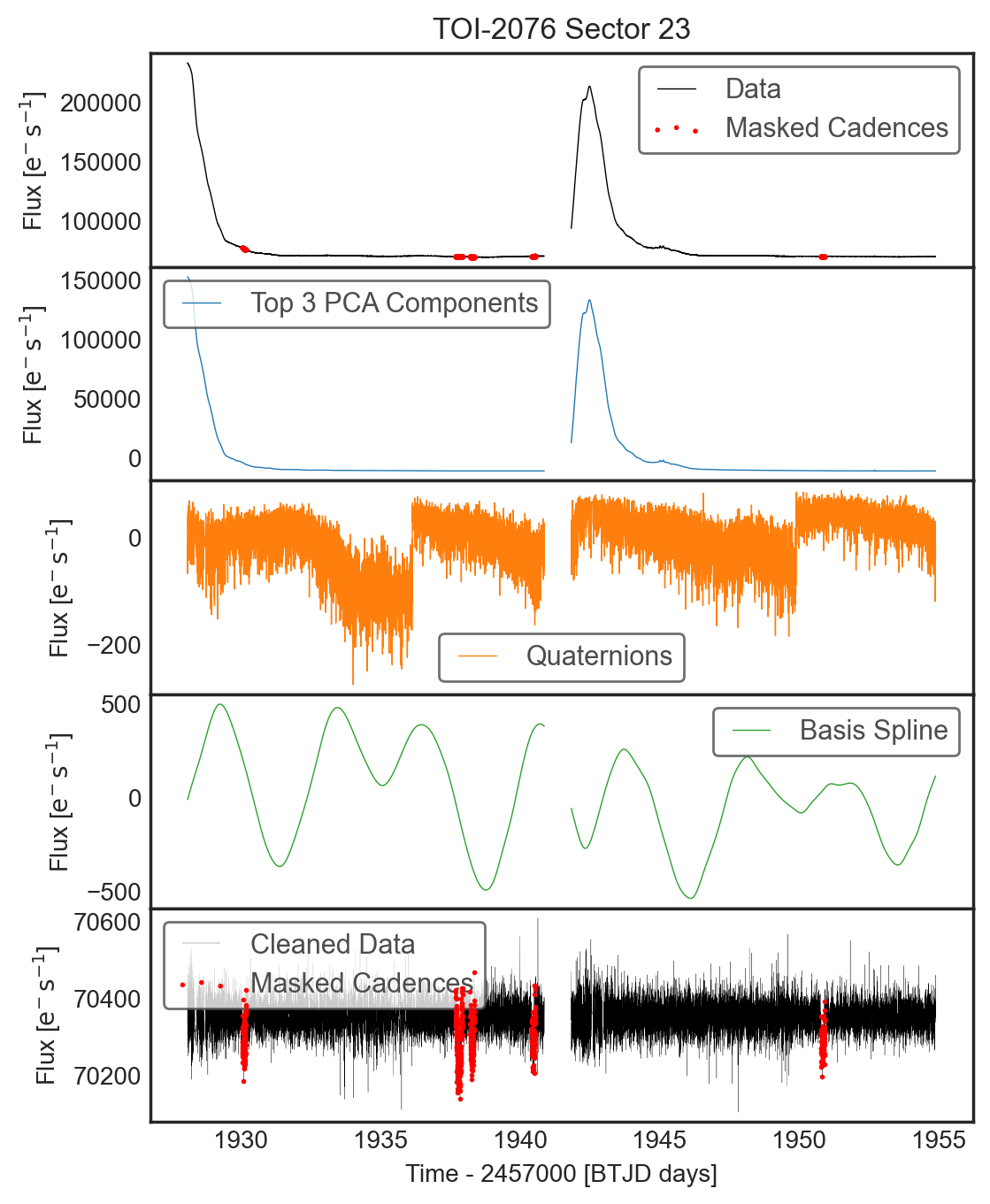}
    \includegraphics[width=0.49\textwidth]{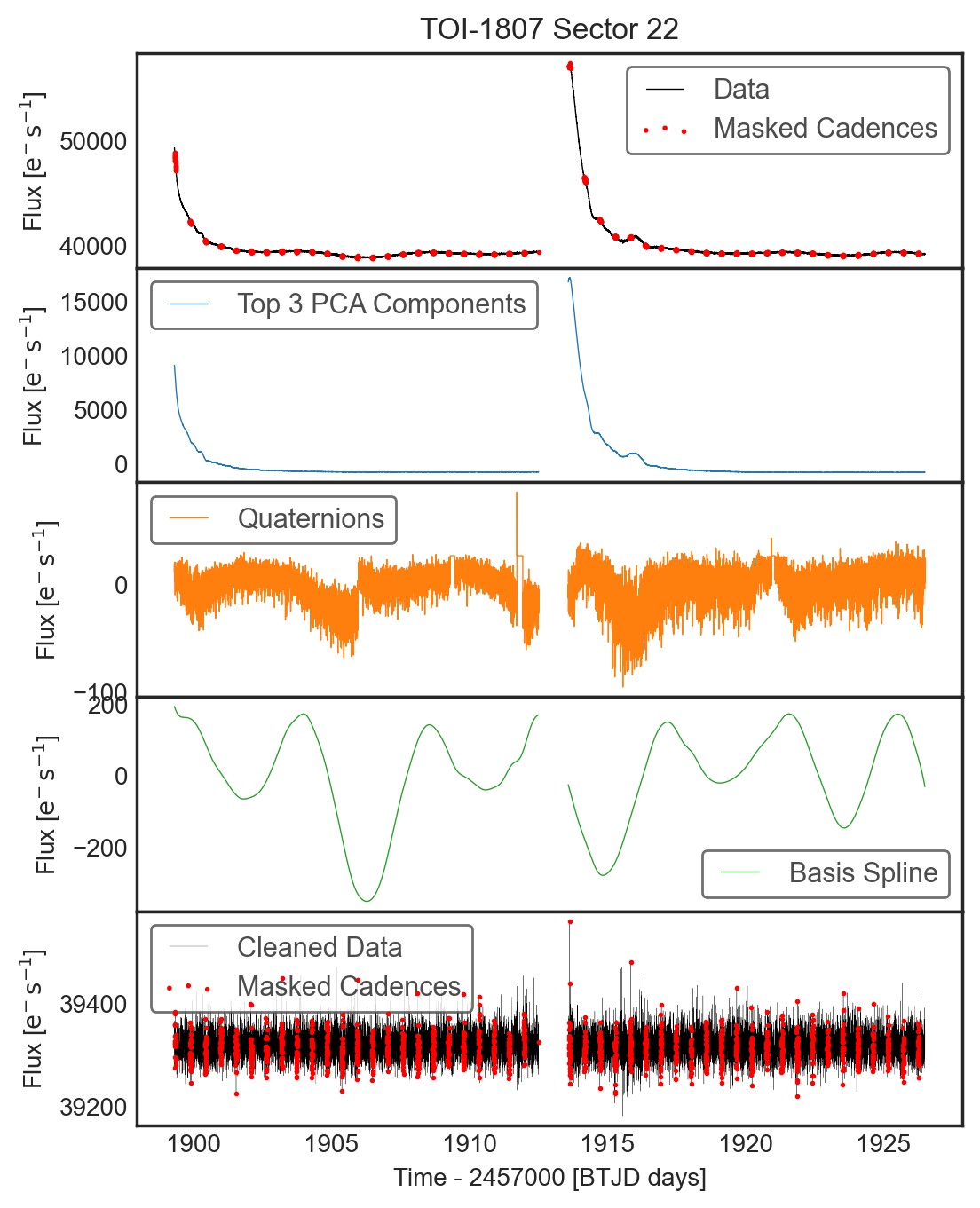}
    \includegraphics[width=0.49\textwidth]{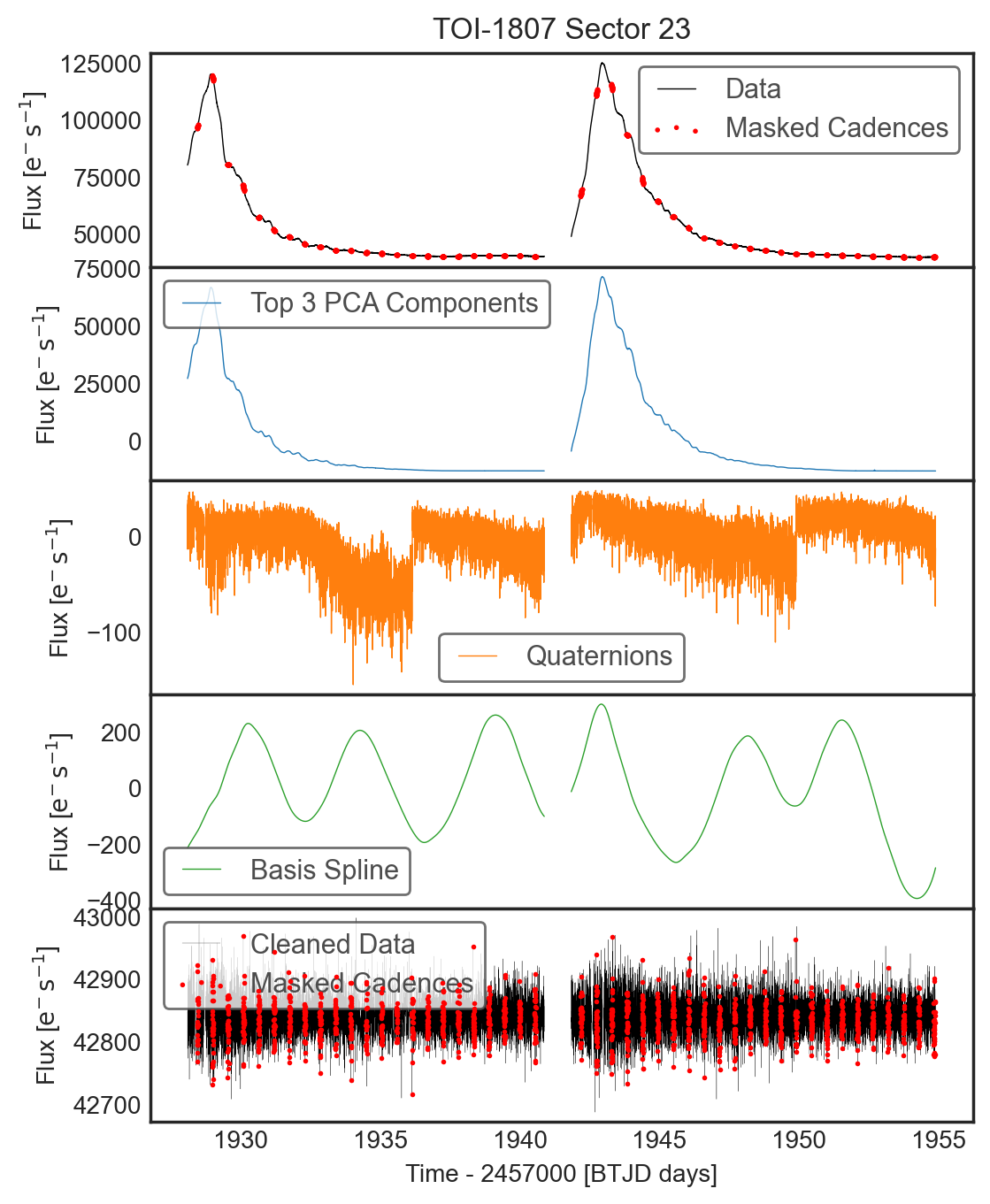}
    \caption{The results from our light curve detrending procedure described in Section ~\ref{sec:tpfs}. Top: \targa in Sector 16 and 23. Bottom: \targb in Sector 22 and Sector 23. Top panels show the original raw data, with no sky background light removed. Red points show cadences containing transits that are masked during our model fit. (\targb{} is a short period planet, and so has many cadences masked). The next three panels show the best fit of each component in our model; 1) the top components from singular value decomposition (SVD) of pixels outside the aperture to fit scattered light background 2) the best fit mean and standard deviation of the quaternions in within a cadence 3) a basis spline to fit the stellar variability. Our resultant light curves are in the final panel. All three of our model components are fit simultaneously.}
    \label{fig:detrending}
\end{figure*}

We create light curves for \targa and \targb using the following procedure. The results of this procedure are shown in Figure~\ref{fig:detrending}.
\begin{itemize}
    \item Using our basic, mean-normalized SAP flux light curves from Section~\ref{sec:photometry}, we estimate periods, transit mid-points and durations for each transiting planet.
    \item We use the TESS Pipeline TPF products for \targa in Sector 16 and 23, and for \targb from Sector 22 and 23, conservatively removing cadences where the quality flags are consistent with "Coarse Point", "Desaturation", or "Argabrightening" (flags 4, 32 and 16) which cause significant outliers. TPFs are delivered with a background light estimate subtracted by the pipeline. We use the \texttt{FLUX\_BKG} keyword in the TPF FITS files to add the TESS Pipeline background correction back into the TPF (resulting in uncorrected, but calibrated TPFs). As discussed above, (this enables us to perform a bespoke background correction, and preserve more data that the pipeline flags as poor quality close to the data downlink).
    \item We build light curves from the TPFs using the pipeline provided apertures. Since these stars are isolated and the \tess\ pipeline estimates that more than 99.9\% of the light in the apertures comes from the target stars (based on the pipeline's crowding metric), contamination from background sources is negligible, and we do not apply a dilution correction. 
    \item We detrend these light curves to remove the background signal, using \texttt{lightkurve}'s \texttt{RegressionCorrector} tool. We model the light curve as a linear combination of 1) the top 3 components of the pixels outside the aperture using singular value decomposition (SVD)
    2) A vector containing i) the mean and ii) the standard deviation of each of the three quaternions (available in the \tess engineering data, see \citealt{vanderburg2019}) during each individual \tess exposure to account for \tess jitter. 3) A basis spline with 80, evenly space knots between the start and end of the sector to capture the stellar variability. We fit this model, using Gaussian priors, masking out cadences that we expect to contain transiting planet signals. 
\end{itemize}

This procedure results in light curves with long term stellar variability removed, while transits remain intact in the dataset. Using the \texttt{estimate\_cdpp} method from \texttt{lightkurve} we estimate the photometric precision of all the light curves to determine the improvement in precision we obtain. The official \tess\ pipeline computes the Combined Differential Photometric Precision (CDPP) metric using a wavelet-based algorithm to calculate the signal-to-noise of the specific waveform of transits of various durations \citep[see][]{cdpp}. In the \texttt{lightkurve} implementation, we use the simpler “sgCDPP proxy algorithm” discussed by \cite{gilland} and \cite{vancleve}. Using this estimate the PDCSAP light curves available in 2020 for \targa and \targb have a sgCDPP of 100 and 164 respectively for a 1 hour transit duration in parts per million (PPM). The procedure we describe here reduces the sgCDPP to 82 and 86 PPM respectively, which indicates a significant reduction in noise. Having improved the precision of the light curves, we re-searched both \targa and \targb light curves to search for any shallower transiting signals using a simple Box Least Squares (BLS), but find no evidence of additional planets.

We use \texttt{lightkurve} and \texttt{astropy} to perform a basic Box Least Squares (BLS) search for transiting signals in the light curves of both targets. We identify 3 transiting objects around \targa with periods of 10.35d, 17.19d and 25.08d, and transit depths of 913$\pm$19ppm, 1906$\pm$28ppm, and 1181$\pm$32. \planb{} transits 4 times during Sector 16 and Sector 23. \planc{} and \pland{} transit once in each in Sector 16, and once each in Sector 23. We identify a single transiting object around \targb during Sector 22 and Sector 23. Using a simple BLS, \planbtwo{} has a period of 0.55d and a transit depth of 271$\pm$11. 

\subsection{Spectroscopic Stellar Parameters}
\label{sec:spectra}

In order to refine the stellar parameters upon which the planetary parameters depend, we fit the stellar spectra and stellar spectral energy distributions (SEDs) for \targa and \targb.

We obtained two reconnaissance spectra of \targa on UT2020-02-20 and UT2020-02-24, using the 1.5\,m Tillinghast Reflector Echelle Spectrograph \citep[TRES;][]{phdthesis} located at the Fred Lawrence Whipple Observatory (FLWO) in Arizona, USA. For \targb, we obtained two spectra on UT2020-05-31 and UT2020-07-01 with the FIbre-fed Echelle Spectrograph \citep[FIES;][]{Telting2014} at the 2.56\,m Nordic Optical Telescope (NOT) in La Palma, Spain, and another spectrum with TRES on UT2020-07-19. TRES has a resolving power of $R\sim$44,000 with wavelength coverage from 3860-9100\,\AA{}, while FIES offers a resolution of $R\sim$67,000 and covers the range 3760-8220\,\AA.

All the spectra are extracted as described in \cite{Buchhave2010}. We derive stellar parameters using the Stellar Parameter Classification tool \citep[SPC,][]{Buchhave2012,Buchhave2014}. SPC compares an observed spectrum against a grid of synthetic spectra based on Kurucz atmospheric models \citep{kurucz}. We analyze each spectrum independently to obtain the effective temperature ($T_\textrm{eff}$), surface gravity ($\log(g)$), metallicity ([m/H], a solar mix of metals rather than Fe alone), and projected rotational velocity ($v$sin$i$). The individually derived parameters agree to within their respective uncertainties, and we report their weighted average: 
\targa has $T_\textrm{eff}$=5227$\pm$50K, $\log(g)$=4.56$\pm$0.10, [m/H]=-0.15$\pm$0.08.
\targb has $T_\textrm{eff}$=4830$\pm$50K, $\log(g)$=4.65$\pm$0.10, [m/H]=-0.09$\pm$0.08, $v$sin$i$=4.3$\pm$0.5 km/s. These values are derived from spectra alone. 
These estimates are used to inform our SED fit in \S \ref{sec:stars}.

\subsection{Spectral Energy Distribution}
\label{sec:stars}
To determined the properties of both host stars, we perform a Spectral Energy Distribution (SED) fit of the broadband photometry from Gaia DR2  \citep{Gaia:2018}, 2MASS \citep{Cutri:2003}, and WISE \citep{Cutri2012, Zacharias:2017} using the publicly available exoplanet fitting suite, \texttt{EXOFASTv2} \citep{Eastman:2013, Eastman:2019}. We place a Gaussian prior on the Gaia DR2 parallax of 23.862$\pm$0.0384 mas for TOI-2076 and 23.488$\pm$0.042 mas for TOI-1807, which have been corrected for the known offset as described in \citet{Gaia:2018}. We also place Gaussian priors on the metallicities determined by analyzing the TRES spectra (see \S \ref{sec:spectra}) and host star ages (0.188$\pm$0.053 Gyr for TOI-2076 and 0.17$\pm$0.04 Gyr for TOI-1807; see \S\ref{sec:gyro}). Using the galactic dust maps from \citet{Schlegel:1998} \& \citet{Schlafly:2011}, we place upper limits on the line of sight extinction of 0.02635 mag (TOI-2076) and 0.0313 mag (TOI-1807). The resulting best fit parameters and the 68\% confidence intervals are shown in Table \ref{tab:seds}.

\begin{table}
\scriptsize
\setlength{\tabcolsep}{2pt}
\centering
\caption{Median values and 68\% confidence interval for global model of TOI-2076 and TOI-1807. \ch{These values are derived through SED fitting.}}
\begin{tabular}{llccc}
  \hline
  \hline
Parameter & Units & Values & Values \\
\hline
&&TOI-2076 & TOI-1807\\
\multicolumn{2}{l}{Stellar Parameters:}\\
$M_*$\dotfill &Mass (\msun)\dotfill &$0.850^{+0.025}_{-0.026}$&$0.750^{+0.025}_{-0.024}$\\
$R_*$\dotfill &Radius (\rsun)\dotfill &$0.761\pm0.016$&$0.680\pm0.015$\\
$L_*$\dotfill &Luminosity (\lsun)\dotfill &$0.3777^{+0.0094}_{-0.0092}$&$0.2135\pm0.0053$\\
$F_{Bol}$\dotfill &Bolometric Flux (cgs)$\times10^{-9}$\dotfill &$6.88\pm0.17$&$3.769\pm0.092$\\
$\rho_*$\dotfill &Density (cgs)\dotfill &$2.72^{+0.17}_{-0.16}$&$3.36^{+0.23}_{-0.21}$\\
$\log{g}$\dotfill &Surface gravity (cgs)\dotfill &$4.605^{+0.018}_{-0.019}$&$4.648^{+0.021}_{-0.020}$\\
$T_{\rm eff}$\dotfill &Effective Temperature (K)\dotfill &$5187^{+54}_{-53}$&$4757^{+51}_{-50}$\\
$[{\rm Fe/H}]$\dotfill &Metallicity (dex)\dotfill &$-0.032^{+0.048}_{-0.047}$&$-0.029^{+0.061}_{-0.043}$\\
$[{\rm Fe/H}]_{0}$\dotfill &Initial Metallicity$^{1}$ \dotfill &$-0.069\pm0.054$&$-0.065^{+0.062}_{-0.053}$\\
$Age$\dotfill &Age (Gyr)\dotfill &$0.204^{+0.053}_{-0.050}$&$0.180^{+0.040}_{-0.038}$\\
$EEP$\dotfill &Equal Evolutionary Phase$^{2}$ \dotfill &$241.3^{+7.5}_{-9.1}$&$228.8^{+6.6}_{-7.6}$\\\
$A_V$\dotfill &V-band extinction (mag)\dotfill &$0.0139^{+0.0087}_{-0.0092}$&$0.017^{+0.010}_{-0.011}$\\
$\sigma_{SED}$\dotfill &SED photometry error scaling \dotfill &$1.03^{+0.42}_{-0.25}$&$1.70^{+0.68}_{-0.41}$\\
$\varpi$\dotfill &Parallax (mas)\dotfill &$23.863^{+0.040}_{-0.039}$&$23.487\pm0.042$\\
$d$\dotfill &Distance (pc)\dotfill &$41.906\pm0.069$&$42.577\pm0.076$\\

$\ch{P_{rot}}$\dotfill &Inferred Rotation Rate(d)\dotfill &$6.84\pm0.58$ &  $7.22\pm0.77$\\

\hline
\label{tab:seds}
\end{tabular}
\begin{flushleft}
 \footnotesize{ \textbf{\textsc{NOTES:}\\}
See Table 3 in \citet{Eastman:2019} for a detailed description of all parameters.\\
$^1$The metallicity of the star at birth\\
$^2$Corresponds to static points in a star's evolutionary history. See \S2 in \citet{Dotter:2016}.\\
}
\end{flushleft}
\label{tab:seds}
\end{table}

\subsection{Planet Model Fit}
\label{sec:modelfit}

We use the \texttt{exoplanet} package\footnote{https://docs.exoplanet.codes/en/stable/} \citep{exoplanet:exoplanet} and \texttt{pymc3} \citep{exoplanet:pymc3} to fit the transit signals, given the best fit stellar parameters derived above, using the light curves from our correction procedure described in Section~\ref{sec:tpfs}. \texttt{exoplanet} is a probabalistic model, which allows us to create distributions for each parameter and jointly model them. Using \texttt{exoplanet} we are able to sample each parameter using MCMC, including any derived parameters (e.g. semi-major axis is derived from period and the stellar properties). In the case of \targa{}, we jointly fit a single set of stellar parameters (i.e. stellar density and limb darkening) and three transiting planets.

\begin{figure}
    \centering
    \includegraphics[width=0.5\textwidth]{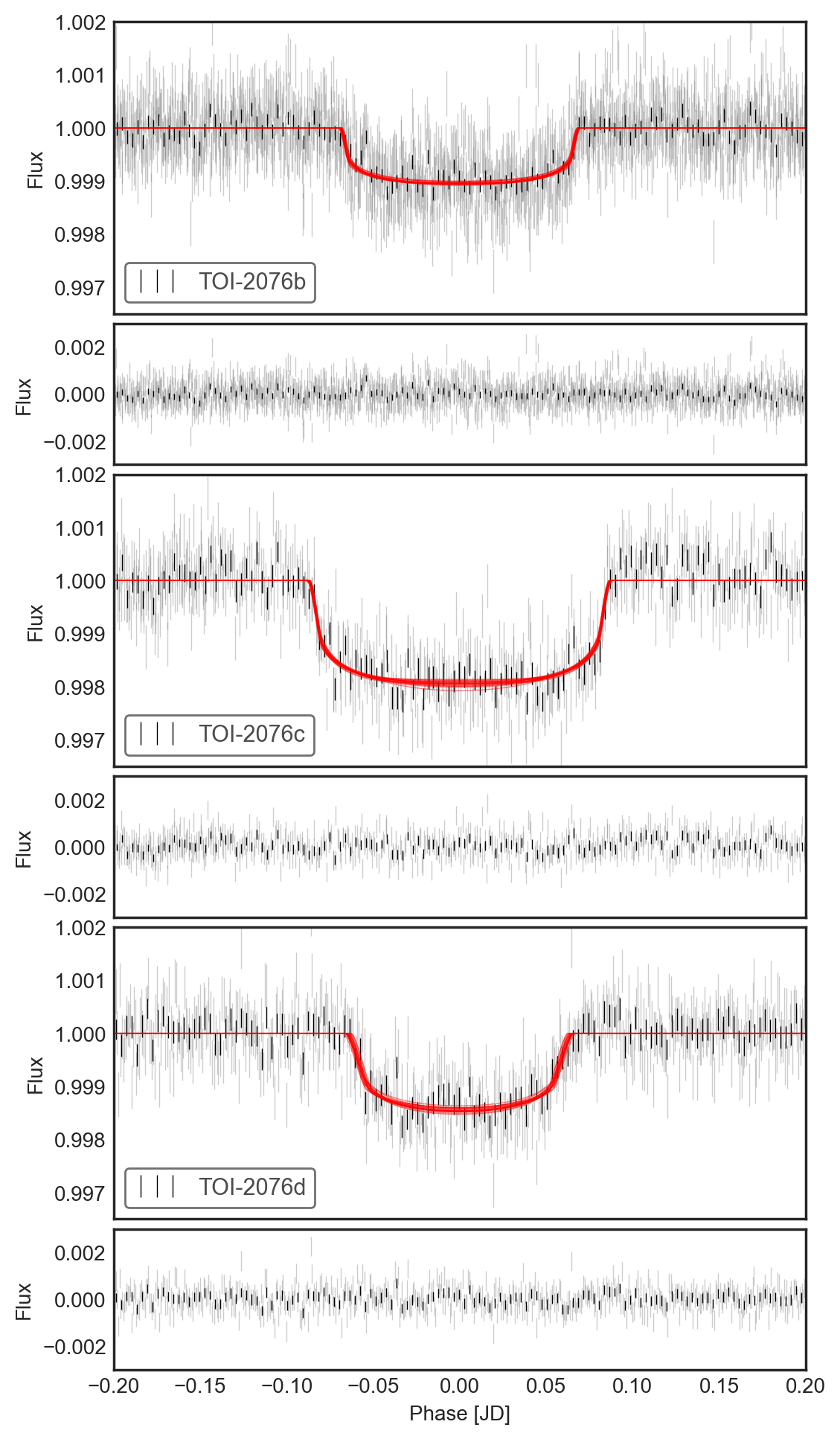}
    \includegraphics[width=0.5\textwidth]{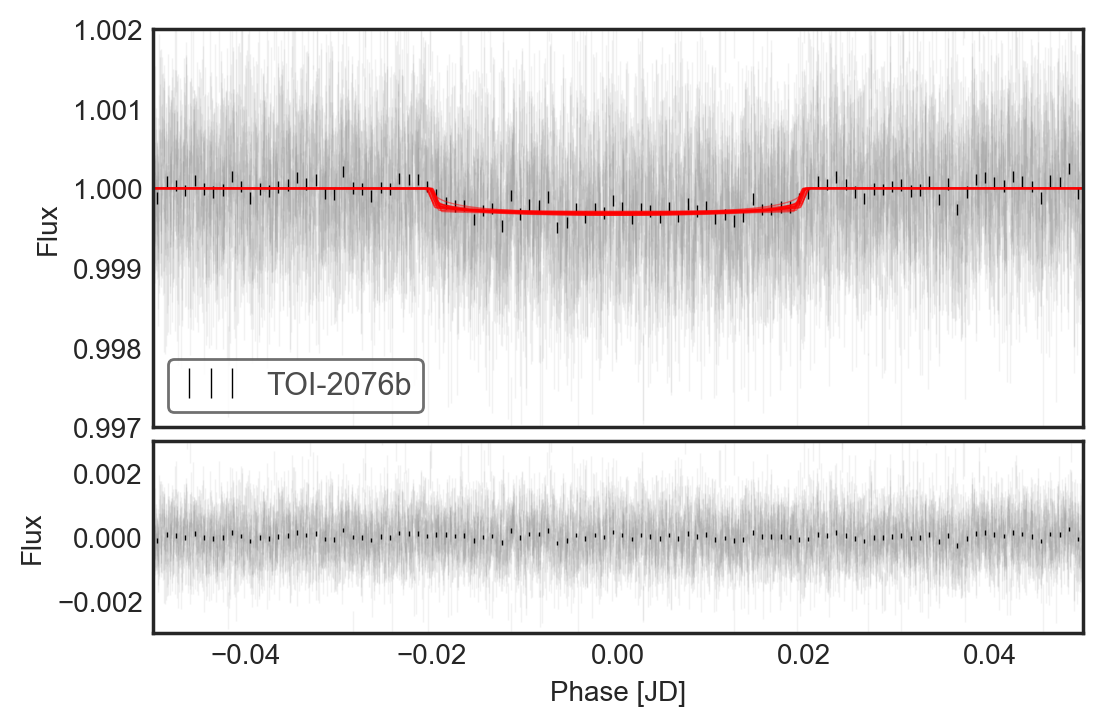}
    \caption{Best fit transit model for each planet. Top: Transit fit for \planb, \planc, and \pland. Bottom: Transit fit for \planbtwo. TESS photometry is shown in black, cleaned using the procedure outlined in Section~\ref{sec:correction}, folded at the best fit period for each planet. Twenty random model samples are shown in red. Parameters of our best fit with errors are shown in Tables~\ref{tab:TOI-2076} and ~\ref{tab:TOI-1807}.}
    \label{fig:transitfit}
\end{figure}

To fit the transiting planets in the dataset, we first remove stellar variability. We use the spline term from our fit to detrend the stellar variability by dividing the light curve by the best fit spline component from Section~\ref{sec:correction}. Since the stellar variability is long period, we assume that the stellar variability can be adequately detrended, and does not require a joint fit with planet parameters. In the case of \targb, we tested a joint fit for stellar variability and transits and found no significant improvement. For \targb, we fit a single planet model, and for \targa we fit a model consisting of three planets, in circular orbits, simultaneously. We assume that eccentricity cannot be measured using these data, as there are relatively few transits of each planet. (We explore eccentricities and period aliases of planets c and d in Section~\ref{sec:periodestimates}.) We fit for period, transit-midpoint, planet radius, impact parameter and limb darkening in our model, and set the starting stellar parameters to those derived above, with Gaussian priors. We find the maximum likelihood fit, and then use an MCMC No-U-Turn Sampler to find errors on each variable. The priors of our model are given in Table~\ref{tab:priors}, and results of this fit are shown in Table~\ref{tab:TOI-2076} and ~\ref{tab:TOI-1807} and Figure~\ref{fig:transitfit}, which shows good agreement with the data. We marginalize over the errors in the stellar parameters from Section~\ref{sec:stars}.

\begin{table*}
\begin{tabular}{llllll}
\toprule
    Parameter &     Distribution &                                   TOI-1807b &                                   TOI-2076b &                                   TOI-2076c &                                   TOI-2076d \\
\hline
  R$_*$ [$R_\odot$] &           Normal &  0.680$R_{\odot}$ $\pm$ 0.015$R_{\odot}$ &  0.761$R_{\odot}$ $\pm$ 0.016$R_{\odot}$ &  0.761$R_{\odot}$ $\pm$ 0.016$R_{\odot}$ &  0.761$R_{\odot}$ $\pm$ 0.016$R_{\odot}$ \\
     $\rho_*$ [cgs] &           Normal &                         3.36 $\pm$ 0.023 &                         2.72 $\pm$ 0.027 &                         2.72 $\pm$ 0.027 &                         2.72 $\pm$ 0.027 \\
              $T_*$ &           Normal &                          4757K $\pm$ 51K &                          5187K $\pm$ 54K &                          5187K $\pm$ 54K &                          5187K $\pm$ 54K \\
                $u$ &     QuadLimbDark$^1$ &                           [0.525, 0.215] &                           [0.525, 0.215] &                           [0.525, 0.215] &                           [0.525, 0.215] \\
       t$_0$ [BTJD] &           Normal &                        1899.34 $\pm$ 0.1 &                        1743.72 $\pm$ 0.1 &                       1748.69  $\pm$ 0.1 &                        1762.66 $\pm$ 0.1 \\
            $P$ [d] &           Normal &                        0.549d $\pm$ 0.1d &                      10.3562 $\pm$ 0.01d &                      17.1932 $\pm$ 0.01d &                      25.0893 $\pm$ 0.01d \\
 $R_P$ [$R_\odot$] &          Uniform &                            0.0001 .. 0.3 &                            0.0001 .. 0.3 &                            0.0001 .. 0.3 &                            0.0001 .. 0.3 \\
                $b$ &  ImpactParameter$^1$ &                                          &                                          &                                          &                                          \\
\hline
\end{tabular}

    \caption{Priors and their distributions for our exoplanet transit model fit. Parameters that are further derived from these parameters within the \texttt{exoplanet} model are not explicitly given priors, but have derived priors. Derived parameters in Table~\ref{tab:TOI-2076} and ~\ref{tab:TOI-1807} are highlighted in bold.\\
    $^1$ The QuadLimbDark and ImpactParameter prior distributions are provided in the \texttt{exoplanet} package. The QuadLimbDark is an uninformative prior based on the implementation discussed in \cite{exoplanet:kipping13}. ImpactParameter is a uniform prior between 0 and 1+$\frac{R_p}{R_*}$}
    \label{tab:priors}
\end{table*}

\begin{table*}


\begin{tabular}{lllll}

\toprule
         Parameter &                     Description & \multicolumn{3}{l}{Value} \\
\hline
     R$_*$\dotfill &      Radius [$R_\odot$]\dotfill &  0.7622 $_{-0.0159}^{+0.0157}$ &   &   \\
  $\rho_*$\dotfill &           Density [cgs]\dotfill &     2.244 $_{-0.058}^{+0.058}$ &   &   \\
     u$_1$\dotfill &  Limb Darkening Coeff 1\dotfill &     0.219 $_{-0.144}^{+0.143}$ &   &   \\
     u$_2$\dotfill &  Limb Darkening Coeff 2\dotfill &     0.451 $_{-0.235}^{+0.228}$ &   &   \\

\toprule

        Parameter &                          Description &                                    b &                                   c &                                     d\\

\hline
R$_P$\dotfill &        Radius [$R_\Earth$]\dotfill &           3.282 $_{-0.043}^{+0.042}$ &           4.438 $_{-0.046}^{+0.046}$ &              4.14 $_{-0.07}^{+0.07}$ \\
    \textbf{ R$_p$/R$_*$}\dotfill &  Planet Radius/Star Radius\dotfill &          0.0395 $_{-0.001}^{+0.001}$ &        0.0534 $_{-0.0013}^{+0.0013}$ &        0.0498 $_{-0.0013}^{+0.0013}$ \\
           P\dotfill &              Period [days]\dotfill &        10.35566 $_{-6e-05}^{+6e-05}$ &        \dotfill &    \dotfill \\
       t$_0$\dotfill &     Transit Mid Point [JD]\dotfill &  2458847.2776 $_{-0.0006}^{+0.0006}$ &  2458834.6615 $_{-0.0005}^{+0.0005}$ &  2458837.9363 $_{-0.0009}^{+0.0009}$ \\
           \textbf{i}\dotfill &     Inclination [$^\circ$]\dotfill &              88.9 $_{-0.11}^{+0.11}$ &             89.84 $_{-0.12}^{+0.12}$ &          88.607 $_{-0.037}^{+0.036}$ \\
           b\dotfill &           Impact Parameter\dotfill &           0.342 $_{-0.033}^{+0.032}$ &              0.07 $_{-0.05}^{+0.05}$ &            0.78 $_{-0.011}^{+0.011}$ \\
         $\mathbf{a}$\dotfill &       Semi-Major Axis [AU]\dotfill &        0.0631 $_{-0.0027}^{+0.0027}$ &        0.0885 $_{-0.0038}^{+0.0038}$ &        0.1138 $_{-0.0049}^{+0.0048}$ \\
     $\mathbf{a/R_*}$\dotfill &    Semi-Major Axis / R$_*$\dotfill &               17.79 $_{-0.4}^{+0.4}$ &                24.9 $_{-0.6}^{+0.6}$ &                32.1 $_{-0.7}^{+0.7}$ \\
    \textbf{t$_{14}$}\dotfill &           Duration [hours]\dotfill &           3.326 $_{-0.036}^{+0.036}$ &           4.215 $_{-0.031}^{+0.031}$ &               3.2 $_{-0.06}^{+0.06}$ \\
    \textbf{T$_{eq}$}\dotfill &       Equilibrium Temp [K]\dotfill &                   870 $_{-13}^{+13}$ &                   734 $_{-11}^{+11}$ &                   648 $_{-10}^{+10}$ \\
 \textbf{t$_{depth}$}\dotfill &              Transit Depth\dotfill &    0.001047 $_{-2.4e-05}^{+2.5e-05}$ &    0.001943 $_{-3.8e-05}^{+3.8e-05}$ &    0.001445 $_{-3.8e-05}^{+3.8e-05}$ \\

\hline


\end{tabular}

    \caption{Best Fit parameters for \targa. Top: Host star parameters. Bottom: Planet parameters. Derived parameters are highlighted in bold. $\pm$ values indicate the 1$\sigma$ errors. In the case of sampled parameters from our transit fit, we quote the 16th and 84th percentiles of our samples, equivalent to 1 $\sigma$ errors. Planets b, c and d are jointly fit at the same time, with shared stellar parameters.} Note periods for planet c and d are omitted, see Section~\ref{sec:periodestimates}
    \label{tab:TOI-2076}
\end{table*}

\begin{table}
\begin{tabular}{lll}

\toprule
         Parameter &                           Description &                         Value \\
\hline
    R$_*$\dotfill &      Radius [$R_\odot$]\dotfill &  0.6802 $_{-0.0145}^{+0.0146}$ \\
  $\rho_*$\dotfill &           Density [cgs]\dotfill &     3.374 $_{-0.233}^{+0.228}$ \\
     u$_1$\dotfill &  Limb Darkening Coeff 1\dotfill &     0.304 $_{-0.225}^{+0.242}$ \\
     u$_2$\dotfill &  Limb Darkening Coeff 2\dotfill &      0.152 $_{-0.285}^{+0.29}$ \\
\toprule

        Parameter &                          Description &                                    b \\

\hline

       R$_P$\dotfill &        Radius [$R_\Earth$]\dotfill &         1.849 $_{-0.043}^{+0.042}$ \\
     \textbf{R$_p$/R$_*$}\dotfill &  Planet Radius/Star Radius\dotfill &      0.0249 $_{-0.0008}^{+0.0008}$ \\
           P\dotfill &              Period [days]\dotfill &      0.549372 $_{-7e-06}^{+7e-06}$ \\
       t$_0$\dotfill &     Transit Mid Point [JD]\dotfill &   2457000.166 $_{-0.026}^{+0.024}$ \\
           \textbf{i}\dotfill &     Inclination [$^\circ$]\dotfill &              77.7 $_{-1.2}^{+1.1}$ \\
           b\dotfill &           Impact Parameter\dotfill &         0.546 $_{-0.038}^{+0.038}$ \\
         $\mathbf{a}$\dotfill &       Semi-Major Axis [AU]\dotfill &   0.00812 $_{-0.00038}^{+0.00037}$ \\
     $\mathbf{a/R_*}$\dotfill &    Semi-Major Axis / R$_*$\dotfill &            2.57 $_{-0.08}^{+0.08}$ \\
    \textbf{t$_{14}$}\dotfill &           Duration [hours]\dotfill &         0.972 $_{-0.015}^{+0.015}$ \\
    \textbf{T$_{eq}$}\dotfill &       Equilibrium Temp [K]\dotfill &                2100 $_{-40}^{+39}$ \\
 \textbf{t$_{depth}$}\dotfill &              Transit Depth\dotfill &  0.000312 $_{-1.5e-05}^{+1.5e-05}$ \\

\hline
\end{tabular}

    \caption{Best Fit parameters for \targb. Top: Host star parameters. Bottom: Planet parameters. Derived parameters are highlighted in bold. $\pm$ values indicate the 1$\sigma$ errors. In the case of sampled parameters from our transit fit, we quote the 16th and 84th percentiles of our samples, equivalent to 1 $\sigma$ errors.}
    \label{tab:TOI-1807}
\end{table}

\subsection{Phase Curve Modeling}
\planb{} is a short period, hot planet with an equillibrium temperature of $>$2000K. Given the high signal to noise light curve of the bright host star, it may be possible to use the \tess data to identify a phase curve; a simple calculation of the maximum surface brightness ratio of \planb{} gives an eclipse depth of $\sim$20ppm. We additionally fit a transit model for \targb with an eclipse and phase curve component, jointly fitting stellar variability. Using this approach, we are unable to detect a significant phase curve using the \tess data. 


We additionally undertook the following search for a phase curve in the TESS Pipeline Products. First, the transits of \targb and the expected occultation events were removed from the observed TESS light curve. The photometry was separated into segments defined by each TESS orbit, then normalized by their average flux offset and detrended using a linear function that best-fit each light curve segment. (We note that detrending each segment by a higher degree polynomial did not significantly alter our results.) Significant stellar variability was removed from the light curve by subtracting the two strongest sinusoidal signals detected in a Lomb-Scargle periodogram of the out-of-transit light curve at 4.34 days and 6.06 days. Finally, the variability corrected out-of-transit light curve was fit with a double harmonic sinusoidal model to search for an atmospheric phase curve signature at the orbital period of \planbtwo{}. The double harmonic sinusoidal model is defined as
\begin{equation}
F(\phi) = A_n + A_r \cos{2\pi\phi} + A_b \sin{2\pi\phi} + A_e \cos{4\pi\phi}
\text{,}
\end{equation}
where $A_n$ is the flux normalization offset and $A_r$, $A_b$, and $A_e$ represent the effects of planetary emission/reflection, Doppler boosting, and ellipsoidal variations, respectively. To determine the significance of the best-fit phase curve model, the reduced $\chi^2$ statistic was compared to that of a horizontal line. 

Regardless of whether we used 1) the TESS Pipeline SAP photometry 2) PDCSAP photometry, 3) a correction for stellar variability, or 4) a higher-order polynomial (up to 10th degree) to detrend the light curve, we did not detect a significant atmospheric phase curve for \planbtwo{}. In all cases, the best-fit phase curve model was either consistent with a horizontal line or exhibited a $<$3$\sigma$ significance phase curve shape that is inconsistent with the expected shape of a planetary atmospheric phase curve.

We conclude that, using the TESS data alone, there is no detectable phase curve for \planbtwo{}. However, \tess data from future cycles may increase signal to noise, or additional data at redder wavelengths, may reveal a phase curve for this planet. \targb will be observed again by \tess in Sector 49, in February 2022.

\subsection{Period Aliases of \planc{} and \pland}
\label{sec:periodestimates}
We find best fit periods for \planc\  and \pland\  of \perc\ and \perd\, respectively.  Our best fit periods reflect the shortest period, in each case, that is consistent with the data. However, due to the long gap between TESS observations, many aliased periods are also fit well by the data. 

To find the best fitting periods for \planc{} and \pland{}, we first recalculate the best fit model for \planall{}, relaxing our assumptions of a Keplerian orbit. Instead, we fit a "simple" orbit, where each planet occults the star, not on a circular orbit, but traveling on a straight path. This occultation is parameterised by the velocity of the planet. By adopting this approach, none of the parameters are forced by our prior knowledge of Keplarian laws (which link, for example, duration and impact parameter), and each parameter (e.g. impact parameter) is only constrained by the data itself. We set up this model such that each planet passes in front of the same star, with the same radius and limb darkening parameters, and use MCMC (e.g. see Section~\ref{sec:modelfit}) to vary all parameters in our model. 

We perform a Monte Carlo analysis combining the posteriors from the simple transit fit with inferences based on both 1) dynamical stability and 2) the window function of allowed orbital periods derived from the observation times of the TESS sectors. This method of constraining orbital periods follows the line of analysis in \citet{Vanderburg2016} and \citet{Becker2019}. For each link of the transit fit posterior, we take parameters for each planet and then numerically solve the following equation for $P$, the planetary orbital period \citep[see][]{Seager2003}:

\begin{multline} \label{eq:duration}
D = \frac{P}{\pi} \arcsin{}\left[\left(\frac{G (M_{*} + m_{p}) P^{2}}{4 \pi^{2}}\right)^{-1/3} \times \right. \\
\left. \sqrt{(r_{P} + R_{*})^2 - (b^2 \times R_{*}^2)} \right] \frac{\sqrt{1-e^2}}{1+e \cos{\varpi}}
\end{multline}

The parameters taken from the observationally-derived posterior include $D$, which is the transit duration of the planet, $r_{p}$, which is the planetary radius, $m_{p}$, which is the planetary mass, $e$, which is the orbital eccentricity, $\varpi$, which is the longitude of periastron, $b$, which is the planet's impact parameter, $R_{*}$ and $M_{*}$, which are the stellar radius and mass.
Additional parameters that cannot be directly derived from the light curve must be computed: the planet mass $m_{p}$ ($<<M_{*}$) is inferred using the mass-radius relation of \citet{angie}, $e$ was chosen using a beta distribution prior with shape parameters $\alpha = 0.867$ and $\beta = 3.03$ \citep[][]{kipping_prior1, kipping_prior2, kipping16}, and then $\varpi_i$ was chosen using Equation 19 of \citet{kipping16}.
Finally, $G$ is defined as the gravitational constant.
For each link of the posterior, we solve Equation \ref{eq:duration} numerically for each planet to derive the orbital period that corresponds to the observed parameters.

Once a set of two orbital periods (one for \planc\ and one for \pland) have been computed from a single link, we check two markers of dynamical instability: whether the chosen initial parameters are Hill unstable \citep{Fabrycky2014}, and whether the computed secular oscillation amplitudes in eccentricity \citep[computed using the Laplace-Lagrange secular disturbing function, see][]{MD99} result in orbits that cross. If either of those conditions are met, the link is thrown out; if not, the computed periods are kept and used to construct a probability density function for orbital periods that are consistent with the data and also likely dynamically stable.
We then combine that with the baseline prior (see Equation 1 of \citealt{Becker2019} and the general form in Equation 2 of \citealt{Dholakia2020}) to construct a final probability density function for each possible orbital period. The baseline prior also corrects this final probability to zero for any orbital period where a third transit should have been observed anywhere in the TESS data.
 
Using this final probability density function for each planet's orbital period, we check each possible orbital period (corresponding to an positive integer number of conjunctions in between the two observed transits) and normalize the probabilities using those discrete values as the only possible orbital periods.
For \pland, the mostly likely orbital period is 25.089 days (with a 60\% probability), which corresponds to a circular orbit. The next most likely orbital period is 29.271 days, followed by 35.125 days and 43.906 days.
For \planc, a secure determination of a best candidate period cannot be made. Orbital periods which have a greater than 10\% chance of being correct given the above analysis include (in order of computed likelihood) 23.641 days, 21.014 days, 27.018 days, 18.913 days, and 17.193 days.
Of these, 18.913 days and 17.193 days had the greatest positive correlation in occurrence with the 25.089 day orbital period for \pland. The 17.193 orbital period for \planc\ also corresponds to a circular orbit.
 
To characterize the full state of the system, it is important to confirm the true orbital periods and subsequently refine the ephemerides and limits on transit timing variations. The determination of \pland{}'s orbital period is likely to be more straightforward, given the strong preference for the 25.089 day solution. \planc{} will be harder to constrain. We discuss ground-based data of \targa in the context of \planc{} in Section~\ref{sec:ground}.

\section{Vetting and Validation}
\label{sec:validation}
In this Section we discuss the validation of the planet candidates around \targa{} and \targb{}. In Section~\ref{sec:contamination} we discuss the constraints on contamination by background objects, using archival data and show that archival data are able to rule out contamination for \targb. In Section~\ref{sec:centroid} we show there are no significant centroid offsets during transit, indicating that \targa and and \targb are both the true sources of the planet signal. In Section~\ref{sec:triceratops}, we use the \texttt{TRICERATOPS} toolkit \citep{triceratops} to show that there is a very small false probability chance in either the case of \targa or \targb. 

We additionally note that \emph{Gaia} DR2 provides the Renormalized Unit Weight Error (RUWE) \citep{ruwe} to determine whether \emph{Gaia} astrometric fits are good. A value significantly above 1 indicates that a single source is not a good fit to the data. \targa has a RUWE of 1.0857, and \targb has a RUWE of 1.07523, suggesting that they are consistent with being single stars.

\subsection{Contamination (Archival Data)}
\label{sec:contamination}
\begin{figure*}
    \centering
    \includegraphics[width=0.95\textwidth]{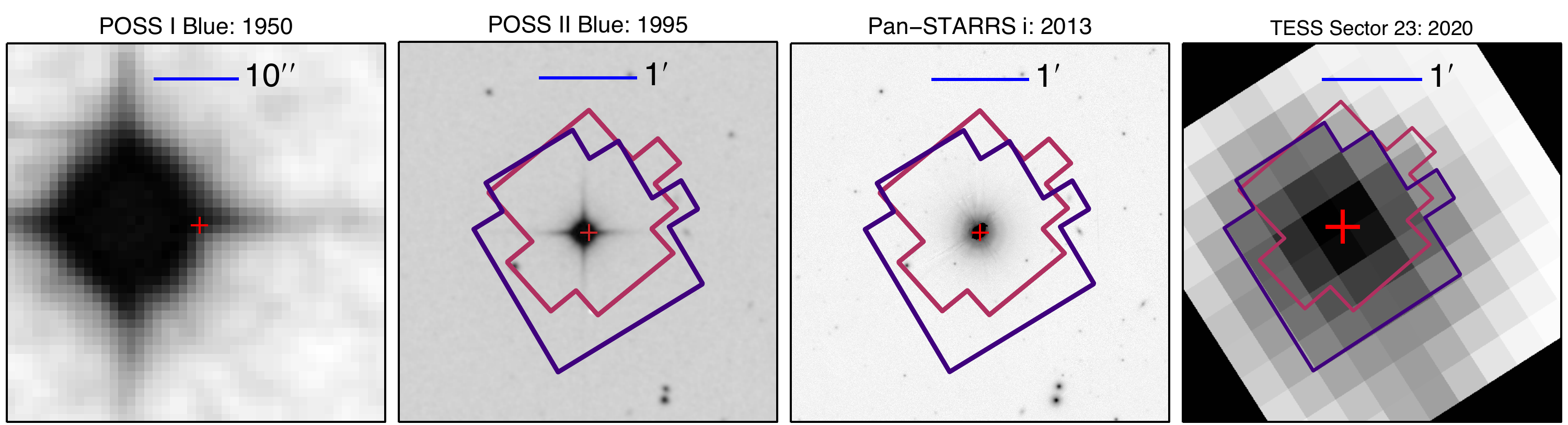}
    \includegraphics[width=0.95\textwidth]{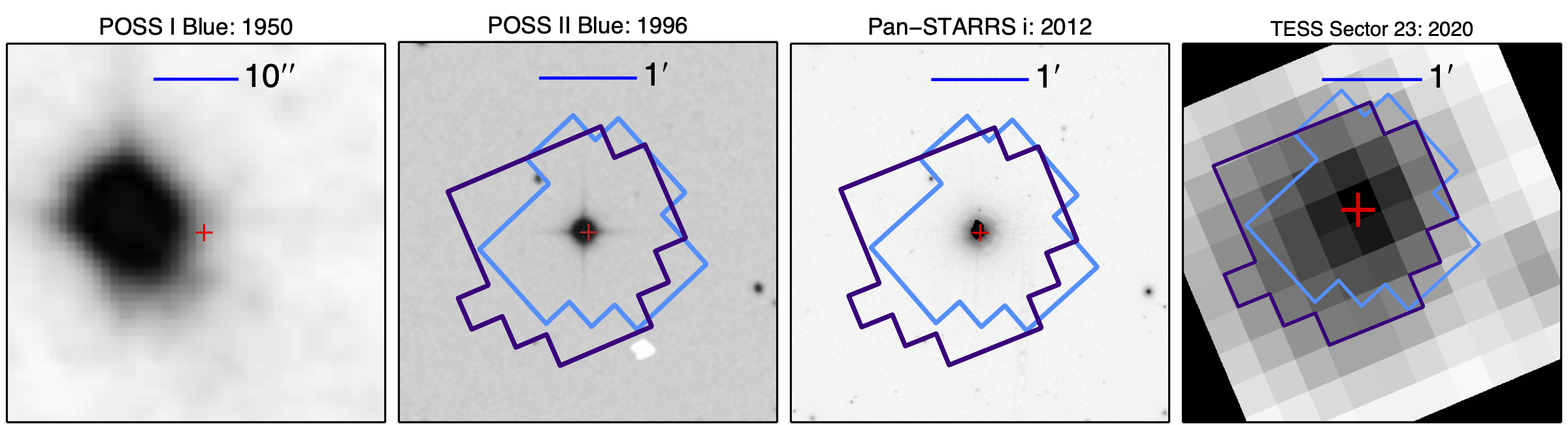}
    \caption{Archival data from Palomar Observatory Sky Survey and PanSTARRS for both \targa (top) and \targb (bottom). The TESS data are also shown. Apertures selected by the \tess pipeline in each sector are shown in purple for Sector 23, blue for Sector 22 and pink for Sector 16. Using archival data we are able to rule out a contaminant for \targa with high confidence. The diffraction spike caused by \targb prevents us from ruling out a faint contaminant using archival data alone. }
    \label{fig:archival}
\end{figure*}

Figure~\ref{fig:archival} shows the potential contamination of \targa and \targb using archival data. We downloaded images from the first and second Palomar Observatory Sky Survey \citep{poss1, poss2}, as well as Pan-STARRS \citep{panstarrs}, and plotted the present-day position of the stars from the TIC \citep{TIC} (propagating the proper motion forward to the time of TESS observations). We overplot the apertures assigned by the SPOC Pipeline that we use to extract the TESS light curves. Owing to the high proper motion of \targa and \targb, the POSS I Blue image shows a significant offset between the centroid of the targets and their present day positions. 

Using the POSS I Blue data, we fit PSFs of stars around both \targa and \targb, using a simple 2D Gaussian model. By evaluating this model at the present day location of both targets, we are able to rule out background contaminants. For \targb, our PSF modeling rules out contaminating targets down to ~20th magnitude. \targa is sufficiently bright in POSS I Blue to cause a significant diffraction spike. Due to this spike, our PSF modeling is unable to rule out the presence of a contaminating source for \targa fainter than 11th magnitude using archival data alone.

We note that in POSS II and Pan-STARRS there are some fainter targets contained within the SPOC pipeline aperture at the edge, and so we additionally perform a centroid test.

\subsection{Contamination (Centroiding)}
\label{sec:centroid}
We perform a simple centroid test on the \tess data of \targa and \targb using the following procedure. 

\begin{itemize}
    \item We estimate the centroid of the pixels within the SPOC Pipeline aperture using a weighted mean (weighted by the flux in each pixel). We propagate uncertainties by sampling from the flux errors for each pixel given by the pipeline.
    \item We correct these centroids for long term trends by removing a smooth trend built by a Gaussian smoothing kernel, with a default width of 21 cadences, using \texttt{astropy}'s convolution module. This removes long term trends due to velocity aberration and focus change during a single TESS observation.
    \item We then compare the X and Y centroid position distribution of cadences with no transits, to cadences containing transits of planets. Using a simple Student t-test, we test whether the means of these distributions are consistent, assuming they have the same variance. We use \texttt{scipy.stats}'s \texttt{ttest\_ind} function to perform this test.
\end{itemize}

\begin{figure*}
    \centering
    \includegraphics[width=0.95\textwidth]{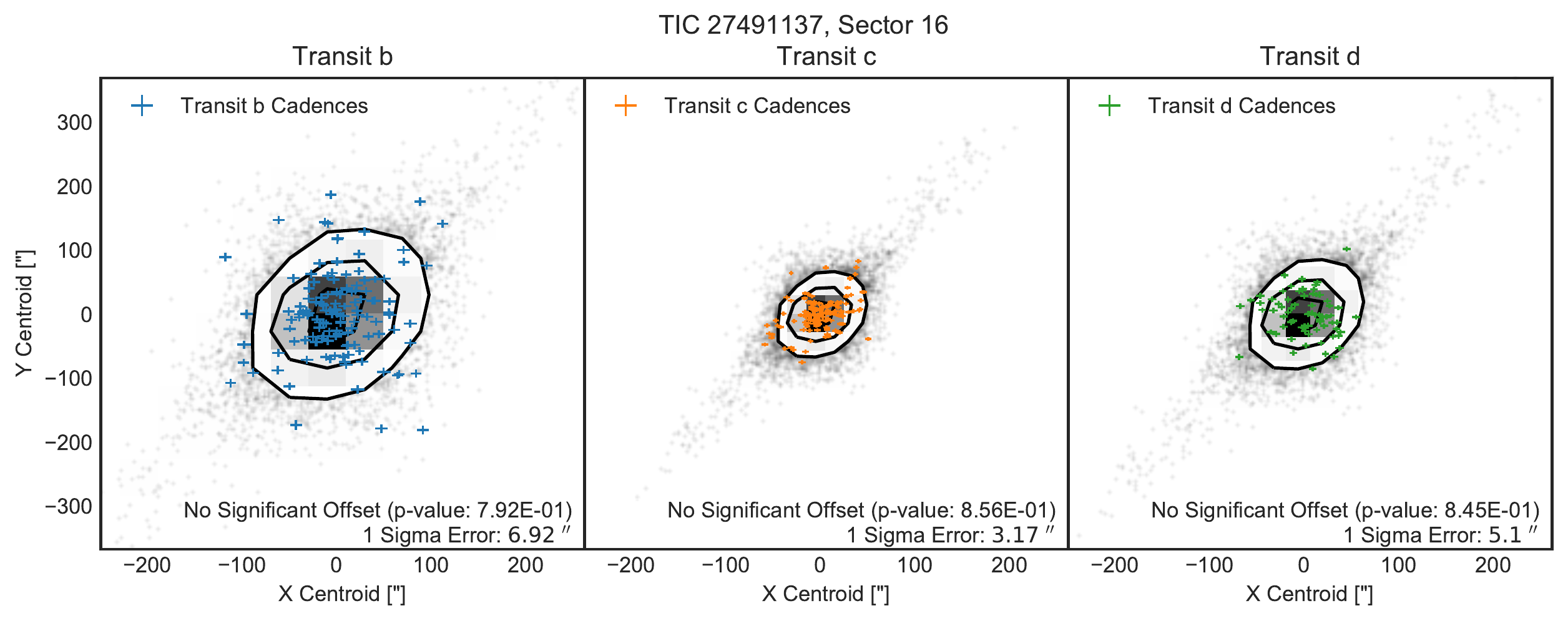}
    \includegraphics[width=0.95\textwidth]{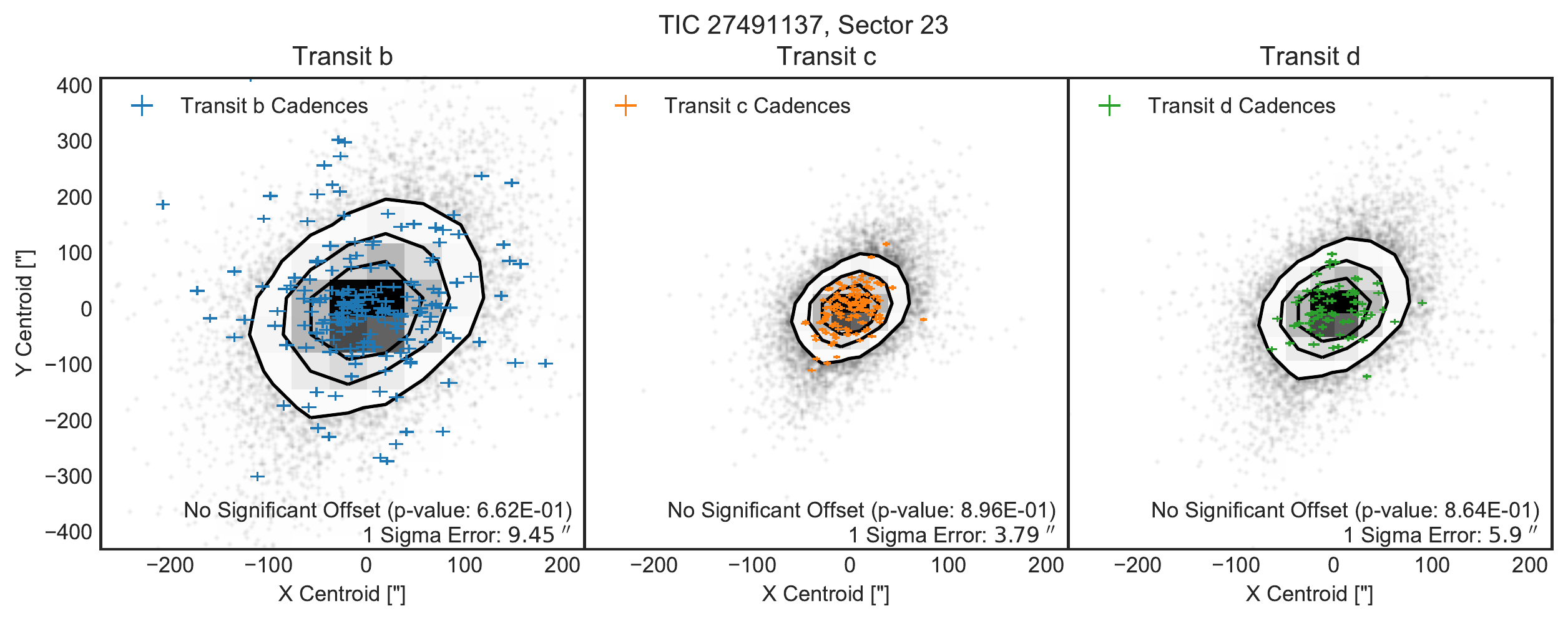}
    \includegraphics[width=0.35\textwidth]{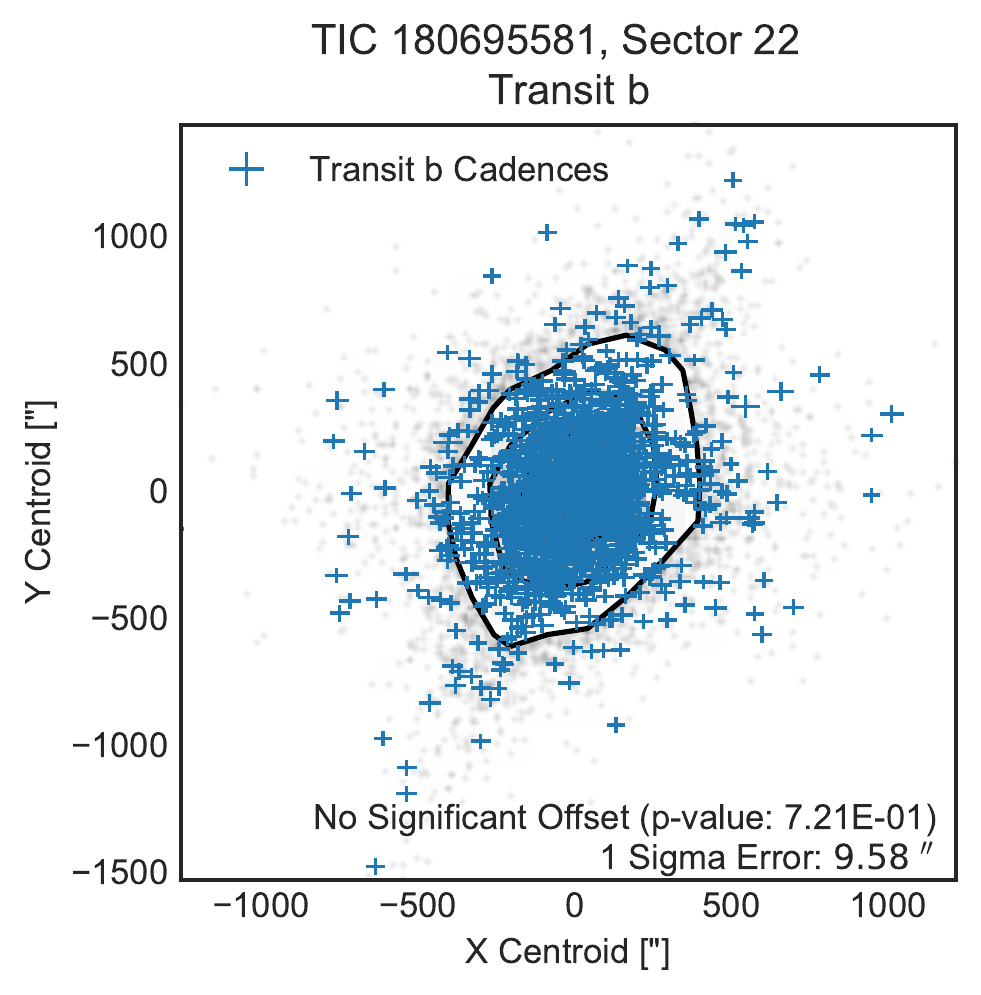}
    \includegraphics[width=0.35\textwidth]{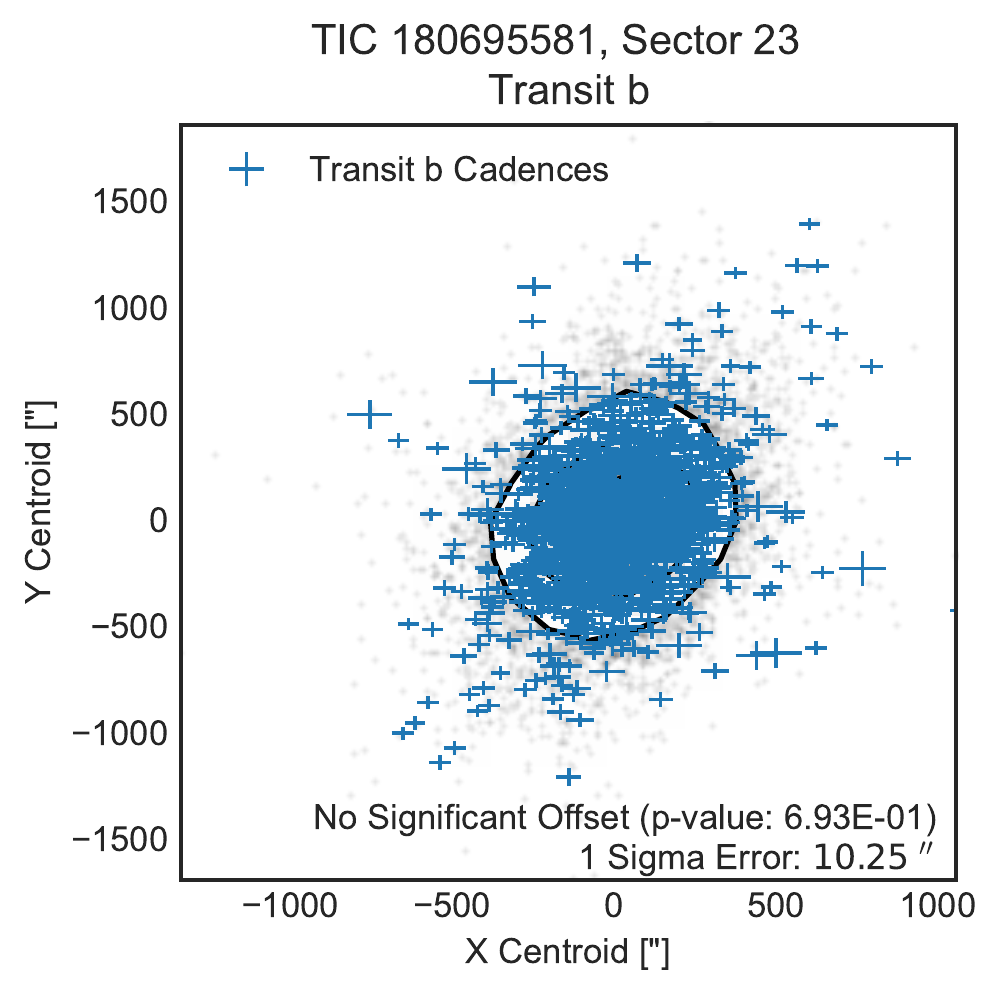}
    \caption{Results of our centroid test described in Section~\ref{sec:centroid}. We estimate the centroid of the flux in X and Y pixel inside the SPOC pipeline aperture using a weighted average, and then perform a Student t-test to identify if there are significant differences between the centroid during transit and out of transit. Grey points and contours show the 2D histogram of X and Y points during cadences where there is no transit. (Bins where there is a high density of points have been converted to a 2D histogram using the \texttt{corner} package\citep{corner}). Top two panels: Centroid test for \targa in Sector 16 and Sector 22. Blue, orange and green points show centroids for cadences that contain planets b, c and d respectively. Bottom: Centroid test for \targb in Sectors 22 and 23. The p-value from the student t-test is given in each panel. We find there is no significant evidence of centroid shifts. 1 $\sigma$ errors on the centroid are given in the lower right corner of each panel, for each planet. 
    }
    \label{fig:centroids}
\end{figure*}

The tool to produce this centroid test is available as an open source project on GitHub\footnote{github.com/ssdatalab/vetting}, and available as a pip installable tool named \texttt{vetting}\footnote{https://pypi.org/project/vetting/}. The results of our centroid test are shown in Figure~\ref{fig:centroids}. We find for all planets, in all sectors, that there is no significant offset in the means of the centroid distributions. We find no significant evidence that there is any change in the target centroid during transits; our student t-test has a p-value of $\gtrsim$0.8 (see Figure~\ref{fig:centroids}) for each transit, in each sector. This shows there is a $\gtrsim$80\% probability that the distributions have the same mean, (i.e. that the centroids during transit are consistent with centroids out of transit.) We calculate the 1$\sigma$ errors in separation from our centroid test for each planet, in each Sector. The distance at which we can rule out blends at the 1 sigma level is given in the corner of each panel of Figure~\ref{fig:centroids}. For \targa we can rule out blends out to distances of 7, 4, and 6 arcseconds at the 1$\sigma$ level for \planb{}, \planc{}, \pland{} respectively. For \targb we can rule out blends out to distances of 10 arcseconds at the 1$\sigma$ level for \planbtwo{}. As such, we find no evidence that the transits originate from background sources, based on the \tess data alone. Further validation with external data sources is discussed below.

\subsection{High-Resolution Imaging Follow-up}
\label{sec:photometricfollowup}

\begin{figure*}
    \centering
    \includegraphics[width=0.48\textwidth]{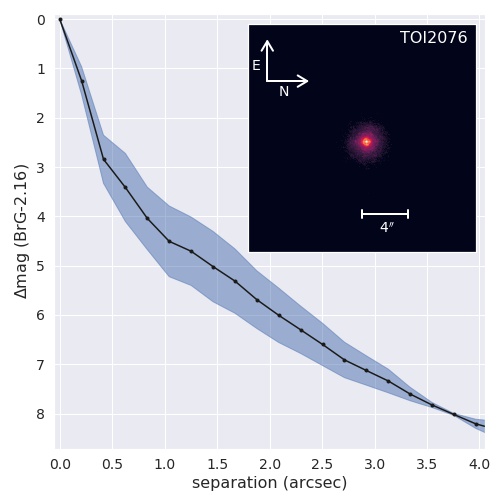}
    \includegraphics[width=0.48\textwidth]{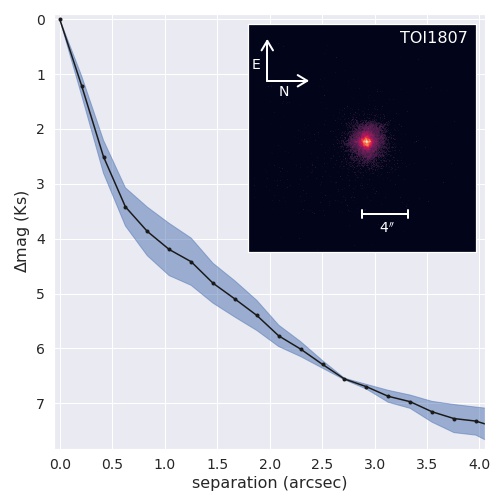}
    \includegraphics[width=0.48\textwidth]{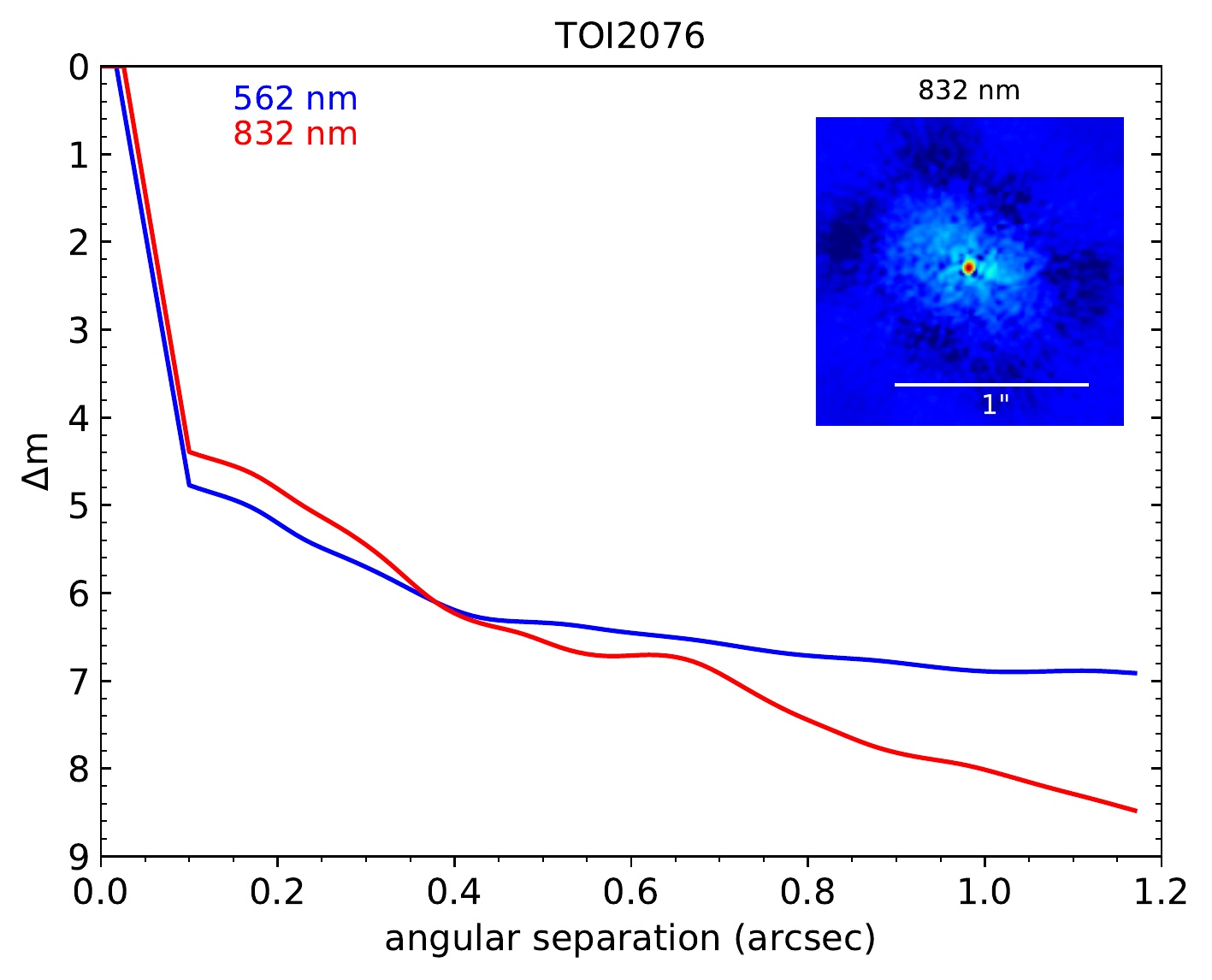}
    \includegraphics[width=0.48\textwidth]{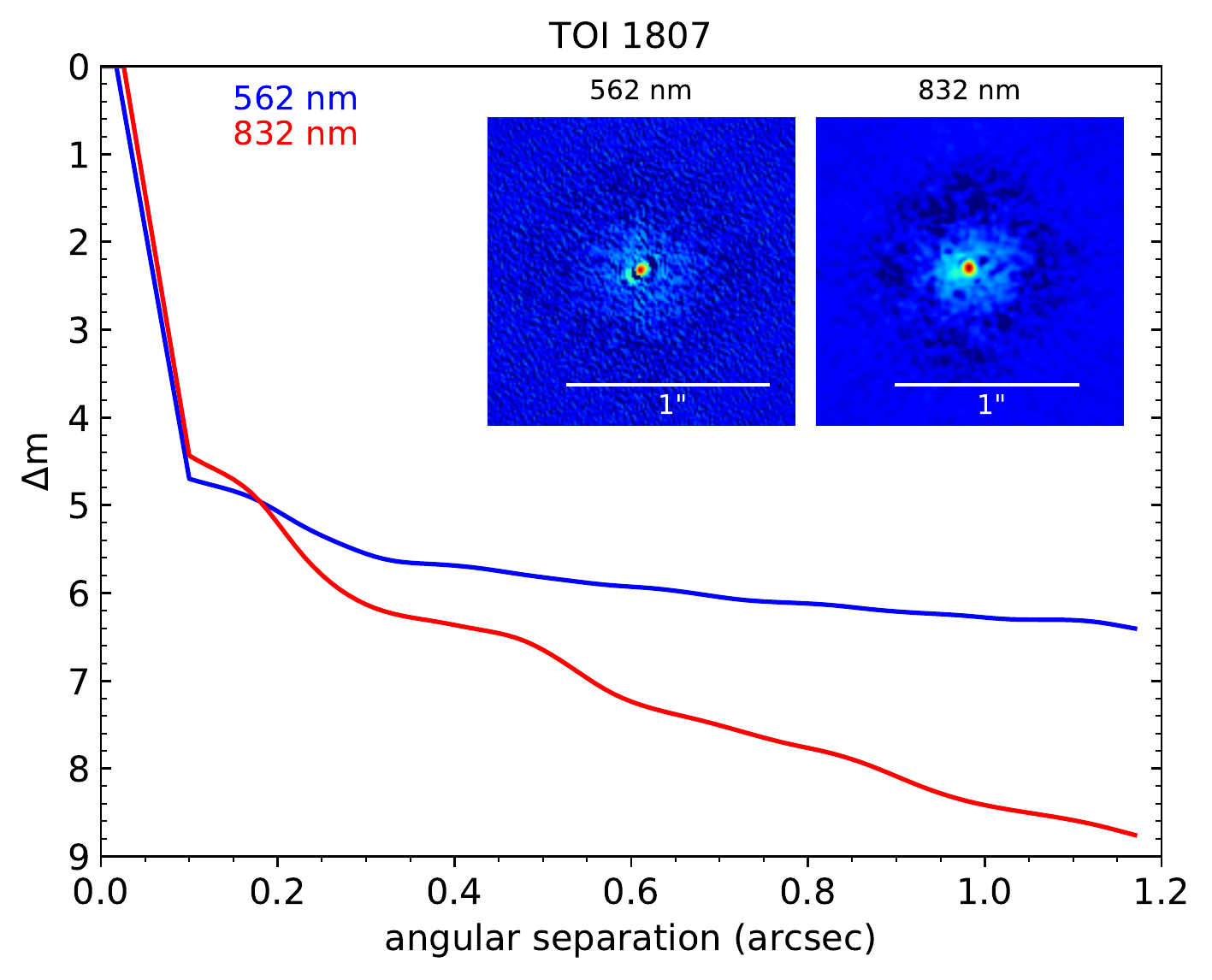}
    \caption{\emph{Top:} Imaging observations and contrast curves taken using the ShARCS instrument at the Shane 3m telescope at  2.167 and 2.150$\mu$m using adaptive optics. The direct image is shown as an inset image, and 4 arcseconds is shown for scale (approximately 1/7th of a TESS pixel). \emph{Bottom:} Imaging observations and contrast curves taken using the ‘Alopeke instrument on the Gemini telescope. The speckle image is shown as an inset image, and 1 arcsecond is shown for scale. Left: \targa Right: \targb. }
    \label{fig:imaging}
\end{figure*}

We observed \targa and \targb on UT 2020 December 2 using the ShARCS camera on the Shane 3-meter telescope at Lick Observatory (see Figure~\ref{fig:imaging}, top row). Observations were taken using the Shane adaptive optics (AO) system in natural guide star mode. We collected our observations using a 4-point dither pattern with a separation of 4$\arcsec$ between each dither position. For \targa, we obtained one sequence of observations in the BrG-2.16-band with exposure times of 15 s, which rules out companions with $\Delta$ mag < 3 at 0$\farcs$5 and companions with $\Delta$ mag < 4.5 at 1$\arcsec$. For \targb, we obtained one sequence of observations in the Ks-band with exposure times of 1.5 s, which rules out companions with $\Delta$ mag < 3 at 0$\farcs$5 and companions with $\Delta$ mag < 4 at 1$\arcsec$. See \cite{savel2020closer} for a detailed description of the observing strategy and reduction prodecure.

We obtained speckle interferometric images of \targa (see Figure~\ref{fig:imaging}, bottomw row) on UT 2021 February 07 using the ‘Alopeke instrument\footnote{https://www.gemini.edu/instrumentation/alopeke-zorro} mounted on the 8 m Gemini North telescope on the summit of Mauna Kea in Hawai’i. We also obtained speckle interferometric images of \targb on UT 2020 June 09. ‘Alopeke simultaneously collects diffraction-limited images at 562 and 832 nm. Our data set consisted of 4 minutes of total integration time taken as sets of 1000 × 0.06 s images followed by the observation of a local PSF standard star. As discussed in \cite{Howell2011}, we combined all images, subjected them to Fourier analysis, and produced reconstructed images from which the 5$\sigma$ contrast curves are derived in each passband. The bottom row in Figure~\ref{fig:imaging} presents the two contrast curves as well as the 832 nm reconstructed image for \targa and \targb. Our measurements reveal \targa and \targb to have no nearby, contaminating stars. For \targa{}, we are confident of our determination of no companions to contrast limits of 5-8 mag, within the spatial limits of 0.7 AU (562nm) to 1.18 AU (832nm) at the inner working angle out to 50 AU at 1.2" (d=42 pc). For \targb{}, we are confident in our determination of no companions of $\Delta$ mag $<$ 3 at 0.5" (21 AU) and companions of $\Delta$ mag $<$ 4 at 1" (42 AU).

\section{Ground based photometry}
\label{sec:ground}

We acquired ground-based time-series follow-up photometry of \planbtwo{} and \planc{} as part of the TESS Follow-up Observing Program (TFOP)\footnote{https://tess.mit.edu/followup}. We used the {\tt TESS Transit Finder}, which is a customized version of the {\tt Tapir} software package \citep{Jensen:2013}, to schedule our transit observations. The photometric data were extracted using {\tt AstroImageJ} \citep{Collins:2017}.

We observed a full transit window of \planbtwo{}, as predicted by the SPOC pipeline analysis of TESS sector 22, on UTC 2020 April 19 in Sloan $i'$ band from the 0.5\,m CDK20N telescope at the University of Louisville Moore Observatory near Louisville, Kentucky. We observed a second full transit window on UT 2020 April 25 in Pan-STARRS $z$-short band from the Las Cumbres Observatory Global Telescope \citep[LCOGT;][]{Brown:2013} 1.0\,m network node at McDonald Observatory. Since the $\sim 378$ ppm event detected by the SPOC pipeline is generally too shallow to detect with ground-based observations, we checked for a faint nearby eclipsing binary (NEB) that could be contaminating the SPOC photometric aperture. To account for possible contamination from the wings of neighboring star PSFs, we searched for NEBs at the positions of Gaia DR2 stars out to $2\farcm5$ from the target star. If fully blended in the SPOC aperture, a neighboring star that is fainter than the target star by 8.6 magnitudes in TESS-band could produce the SPOC-reported flux deficit at mid-transit (assuming a 100\% eclipse). To account for possible delta-magnitude differences between TESS-band and Sloan $i'$ band and Pan-STARRS $z$-short band, we included an extra 0.5 magnitudes fainter (down to \textit{TESS}-band magnitude 17.7). We visually compared the light curves of the 4 nearby stars that meet our search criteria with models that indicate the timing and depth needed to produce the $\sim 400$ ppm event in the SPOC photometric aperture. We found no evidence of an NEB that might be responsible for the SPOC detection. By a process of elimination, we conclude that the transit is likely occurring in TOI-1807, or a star so close to TOI-1807 that it was not detected by Gaia DR2 and too faint to be detected by high resolution imaging.

We observed a predicted egress of \planc{} on UTC 2020 December 29 in Pan-STARRS $z$-short band from the LCOGT 1.0\,m node at McDonald Observatory. The observation would contain a transit egress of \planc{}, if the period were the shortest period estimate derived in Section~\ref{sec:periodestimates} (\perc). The \planc{} observation was moderately defocused, resulting in a typical point source full-width at half-maximum of $\sim 7\arcsec$, and used 15 second exposures. A photometric aperture radius of $\sim12\arcsec$ was used to extract the differential photometry, resulting in $\sim 870$ ppm model residuals in 5 minute bins. The photometric aperture is not contaminated with flux from any known Gaia DR2 neighboring stars. We recover a a $\sim 2000$ ppm egress using LCO data alone.  The follow-up light curve data are available at ExoFOP-TESS\footnote{https://exofop.ipac.caltech.edu/tess}.

\begin{figure}
    \centering
    \includegraphics[width=0.5\textwidth]{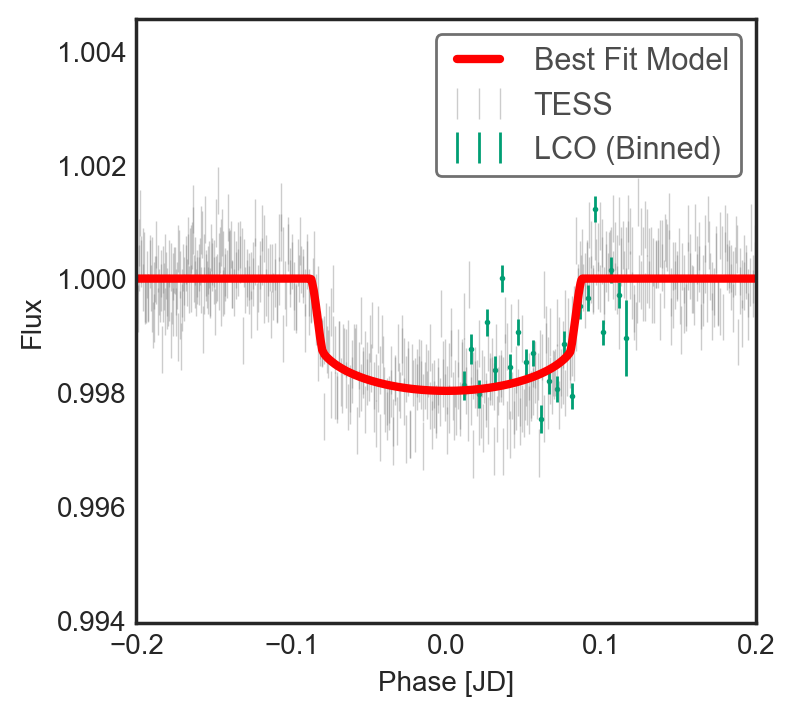}
    \caption{\planc{} observations from \tess, folded at 17.1936d, with ground based LCO data binned using a median to a cadence of 8 minutes. We perform a joint fit of the transit model and the instrument systematics for the ground-based data, jointly detrending against the measured airmass of the LCO observation. When comparing our fit for all periods quoted in Section~\ref{sec:periodestimates}, we find modest evidence that a period of 17.1936 provides the best fit to the data. Given the multiple periods that are equally likely from Section~\ref{sec:periodestimates}, we suggest may be close to an orbital resonance with \planb{}, but further data is needed for a firm detection.}
    \label{fig:lco}
\end{figure}

We jointly fit the \tess data for \planc{} with the ground based LCO data, fitting every period that is consistent with the data derived in Section~\ref{sec:periodestimates}. We simultaneously detrend the LCO data against the reported airmass for the observation, and fit for a variable mean offset. We calculate the reduced chi squared fit of the model to the data, and find a slight preference for the period of 17.1936d. Figure~\ref{fig:lco} shows the best fitting model with the LCO data for \planc{}. Given that we were able to obtain a single egress event, we find moderate evidence that the period of 17.1936d is the correct period for \planc{}. If this is the correct period for \planc{}, this would put \planc{} and \planb{} very close to an orbital resonance of 5:3. Further data is needed to well constrain the period of \planc{}.

\subsection{MuSCAT2 observations}

MuSCAT2 (\citealp{Narita2019}) is a multi-colour optical camera mounted on the 1.52 m Telescopio Carlos S\'{a}nchez (TCS) at Teide Observatory, Tenerife, Spain. The instrument is able to obtain simultaneous observations in four bands: \textit{Sloan-g}, \textit{Sloan-r}, \textit{Sloan-i}, and \textit{Sloan-$z_s$}. The field of view of MuSCAT2 is $7.4 \times 7.4$ arcmin$^2$ with a pixel scale of 0.44 arcsec per pixel. With read out times between 1 and 4 seconds, MuSCAT2 an ideal instrument for transit follow-up and time-series observations in general. We observed two primary transits of TOI-1807.01 using MuSCAT2, using four bands on the nights of 8 and 13 of May 2020. For each night the field of view was slightly offset from the center in order to observe a bright reference star north of the target. The telescope was defocused and the exposure times for each band were set to avoid saturation of the target star. The data was reduced using standard procedures, and the photometry and transit model fit (including systematic effects) was done by the MuSCAT2 pipeline (for details see \citealp{Parviainen2019}, \citealp{Parviainen2020}). In both nights we could not detect the transit on target due to the shallow depth of the transit and the scatter in the light curves, nonetheless the MuSCAT2 data were useful to discard the presence of eclipsing binaries inside the TESS aperture.

\subsection{Statistical Validation}
\label{sec:triceratops}

In addition to the vetting performed above, we statistically validate each target to provide strong evidence of each being a bona fide planet. We do so using \texttt{triceratops} \citep{triceratops, 2020ascl.soft02004G} and \texttt{vespa} \citep{vespa}, algorithms that rule out astrophysical false positives by calculating and comparing the probabilities of various transit-producing scenarios. \texttt{triceratops} is a tool specifically designed for \tess observations, and considers transit scenarios originating from the target star and sources unresolved with the target star, in addition to transit scenarios originating from nearby resolved stars. Because of the low spatial resolution of TESS and the resulting flux contamination from nearby stars, the assumption that the transit originates from within the resolution limits of the target star is not valid for many planet candidates, so tools like \texttt{vespa} \citep{vespa}, (which was originally designed to validate planet candidates from Kepler and later adapted to TESS) are less widely applicable. \texttt{vespa} operates assuming that the transit originates from within the resolution limits of the target star, and therefore cannot be used for many TESS planet candidates. However, because the photometric follow-up described in Section~\ref{sec:photometricfollowup} rules out nearby resolved stars as transit sources for both \targb and \targa, both of these tools can be used to validate planet candidates in these systems.

As additional constraint in our calculations, we fold in the high-resolution imaging follow-up observations discussed in Section~\ref{sec:photometricfollowup}. Because these observations reveal no previously unresolved companions within their detection limits, incorporating the follow-up reduces the calculated probability of the transit originating from a bound or chance-aligned star within the resolution limits of the target star, thereby reducing the overall false positive probability (FPP) for each target. The ground based follow-up presented above directly informs our statistical validation. Below, we present the results of this analysis for each planet candidate.

\subsubsection{\targa}

We run \texttt{triceratops} for the three planet candidates around \targa 20 times each and calculate the mean and standard deviation of the resulting distributions of FPPs. We find ${\rm FPP} = (2.2 \pm 9.6) \times 10^{-6}$, ${\rm FPP} = (2.2 \pm 9.7) \times 10^{-15}$, and ${\rm FPP} = (1.2 \pm 5.1) \times 10^{-9}$ for planets b, c, and d, respectively. We also run \texttt{vespa} a single time for each planet candidate and find ${\rm FPP} = 4.6 \times 10^{-13}$, ${\rm FPP} = 3.8 \times 10^{-3}$, and ${\rm FPP} = 6.6 \times 10^{-10}$, respectively. These probabilities are low enough to consider the three planets validated.

\subsubsection{\targb}

We run \texttt{triceratops} for the planet candidate around \targb 20 times and calculate the mean and standard deviation of the resulting distribution of FPPs. We find ${\rm FPP} = (6.7 \pm 9.5) \times 10^{-6}$, which is below the threshold of ${\rm FPP} = 0.015$ required to validate a planet candidate with this tool. We run \texttt{vespa} a single time to ensure that the two tools provide the same result. With vespa, we find ${\rm FPP} = 1.4 \times 10^{-13}$. With these results strongly suggesting that the planet candidate is a bona fide planet, we consider the planet to be validated.

\section{Estimating the age of \targb and \targa}
\label{sec:age_estimate}

We make use of a number of indicators to estimate the ages of \targa and \targb. Young stars retain much of the angular momentum from their formation. As a result of the rapid rotation, young stars also exhibit extensive spot coverage and chromospheric activity. As such, for young Sun-like stars, we can often estimate their ages by their rotation periods, as measured from the light curve, and from the chromospheric activity indicators, such as core emission in the Calcium II lines, and their UV and X-ray fluxes. We describe each of these indicators in the sections below.

\begin{figure}
\includegraphics[width=0.5\textwidth]{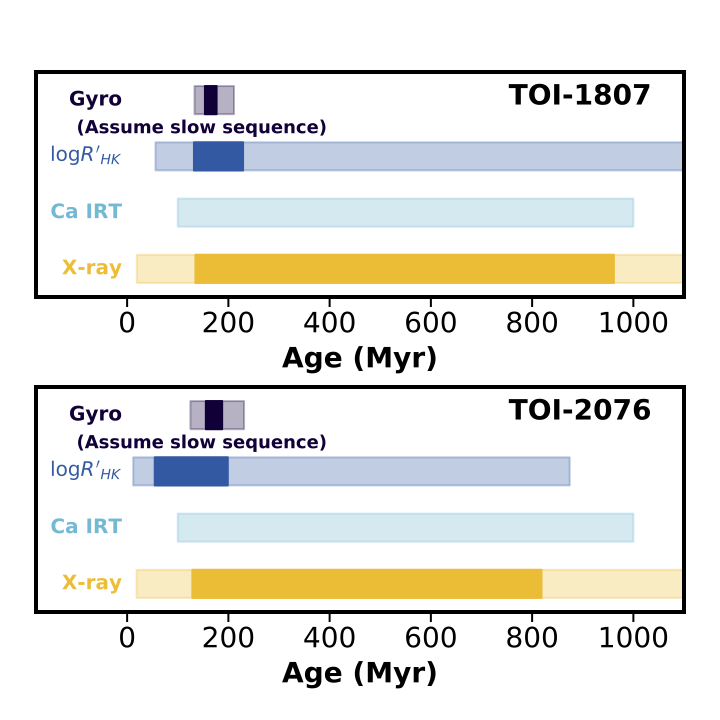}
\caption{A summary of the age estimates from various activity and rotational signatures of \targa and \targb. The $1\sigma$ estimates are shown by the solid region, and the $3\sigma$ estimates by the lightened region. We adopt the gyrochronology ages for both systems for the remainder of the analysis. }
\label{fig:age_summary}
\end{figure}

Figure~\ref{fig:age_summary} presents a summary of the quantitative age estimates we provide. We adopt the $3\sigma$ gyrochronology age estimates of $130-210$ Myr for \targb, and $125-230$ Myr for \targa, in our analyses. We show below that each of the other activity and spectroscopic indicator supports these gyrochronology estimates. We caution, however, that estimating the ages of single stars is always rife with caveats, and the estimates we provide should be taken with the necessary caution as is appropriate for their uncertainties. 

\subsection{Gyrochronology}
\label{sec:gyro}

Sun-like stars with convective envelopes and radiative cores spin down over their main-sequence lifetimes as mass is lost in the form of stellar wind. By comparing the rotation periods of our target stars against members of clusters and moving groups with known ages, we can place constraints on their ages. \targb and \targa exhibit significant spot-induced rotational modulation in their light curves. We make use of the \emph{TESS} continuous light curves and archival ground-based multi-year observations to estimate the rotation periods of these stars.

\begin{figure}
\includegraphics[width=0.5\textwidth]{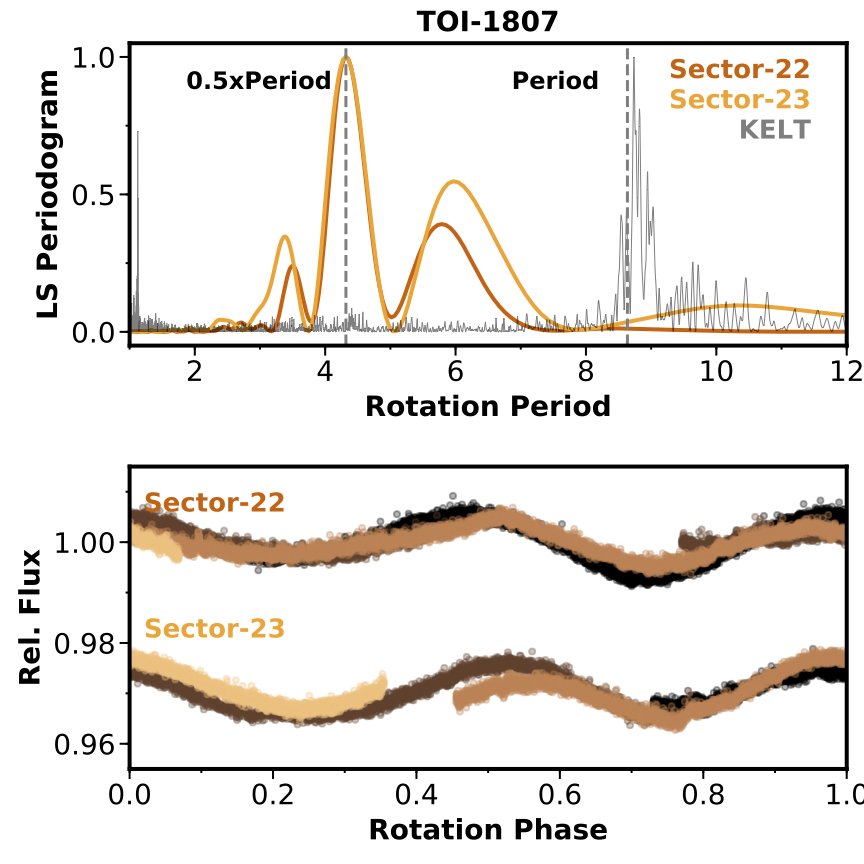}
\includegraphics[width=0.5\textwidth]{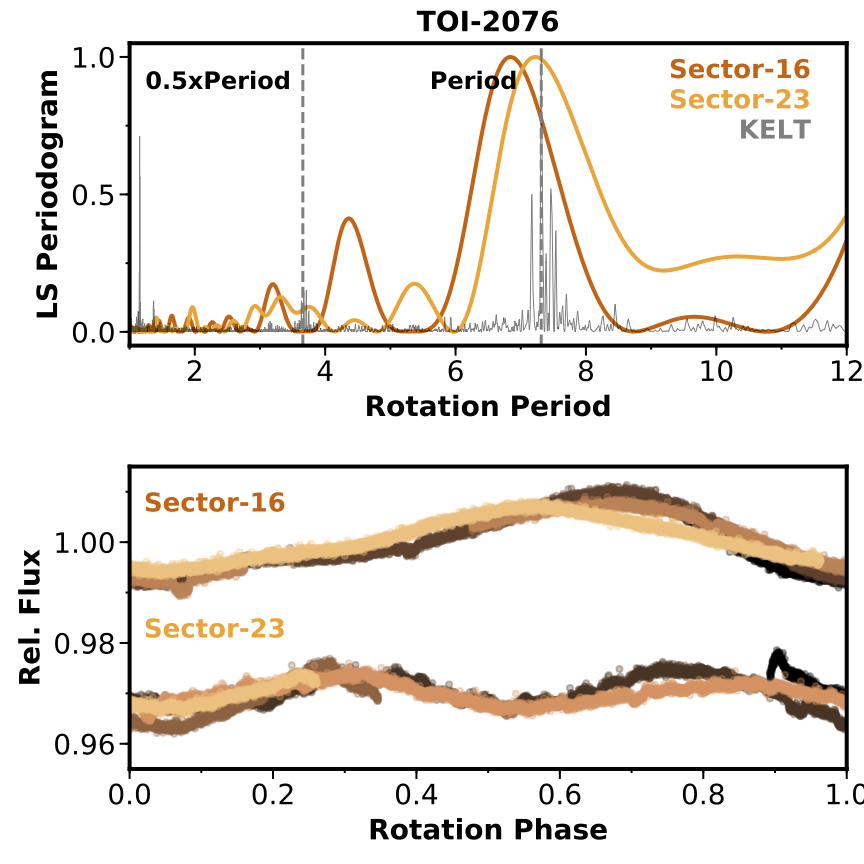}
\caption{\targb and 2076 received continuous photometric observations from \emph{TESS} and the ground-based KELT survey. (Top) The Lomb-Scargle periodograms and rotationally phased light curves of \targb are shown. The periodogram from each \emph{TESS} sector, and that from the KELT observations, are shown individually. (Bottom): The Lomb-Scargle periodogram and phased light curves for \targa are shown. }
\label{fig:rotlc}
\end{figure}

\targa received two sectors of \emph{TESS} observations over $\sim$28-day segments in September 2019 and March 2020, with significant spot evolution between the two separate sets of observations. We find a rotation period of $6.84\pm0.58$\,d and $7.22\pm0.77$\,d during Sectors 16 and 23 respectively (see Figure~\ref{fig:rotlc}). The uncertainties are estimated based on the full width at half maximum of the Lomb-Scargle periodogram peaks for each sector of observations. In addition, we made use of 8 years of light curves from KELT \citep{kelt1, kelt2, kelt3}, spanning between December 2006 to December 2014. A Lomb-Scargle periodogram showed a peak at 7.31\,d, consistent with that measured from the \emph{TESS} observations. Taking the mean and the scatter in the measured periods between TESS and KELT, we get a rotation period for \targa of $7.27\pm0.23$ days. 

\targb received two sectors of continuous \emph{TESS} observations over a period of 54 days, showing consistent stellar variability at the 2\% level. The Lomb-Scargle periodograms for each sector of the \emph{TESS} observations are shown in Figure~\ref{fig:rotlc}. Our initial analysis yielded a highest peak in the periodogram of $4.32\pm0.25$ and $4.317\pm0.26$ days for Sectors 22 and 23 respectively. However, further analysis of the long duration monitoring from the ground based KELT survey showed that the \emph{TESS} period peak is actually 1/2 that of the true rotation period. \targb was observed by the KELT survey from December 2006 to December 2014. The periodogram derived from these observations is also shown in Figure~\ref{fig:rotlc}, with a best matching period of 8.737 days, $2\times$ that from the \emph{TESS} light curves. Given the extensive coverage from the KELT survey, and the rapid evolution expected for such young stars, we adopt a period of $8.670 \pm 0.048$ days for \targb.

\begin{figure}
\includegraphics[width=0.5\textwidth]{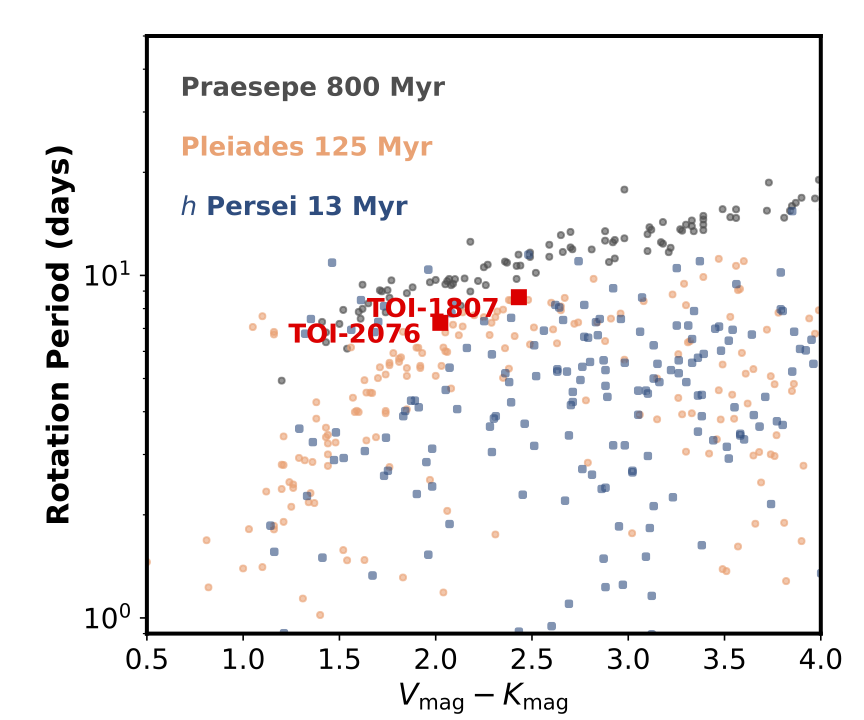}
\includegraphics[width=0.5\textwidth]{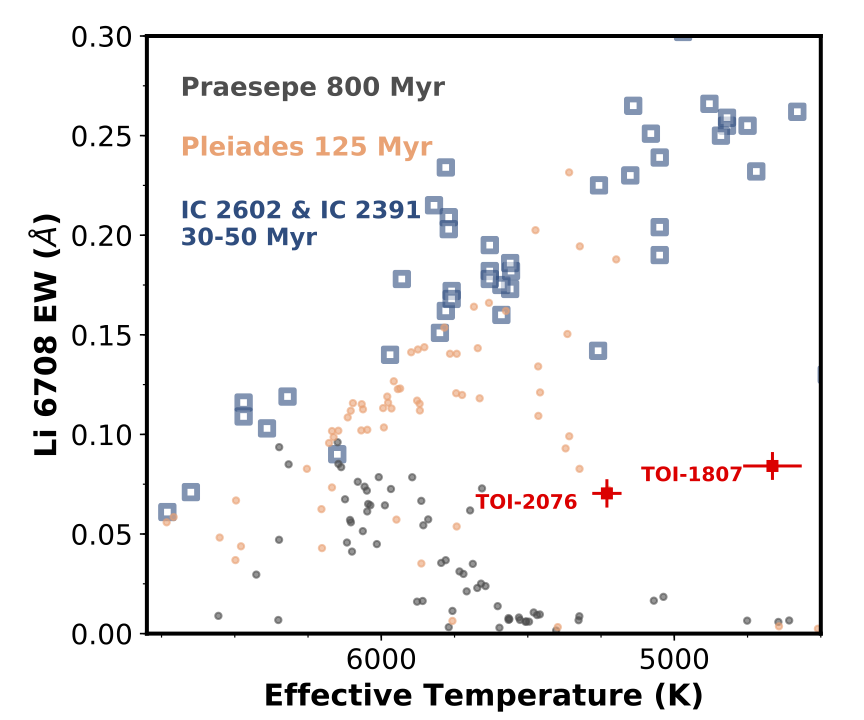}
\caption{Comparison of properties of \targa\ and \targb\ against stars from well characterized clusters and moving groups. \textit{Top:} Comparison of rotation periods. The distribution of rotation periods are shown against the $V-K$ color of each star. Well characterized stars from the 13 Myr old $h$ Persei cluster \citet{2013A&A...560A..13M} are marked in blue, after de-reddening with 3D dust maps from \emph{dustmap} \citep{2018JOSS....3..695M,2019ApJ...887...93G}. Stars from the 125 Myr old Pleiades cluster \citet{2016AJ....152..113R} are shown in orange, and from the 800 Myr old Praesepe cluster \citet{2017ApJ...839...92R} in grey. The periods of \targb and 2076 best resemble the Pleiades distribution, agreeing with our gyrochronology estimates for these stars. \textit{Bottom:} Comparison of Lithium abundances measured using high resolution spectroscopy from TRES. Lithium abundances show that \targa and \targb are both significantly younger than stars in Praesepe, and significantly older than stars in the IC 2602 and IC 2391 clusters.}
\label{fig:rot_vs_age}
\end{figure}

Figure~\ref{fig:rot_vs_age} compares the colors and rotation periods of \targa and \targb against members of well characterized moving groups and clusters. The target stars fall along the slow-rotation sequence of the 125 Myr old Pleiades cluster. Adopting the gyrochronology relationship from \citet{2007ApJ...669.1167B}, we find an estimated $3\sigma$ age range of $130-210$ Myr for \targb, and  $125-230$ Myr for \targa. To test the robustness of these estimates against the  specific calibration, we also apply the rotation-age relationship from \citet{2008ApJ...687.1264M}, and derive consistent age ranges of $135-205$ Myr for \targb, and $191-423$ Myr for \targa. 

\subsection{Stellar activity}
\label{sec:activity}

As a result of the rapid rotation, young stars exhibit significant chromospheric emission visible in the X-ray and specific activity-sensitive optical features. 

\targb and \targa are X-ray sources in the ROSAT All sky survey \citep{2000yCat.9029....0V}. We convert the X-ray fluxes to X-ray luminosities via the calibration from \citet{1995ApJ...450..401F}, and place age limits from these X-ray luminosities via \citet{2008ApJ...687.1264M} (Equation A3). \targb has an X-ray luminosity of $\log L_\mathrm{X}/L_\mathrm{Bol} = -4.53 \pm 0.24$, and an estimated $3\sigma$ age lower limit of $>19 Myr$. Similarly, \targa has an X-ray luminosity of $\log L_\mathrm{X}/L_\mathrm{Bol} = -4.49 \pm 0.16$, corresponding to a $3\sigma$ age lower limit of $>18$ Myr.

Similarly, chromospheric emission in the cores of the Calcium II lines are also qualitatively informative on the ages of systems. There is significant core emission in the Calcium II H and K lines, as well as the Calcium II infrared triplet lines from the TRES spectra of \targb and \targa. 

Using the calibrations provided in \citet{Zhou2021}, we measured equivalent widths for the core emission in the Calcium II H,\&K lines, and convert them to the Mount Wilson Observatory HK Project \citep{1978ApJ...226..379W,1978PASP...90..267V,1991ApJS...76..383D,1995ApJ...438..269B} $S_{HK}$ indices for both target stars. We measure $S_{HK} = 1.008\pm0.074$ and $S_{HK} = 0.776\pm0.090$ for \targb and 2076 respectively; these were converted to the bolometric flux ratios of $\log R'_{HK} = -4.409 \pm 0.033$ and $\log R'_{HK} = -4.271 \pm 0.056$ respectively. 

Like X-ray, the level of Calcium II core emission is related to the rotation, and therefore age, of the target stars. We make use of the calibration offered by \citet{2008ApJ...687.1264M} (Equation 3) to yield $3\sigma $ age ranges of 60-1800 Myr for \targb, and 12-870 Myr for \targa. 

Similarly, we also follow \citet{Zhou2021} and measured the levels of core emission in the Calcium II infrared triplet lines, finding equivalent widths of $0.36 \pm 0.01$\,\AA{} for \targb and $0.33\pm0.01$\,\AA{} for \targa. Using the qualitative relationships provided in \citet{2017ApJ...835...61Z}, these core emissions are consistent with stars with ages between 100-1000 Myr of age.

\subsection{Lithium absorption}
\label{sec:lithium}

Lithium is rapidly depleted in the envelope of Sun-like stars within the first few hundred million years post formation, as it is convectively mixed into the core and destroyed through proton collisions. The Lithium 6708 \AA{}\ line is therefore often a reliable and easily accessible indicator of youth for young Sun-like stars. Both \targb and \targa exhibit significant Lithium absorption features. We measured Li 6708\,\AA{}\ equivalent widths for these target stars using the high resolution observations from the TRES facility, as per the techniques described in \citet{Zhou2021}, with equivalent widths of $0.0841 \pm 0.0070$~\AA\ and $0.0703 \pm 0.0071$~\AA\ for \targb and 2076 respectively. 

Figure~\ref{fig:rot_vs_age} places the Lithium absorption strength measured for \targb and 2076 into context with other well characterized clusters. As the Lithium absorption strength is dependent on a large number of additional factors, such as rotational evolution and metallicity, we do not derive quantitative ages from the equivalent width measurements. It is clear, however, that these target stars have ages significantly younger than stars in the 800 Myr old Praesepe cluster, and significantly older than the 50 Myr old clusters IC2602 and IC2391.


\section{Discussion}
\label{sec:discussion}


The planets transiting TOI-1807 and TOI-2076 are valuable benchmarks for studying the evolution of small planets. Transiting planets around young ($<$1~Gyr) stars are still relatively rare, and it remains to be seen if this is due to the scarcity of young stars amenable to transit searches, an age-dependence to detection efficiency and/or planet occurrence rates, a lack of precise and accurate ages for planet hosts, or some combination of these effects.    

An especially compelling use case provided by young transiting planets is the possibility of constraining models of radial contraction and atmospheric loss \citep[e.g.][]{OwenWu2013, LopezFortney2013, Jin2014, Chen2016, Ginzburg2016}. For example, one challenge in modeling the atmospheric evolution of planets with a photoevaporation model is the unknown X-ray and extreme ultraviolet (XUV) evolution of the host star \citep[e.g.][]{Kubyshkina2019b,Kubyshkina2019a, OwenCamposEstrada2020}. This is because uncertainties in the time-integrated XUV exposure of a given planet are larger for stars with older and less precise ages, which could have had a wide range of XUV luminosities early in their lives. The X-ray and UV luminosities of nearby, young planet hosts can be directly measured and, provided some knowledge of the stellar age and planetary masses, allow for detailed modeling of the past \citep[e.g.][]{Owen2020} and future \citep[e.g.][]{Poppenhaeger2021} evolution of a planetary system.  

In this context, the most intriguing observations about the TOI-2076 system are the relatively large planet sizes (b, c, and d have radii of 3.2, 4.5 and 4.0 $R_{\earth}$ respectively). All of the transiting planets detected around pre-main sequence stars appear to have anomalously large sizes when compared to exoplanets around field stars, while planets with ages of 0.5--1~Gyr appear to have sizes that are more consistent with those of the field population \citep[see e.g.][and references therein]{Mann2017, David2019b, Livingston2019, Bouma2020, Tofflemire2021}. It remains to be seen whether this size-age correlation is astrophysical or due to lower detection efficiencies for young stars \citep[e.g.][]{Zhou2021}. The TOI-1807 and TOI-2076 systems exist at an intermediate age (0.1--0.5~Gyr) when the most dramatic effects of photoevaporation are expected to be complete \citep{Jackson2012}, though mass loss may proceed further over gigayear timescales for some planets through either core-cooling \citep{Gupta2020} or photoevaporation \citep{Rogers2021}. 

\begin{figure}
    \centering
    \includegraphics[width=\linewidth]{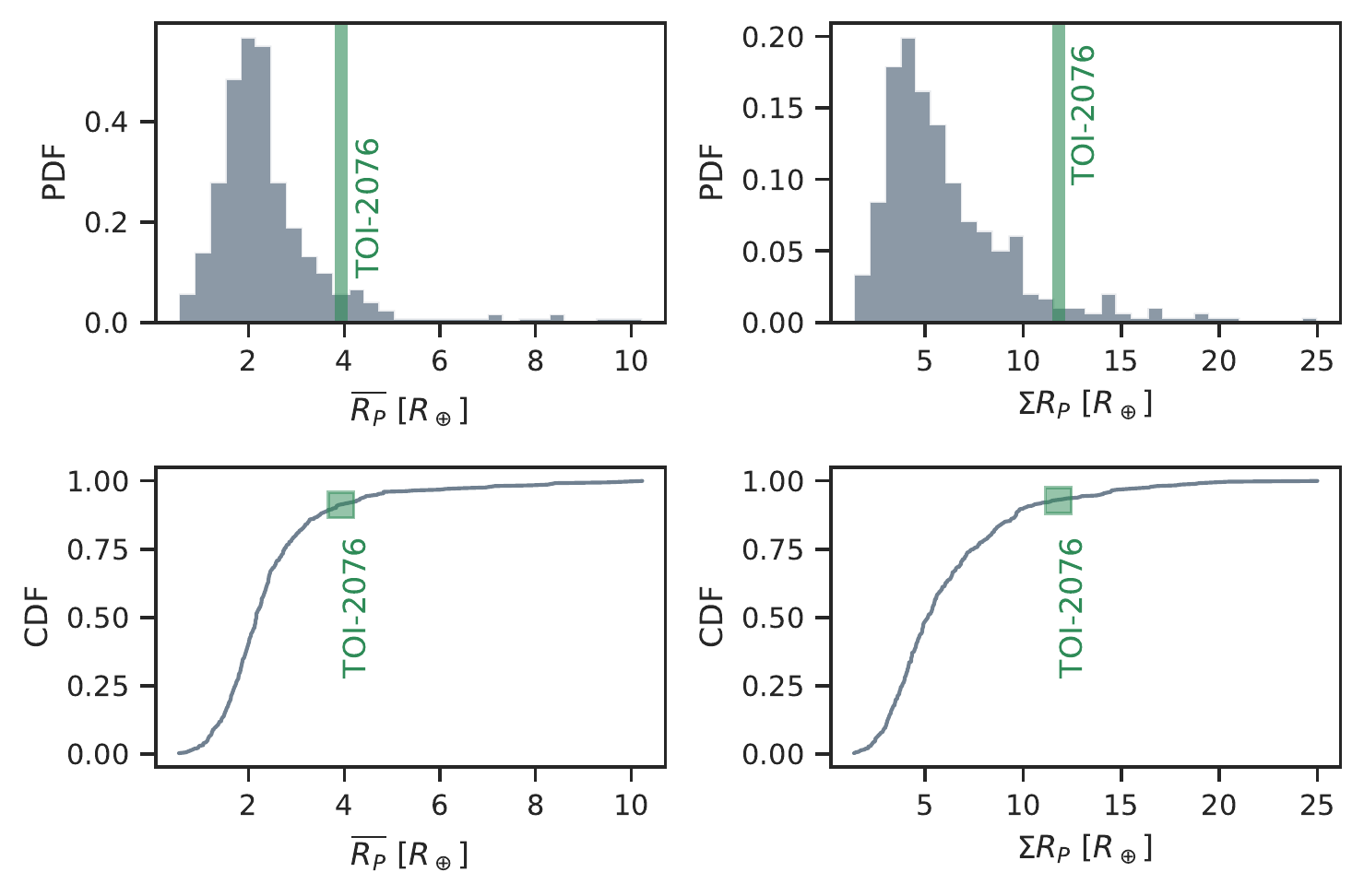}
    \caption{Probability distribution functions (top row) and cumulative distribution functions (bottom row) for the average planet size (left column) and sum of planet sizes in confirmed multi-transiting systems around GK stars.}
    \label{fig:toi2076}
\end{figure}

To place the \targa{} system in the broader context of multi-planet systems we queried the NASA Exoplanet Archive \citep{Akeson2013} for all confirmed, multi-transiting systems with GK host stars (4000~K~$<$~\teff~$<$~6000~K). For each system we computed the average planet radius and the sum of planetary radii (regardless of how many planets were in each system). We then compared the equivalent values for TOI-2076 to the empirical probability distribution functions (PDFs) and cumulative distribution functions (CDFs) of the Exoplanet Archive sample (Fig.~\ref{fig:toi2076}). We found that the average planet radius and sum of planetary radii in the TOI-2076 system are larger than 91\% and 93\% of the equivalent values in confirmed multi-planet systems, respectively. While we can not prove a causal link, it is intriguing that the TOI-2076 system extends the trend of large planetary radii observed in other young systems.


TOI-1807~b is particularly interesting as it belongs to a distinct class of planets known as ultra-short period planets \citep[USPs, see][for a review]{Winn2018}. USPs are intrinsically rare, with an occurrence rate of $\lesssim$0.5--1\% around G and K-type stars \citep{SanchisOjeda2014, Steffen2016}. Despite being about equally as rare as hot Jupiters, USPs are almost certainly unrelated to their more massive and distant cousins: they lack a strong preference for metal-rich hosts \citep{Winn2017}, they almost always occur in compact multi-planet systems \citep{SanchisOjeda2014, Adams2020}, and they are more common around lower-mass stars \citep{SanchisOjeda2014}. All of these trends run counter to what has been established for hot Jupiters. That being established, USPs may well be the remnant cores of sub-Neptunes.  

Several lines of evidence suggest that USPs, including TOI-1807 b, did not form in their current orbits, likely underwent inward migration, and are the result of non-standard evolution. This evidence includes (1) present-day USP orbits lie interior to the dust sublimation radii of typical protoplanetary disks \citep[e.g.][]{Muzerolle2003, Eisner2005}, (2) the observed period ratios between USPs and neighboring planets are much larger than the period ratios typically observed in multi-transiting systems \citep{Steffen2013}, (3) the planet occurrence rate is a steeper function of period inside 1 day relative to the rates in the 1-10 day, or 10-100 day range \citep{LeeChiang2017}, (4) USPs occur in multi-planet systems with larger-than-average mutual inclinations \citep{Dai2018}, and (5) well-characterized USPs are always smaller than 2~$R_\oplus$, having densities consistent with rocky compositions \citep{Dai2019}. The size cut-off of USPs is seen as potential evidence that some experienced atmospheric loss.

\citet{Millholland2020} provided a recent review of the most promising theories for how USPs arrived on their observed orbits, all of which involve tidal dissipation and the accompanying orbital decay. Briefly, these theories can be summarized as: (1) in situ formation near the inner edge of the protoplanetary disk, followed by tidal dissipation \textit{in the star} \citep{LeeChiang2017}, and (2) planet-planet interactions followed by tidal dissipation \textit{in the planet} driven by either the planet's orbital eccentricity \citep{Schlaufman2010, Petrovich2019, Pu2019} or the planetary obliquity \citep[the angle between the planet's spin axis and orbital angular momentum vector][]{Millholland2020}. The latter class of theories naturally account for the high planet multiplicity and mutual inclinations in USP systems.

As the youngest USP detected to date, TOI-1807~b places stringent limits on theories for the formation and evolution of these rare planets. The discovery of an USP around a young star is compatible with a fast formation channel, which is also suggested by a comparison of galactic velocity dispersions between USP hosts and field stars \citep{HamerSchlaufman2020}. If stellar activity can be mitigated, radial velocity follow-up of TOI-1807 should lead to the discovery of the additional non-transiting planets that likely exist; this would help piece together a coherent picture of the past dynamics of the system which may have driven TOI-1807~b inwards.

To place TOI-1807~b in a broader observational context we computed the JWST Emission Spectroscopy Metric \citep[ESM,][]{TSM} for all confirmed USPs ($P<$1~day, $R_P<$2~$R_\oplus$) on the NASA Exoplanet Archive, assuming the Bond albedo of Earth ($A_B$=0.306). We found that TOI-1807~b is the third most favorable USP for the detection of mid-IR thermal emission (Table~\ref{tab:usp-esm} and Fig.~\ref{fig:youngexplainer}). Notably, the two planets which rank more favorably, 55~Cnc~e and LHS~3844~b, have securely detected mid-IR phase curves and secondary eclipses \citep{Demory2016, Kreidberg2019} while K2-141~b, which ranks below TOI-1807~b, has a detected phase curve and secondary eclipse from K2 optical photometry \citep{Malavolta2018}.

Thus, the TOI-1807 system offers an opportunity to study a small, likely rocky planet shortly after its formation and perhaps after recently losing its atmosphere. The youth of TOI-1807~b makes it an even more compelling target for secondary eclipse spectroscopy, as the luminosity of the planet's cooling core may be an order of magnitude higher than it would be at older ($>1$~Gyr) ages \citep{Linder2019}. Finally, as a candidate ``lava world" \citep{Chao2020}, TOI-1807~b presents an opportunity to study the early evolution of these poorly-understood objects. 

\begin{deluxetable}{cc}
\tablecaption{JWST Emission Spectroscopy Metric for the most favorable ultra-short period planets \label{tab:usp-esm}}
\tablewidth{0pt}
\tablehead{
\colhead{Planet name} & \colhead{ESM}
}
\startdata
55~Cnc~e & 101.0 \\ 
LHS~3844~b &  51.4 \\
\textbf{TOI-1807~b} & \textbf{36.9} \\
GJ~1252~b & 26.6 \\
LTT~3780~b & 23.3 \\
K2-141~b & 21.5 \\
HD~3167~b & 20.0 \\
LP~791-18~b & 12.2 \\
TOI-561~b & 11.5 \\
K2-131~b & 9.7 \\
\enddata
\end{deluxetable}


TOI-2076 and TOI-1807 are coeval and comoving; these young stars likely formed together, though we their large physical separation (>9 pc) suggests they are not bound. Theoretical studies show that very close stellar companions can have a significant effect on planet formation; close companions can 1) truncate the proto-planetary disk, preventing planetary formation \citep{jang-condell2015}, 2)  trigger the migration of giant planets, 3) eject smaller planets, and 4) disperse the disk before or during planetary formation \citep{cieza2009}. Systems as widely separated as these two stars essentially evolve as single stars, and we know little of their formation processes and any interrelationship that may be present. The detection of transiting planets in both \targa and \targb reveals that the planetary orbital planes are co-aligned, which hints at a common formation process whereby the both components maintain a nearly-edge on inclination to our line of sight. There is some initial evidence of such alignment between planetary orbits and the orbits of their binary hosts \citep{colton2021} with more evidence to come from high-resolution imaging studies such as \cite{howell2021}.

\subsection{Opportunities for Follow-up Observations}
We have reported the detection and validation of \planall{} and \planbtwo{}. These systems are extremely valuable to the community. The youth of the host stars place \targa and \targb in a valuable parameter space. The proximity of the host stars (40pc) could make these targets excellent candidates for follow up with direct imaging surveys, to search for longer period companions. 

\begin{figure}
    \centering
    \includegraphics[width=0.5\textwidth]{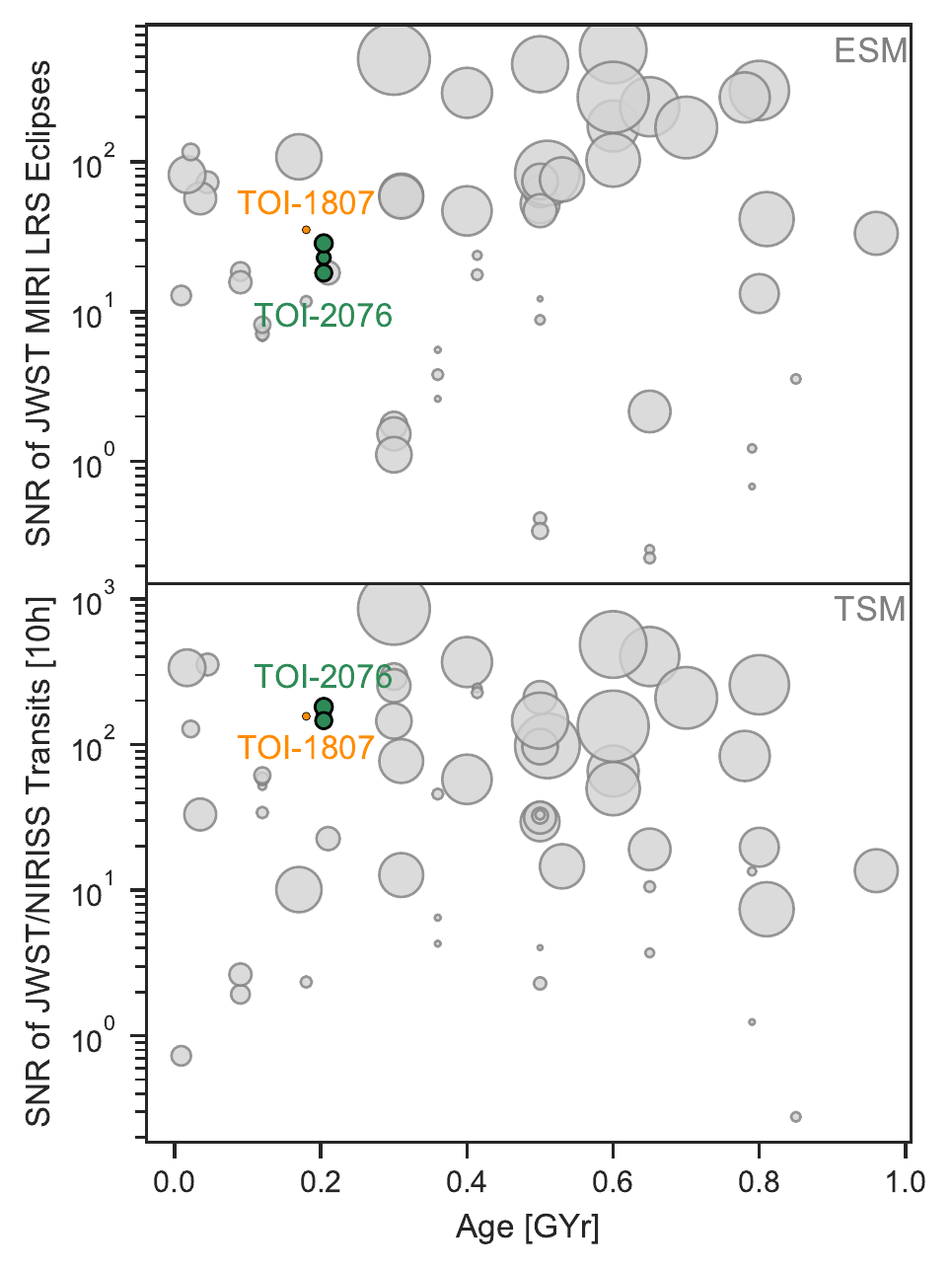}
    \caption{The Emission Spectroscopy Metric (ESM) and Transmission Spectroscopy Metric (TSM) from \cite{TSM} for the sample of confirmed, young, transiting exoplanets (grey), highlighting \targa and \targb. Points are scaled to represent the relative sizes of each planet. Top: ESM as a function of stellar age, not accounting for any residual heat due to formation. \targb Shows a high signal to noise value, pointing to a possible detectable secondary eclipse, despite \planbtwo{} being a small planet. Bottom: TSM as a function of age. All planets show a high TSM compared to other known young transiting planets, indicating these are excellent candidates for followup with JWST.}
    \label{fig:youngexplainer}
\end{figure}

The bright, small host stars also provide an unparalleled opportunity to observe small, young planets in both transmission and emission using the James Webb Space Telescope (JWST) close to a crucial transition age in planet formation. Figure~\ref{fig:youngexplainer} shows the Emission Spectroscopy Metric (ESM) and Transmission Spectroscopy Metric (TSM) from \cite{TSM} for the current sample of confirmed, young transiting planets\footnote{\texttt{https://exoplanetarchive.ipac.caltech.edu/} accessed Jan 2021}. The ESM provides an estimate of the signal to noise ratio of a secondary eclipse in JWST's MIRI LRS bandpass, and the TSM provides the signal to noise ratio of a ten hour observation in JWST's NIRISS, not accounting for the presence of clouds. These values do not account for any residual energy from planet formation, and only account for the atmosphere signal due to heating at the equilibrium temperature of the planet. \targa and \targb are highlighted. We note that 1) there are few known transiting planets close to the $\sim$100Myr age 2) \planbtwo{} has the most observable emission of any small, young planet. 3) \planb{}, \planc{}, \pland{}, and \planbtwo{} are all excellent candidates for transmission spectroscopy with JWST, providing enough signal to noise for an atmosphere detection with just one transit.



One crucial step towards effective atmospheric characterization is obtaining mass measurements for the planets. 
The brightness of \targa and \targb makes them amenable to ground-based radial velocity (RV) follow-up, though the significant stellar activity may make detection more difficult. 
As a first-order guess, we estimate the planet masses using the probabilistic radius-to-mass conversion from \citet{Chen2017} and calculate the expected RV semi-amplitude for a zero-eccentricity orbit. 
This yields an expected semi-amplitude $K = 2.9^{+2.3}_{-0.8}$ m s$^{-1}$ for \planbtwo{} and $K = 3.2^{+2.3}_{-1.4}$ m s$^{-1}$ for \planb{}. The RV signal strength is even less certain for \planc{} and \pland{} due to their unknown orbital periods, but based on the likely periods given in Section~\ref{sec:periodestimates}, their RV semi-amplitudes should be on the order of 3-4 m s$^{-1}$ as well. 

Since both targets are young and active stars, they are likely to exhibit RV jitter on order of 10s to 100s of m s$^{-1}$, well in excess of the photon noise limit for a typical RV spectrograph \citep{Luhn2020}. 
The primary challenge in measuring the planet masses through RV is then mitigating the stellar activity, particularly since it is likely to be larger in amplitude than the Doppler signal. 
In this sense \planbtwo{} is the most promising target for mass follow-up, as its ultra-short orbital period suggests that the Keplerian signal will be separable from activity at the rotation period of the star. 
As demonstrated by recent Rossiter-McLaughlin (RM) effect measurements on very young and active stars, it is feasible to measure short-duration RV signals on timescales of hours even in the presence of high-amplitude stellar variability on longer timescales of days \citep{Montet2020, Zhou2020}.
RV follow-up of \planall{} will be more challenging, since the planetary orbital periods are comparable to the stellar rotation and the complexity of the multi-planet RV signal requires a larger number of observations. 
That being said, recent work on other young systems like K2-100 and AU Mic has effectively employed stellar activity models to extract RV constraints in the face of considerable activity \citep{Barragan2019, Klein2021}, and these targets are prime examples of the importance of developing such methods.

We note also that similar RV amplitudes are expected for the RM effect of these planets. 
The spin-orbit alignment measurement enabled by RM measurements would be particularly valuable for constraining the formation and migration histories of these planets. 
\planc{} is expected to be the best RM target, with an amplitude on the order of 10 m s$^{-1}$ \citep{Triaud2018}, possibly within reach of modern observations. This amplitude is not sensitive to the unknown orbital period of the planet, although refined ephemerides will of course be necessary to obtain the required in-transit observations. 

A previous study of ultra-short-period planets by \citet{SanchisOjeda2014} concluded that USP planets often have longer period coplanar companions in the period range $\leq 50$ days. Because the transit probability of USP planets in these multi-planet systems is significantly higher than the than longer period companions, systems with a single transiting USP planet are likely to also have non-transiting outer planets. While we identified no longer-period transiting planets in the \targb\ system, radial velocity measurements of \targb, and perhaps future direct imaging observations, may reveal additional planets in this system. 

Although the known planets detected in transit are too close to their stars to be directly imaged, giant planets in the outer reaches of the TOI-2076 and TOI-1807 systems may be more detectable. Due to the young age of the stars, giant planets would still be cooling from formation and would therefore appear brighter at infrared wavelengths than mature Jovian planets \citep[e.g.][]{burrows}. Depending on their masses, ages, formation conditions, and cooling rate, massive Jovian planets orbiting TOI-2076 and TOI-1807 at separations comparable to Saturn, Uranus, or Neptune (10 - 30 AU) could be within reach of current and upcoming instruments \citep[e.g.][]{bowler, lacy}. 

\section{Conclusions}
In this paper we have presented and validated two systems of planets around two young, comoving stars. These planets could provide a unique opportunity for further study by characterizing atmospheres in transmission, emission, and phase curves in the immediate future. The host star variability may make radial velocity observations challenging, but in the case of \planbtwo{} we expect mass measurements to be accessible.  Their close proximity to earth could make them excellent candidates for direct imaging. In the case of the USP \planbtwo{}, we may expect further, long period planets to be present. The potential for a joint formation history of these two host stars make them a unique opportunity to intercompare planet systems with the same starting conditions, but different outcomes. We suggest \targa and \targb are exceptional candidates for further follow-up, and to further our understanding of young planets. 


\section{Acknowledgements}

This research made use of \texttt{lightkurve}, a Python package for Kepler and TESS data analysis \citep{Lightkurve}. This research made use of Astropy,\footnote{http://www.astropy.org} a community-developed core Python package for Astronomy \citep{astropy:2013, astropy:2018}.
This research made use of exoplanet \citep{exoplanet:exoplanet} and its
dependencies \citep{exoplanet:agol19, astropy:2013, astropy:2018, exoplanet:kipping13, exoplanet:luger18, exoplanet:pymc3,
exoplanet:theano}.
This paper includes data collected with the TESS mission, obtained from the MAST data archive at the Space Telescope Science Institute (STScI). Funding for the TESS mission is provided by the NASA Explorer Program. We acknowledge the use of public TESS data from pipelines at the TESS Science Office and at the TESS Science Processing Operations Center.
This research has made use of the Exoplanet Follow-up Observation Program website, which is operated by the California Institute of Technology, under contract with the National Aeronautics and Space Administration under the Exoplanet Exploration Program. STScI is operated by the Association of Universities for Research in Astronomy, Inc., under NASA contract NAS 5–26555.
The National Geographic Society - Palomar Observatory Sky Atlas (POSS-I) was made by the California Institute of Technology with grants from the National Geographic Society.
The Second Palomar Observatory Sky Survey (POSS-II) was made by the California Institute of Technology with funds from the National Science Foundation, the National Geographic Society, the Sloan Foundation, the Samuel Oschin Foundation, and the Eastman Kodak Corporation.
Resources supporting this work were provided by the NASA High-End Computing (HEC) Program through the NASA Advanced Supercomputing (NAS) Division at Ames Research Center for the production of the SPOC data products.
Funding for this work for CH is provided by grant number 80NSSC20K0874, through NASA ROSES. T.J.D. acknowledges support for this work from the TESS Guest Investigator program under NASA grant 80NSSC20K0631.
This work is partly supported by JSPS KAKENHI Grant Numbers JP18H01265 and JP18H05439, and JST PRESTO Grant Number JPMJPR1775, and a University Research Support Grant from the National Astronomical Observatory of Japan (NAOJ).
This work makes use of observations from the LCOGT network. Part of the LCOGT telescope time was granted by NOIRLab through the Mid-Scale Innovations Program (MSIP). MSIP is funded by NSF. Some of the observations in the paper made use of the High-Resolution Imaging instrument ‘Alopeke. ‘Alopeke was funded by the NASA Exoplanet Exploration Program and built at the NASA Ames Research Center by Steve B. Howell, Nic Scott, Elliott P. Horch, and Emmett Quigley. ‘Alopeke was mounted on the Gemini North (and/or South) telescope of the international Gemini Observatory, a program of NSF’s OIR Lab, which is managed by the Association of Universities for Research in Astronomy (AURA) under a cooperative agreement with the National Science Foundation on behalf of the Gemini partnership: the National Science Foundation (United States), National Research Council (Canada), Agencia Nacional de Investigación y Desarrollo (Chile), Ministerio de Ciencia, Tecnología e Innovación (Argentina), Ministério da Ciência, Tecnologia, Inovações e Comunicações (Brazil), and Korea Astronomy and Space Science Institute (Republic of Korea).
This publication makes use of data products from the Wide-field Infrared Survey Explorer, which is a joint project of the University of California, Los Angeles, and the Jet Propulsion Laboratory/California Institute of Technology, funded by the National Aeronautics and Space Administration.
This article is based on observations made with the MuSCAT2 instrument, developed by ABC, at Telescopio Carlos Sánchez operated on the island of Tenerife by the IAC in the Spanish Observatorio del Teide.
This paper is partially based on observations made with the Nordic Optical Telescope, operated by the Nordic Optical Telescope Scientific Association at the Observatorio del Roque de los Muchachos, La Palma, Spain, of the Instituto de Astrofisica de Canarias.
D.J.S. is supported as an Eberly Research Fellow by the Eberly College of Science at the Pennsylvania State University. The Center for Exoplanets and Habitable Worlds is supported by the Pennsylvania State University, the Eberly College of Science, and the Pennsylvania Space Grant Consortium.
JNW thanks the Heising-Simons foundation for support.

\facilities{TESS, LCOGT, Gemini Observatory, Lick Observatory, FLWO}
    \software{lightkurve \citep{Lightkurve}, exoplanet \citep{exoplanet:exoplanet}, astropy \citep{astropy:2013, astropy:2018},  AstroImageJ \citep{Collins:2017}, TAPIR \citep{Jensen:2013}, triceratops \citep{triceratops}, vespa \citep{vespa}, vetting, pyia \citep{pyia}}


\bibliographystyle{mnras}
\bibliography{bibli}

\begin{thebibliography}{}
\makeatletter
\relax
\def\mn@urlcharsother{\let\do\@makeother \do\$\do\&\do\#\do\^\do\_\do\%\do\~}
\def\mn@doi{\begingroup\mn@urlcharsother \@ifnextchar [ {\mn@doi@}
  {\mn@doi@[]}}
\def\mn@doi@[#1]#2{\def\@tempa{#1}\ifx\@tempa\@empty \href
  {http://dx.doi.org/#2} {doi:#2}\else \href {http://dx.doi.org/#2} {#1}\fi
  \endgroup}
\def\mn@eprint#1#2{\mn@eprint@#1:#2::\@nil}
\def\mn@eprint@arXiv#1{\href {http://arxiv.org/abs/#1} {{\tt arXiv:#1}}}
\def\mn@eprint@dblp#1{\href {http://dblp.uni-trier.de/rec/bibtex/#1.xml}
  {dblp:#1}}
\def\mn@eprint@#1:#2:#3:#4\@nil{\def\@tempa {#1}\def\@tempb {#2}\def\@tempc
  {#3}\ifx \@tempc \@empty \let \@tempc \@tempb \let \@tempb \@tempa \fi \ifx
  \@tempb \@empty \def\@tempb {arXiv}\fi \@ifundefined
  {mn@eprint@\@tempb}{\@tempb:\@tempc}{\expandafter \expandafter \csname
  mn@eprint@\@tempb\endcsname \expandafter{\@tempc}}}

\bibitem[\protect\citeauthoryear{{Adams} et~al.,}{{Adams}
  et~al.}{2020}]{Adams2020}
{Adams} E.~R.,  et~al., 2020, arXiv e-prints, \href
  {https://ui.adsabs.harvard.edu/abs/2020arXiv201111698A} {p. arXiv:2011.11698}

\bibitem[\protect\citeauthoryear{{Agol}, {Luger}  \& {Foreman-Mackey}}{{Agol}
  et~al.}{2020}]{exoplanet:agol19}
{Agol} E.,  {Luger} R.,   {Foreman-Mackey} D.,  2020, \mn@doi [\aj]
  {10.3847/1538-3881/ab4fee}, \href
  {https://ui.adsabs.harvard.edu/abs/2020AJ....159..123A} {159, 123}

\bibitem[\protect\citeauthoryear{{Akeson} et~al.,}{{Akeson}
  et~al.}{2013}]{Akeson2013}
{Akeson} R.~L.,  et~al., 2013, \mn@doi [\pasp] {10.1086/672273}, \href
  {https://ui.adsabs.harvard.edu/abs/2013PASP..125..989A} {125, 989}

\bibitem[\protect\citeauthoryear{{Astropy Collaboration} et~al.,}{{Astropy
  Collaboration} et~al.}{2013}]{astropy:2013}
{Astropy Collaboration} et~al., 2013, \mn@doi [\aap]
  {10.1051/0004-6361/201322068}, \href
  {http://adsabs.harvard.edu/abs/2013A%26A...558A..33A} {558, A33}

\bibitem[\protect\citeauthoryear{{Astropy Collaboration} et~al.,}{{Astropy
  Collaboration} et~al.}{2018}]{astropy:2018}
{Astropy Collaboration} et~al., 2018, \mn@doi [aj] {10.3847/1538-3881/aabc4f},
  \href {https://ui.adsabs.harvard.edu/abs/2018AJ....156..123A} {156, 123}

\bibitem[\protect\citeauthoryear{{Baliunas} et~al.,}{{Baliunas}
  et~al.}{1995}]{1995ApJ...438..269B}
{Baliunas} S.~L.,  et~al., 1995, \mn@doi [\apj] {10.1086/175072}, \href
  {https://ui.adsabs.harvard.edu/abs/1995ApJ...438..269B} {438, 269}

\bibitem[\protect\citeauthoryear{{Barnes}}{{Barnes}}{2007}]{2007ApJ...669.1167B}
{Barnes} S.~A.,  2007, \mn@doi [\apj] {10.1086/519295}, \href
  {https://ui.adsabs.harvard.edu/abs/2007ApJ...669.1167B} {669, 1167}

\bibitem[\protect\citeauthoryear{{Barrag{\'a}n} et~al.,}{{Barrag{\'a}n}
  et~al.}{2019}]{Barragan2019}
{Barrag{\'a}n} O.,  et~al., 2019, \mn@doi [\mnras] {10.1093/mnras/stz2569},
  \href {https://ui.adsabs.harvard.edu/abs/2019MNRAS.490..698B} {490, 698}

\bibitem[\protect\citeauthoryear{Battley, Pollacco  \& Armstrong}{Battley
  et~al.}{2020}]{Battley_2020}
Battley M.~P.,  Pollacco D.,   Armstrong D.~J.,  2020, \mn@doi [Monthly Notices
  of the Royal Astronomical Society] {10.1093/mnras/staa1626}, 496, 1197–1216

\bibitem[\protect\citeauthoryear{{Becker} et~al.,}{{Becker}
  et~al.}{2019}]{Becker2019}
{Becker} J.~C.,  et~al., 2019, \mn@doi [\aj] {10.3847/1538-3881/aaf0a2}, \href
  {https://ui.adsabs.harvard.edu/abs/2019AJ....157...19B} {157, 19}

\bibitem[\protect\citeauthoryear{{Berger}, {Huber}, {Gaidos}, {van Saders}  \&
  {Weiss}}{{Berger} et~al.}{2020}]{Berger2020}
{Berger} T.~A.,  {Huber} D.,  {Gaidos} E.,  {van Saders} J.~L.,   {Weiss}
  L.~M.,  2020, \mn@doi [\aj] {10.3847/1538-3881/aba18a}, \href
  {https://ui.adsabs.harvard.edu/abs/2020AJ....160..108B} {160, 108}

\bibitem[\protect\citeauthoryear{{Bouma} et~al.,}{{Bouma}
  et~al.}{2020}]{Bouma2020}
{Bouma} L.~G.,  et~al., 2020, \mn@doi [\aj] {10.3847/1538-3881/abb9ab}, \href
  {https://ui.adsabs.harvard.edu/abs/2020AJ....160..239B} {160, 239}

\bibitem[\protect\citeauthoryear{{Bowler}}{{Bowler}}{2016}]{bowler}
{Bowler} B.~P.,  2016, \mn@doi [\pasp] {10.1088/1538-3873/128/968/102001},
  \href {https://ui.adsabs.harvard.edu/abs/2016PASP..128j2001B} {128, 102001}

\bibitem[\protect\citeauthoryear{{Brown} et~al.,}{{Brown}
  et~al.}{2013}]{Brown:2013}
{Brown} T.~M.,  et~al., 2013, \mn@doi [Publications of the Astronomical Society
  of the Pacific] {10.1086/673168}, \href
  {https://ui.adsabs.harvard.edu/\#abs/2013PASP..125.1031B} {125, 1031}

\bibitem[\protect\citeauthoryear{{Buchhave} et~al.,}{{Buchhave}
  et~al.}{2010}]{Buchhave2010}
{Buchhave} L.~A.,  et~al., 2010, \mn@doi [\apj] {10.1088/0004-637X/720/2/1118},
  \href {https://ui.adsabs.harvard.edu/abs/2010ApJ...720.1118B} {720, 1118}

\bibitem[\protect\citeauthoryear{{Buchhave} et~al.,}{{Buchhave}
  et~al.}{2012}]{Buchhave2012}
{Buchhave} L.~A.,  et~al., 2012, \mn@doi [\nat] {10.1038/nature11121}, \href
  {https://ui.adsabs.harvard.edu/abs/2012Natur.486..375B} {486, 375}

\bibitem[\protect\citeauthoryear{{Buchhave} et~al.,}{{Buchhave}
  et~al.}{2014}]{Buchhave2014}
{Buchhave} L.~A.,  et~al., 2014, \mn@doi [\nat] {10.1038/nature13254}, \href
  {https://ui.adsabs.harvard.edu/abs/2014Natur.509..593B} {509, 593}

\bibitem[\protect\citeauthoryear{{Burrows} et~al.,}{{Burrows}
  et~al.}{1997}]{burrows}
{Burrows} A.,  et~al., 1997, \mn@doi [\apj] {10.1086/305002}, \href
  {https://ui.adsabs.harvard.edu/abs/1997ApJ...491..856B} {491, 856}

\bibitem[\protect\citeauthoryear{{Carter} et~al.,}{{Carter}
  et~al.}{2012}]{Carter2012}
{Carter} J.~A.,  et~al., 2012, \mn@doi [Science] {10.1126/science.1223269},
  \href {https://ui.adsabs.harvard.edu/abs/2012Sci...337..556C} {337, 556}

\bibitem[\protect\citeauthoryear{{Chambers} et~al.,}{{Chambers}
  et~al.}{2016}]{panstarrs}
{Chambers} K.~C.,  et~al., 2016, arXiv e-prints, \href
  {https://ui.adsabs.harvard.edu/abs/2016arXiv161205560C} {p. arXiv:1612.05560}

\bibitem[\protect\citeauthoryear{{Chao}, {deGraffenried}, {Lach}, {Nelson},
  {Truax}  \& {Gaidos}}{{Chao} et~al.}{2020}]{Chao2020}
{Chao} K.-H.,  {deGraffenried} R.,  {Lach} M.,  {Nelson} W.,  {Truax} K.,
  {Gaidos} E.,  2020, arXiv e-prints, \href
  {https://ui.adsabs.harvard.edu/abs/2020arXiv201207337C} {p. arXiv:2012.07337}

\bibitem[\protect\citeauthoryear{{Chen} \& {Kipping}}{{Chen} \&
  {Kipping}}{2017}]{Chen2017}
{Chen} J.,  {Kipping} D.,  2017, \mn@doi [\apj] {10.3847/1538-4357/834/1/17},
  \href {https://ui.adsabs.harvard.edu/abs/2017ApJ...834...17C} {834, 17}

\bibitem[\protect\citeauthoryear{{Chen} \& {Rogers}}{{Chen} \&
  {Rogers}}{2016}]{Chen2016}
{Chen} H.,  {Rogers} L.~A.,  2016, \mn@doi [\apj]
  {10.3847/0004-637X/831/2/180}, \href
  {https://ui.adsabs.harvard.edu/abs/2016ApJ...831..180C} {831, 180}

\bibitem[\protect\citeauthoryear{{Christiansen} et~al.,}{{Christiansen}
  et~al.}{2012}]{cdpp}
{Christiansen} J.~L.,  et~al., 2012, \mn@doi [\pasp] {10.1086/668847}, \href
  {https://ui.adsabs.harvard.edu/abs/2012PASP..124.1279C} {124, 1279}

\bibitem[\protect\citeauthoryear{{Cieza} et~al.,}{{Cieza}
  et~al.}{2009}]{cieza2009}
{Cieza} L.~A.,  et~al., 2009, \mn@doi [\apjl] {10.1088/0004-637X/696/1/L84},
  \href {https://ui.adsabs.harvard.edu/abs/2009ApJ...696L..84C} {696, L84}

\bibitem[\protect\citeauthoryear{{Collins}, {Kielkopf}, {Stassun}  \&
  {Hessman}}{{Collins} et~al.}{2017}]{Collins:2017}
{Collins} K.~A.,  {Kielkopf} J.~F.,  {Stassun} K.~G.,   {Hessman} F.~V.,  2017,
  \mn@doi [\aj] {10.3847/1538-3881/153/2/77}, \href
  {http://adsabs.harvard.edu/abs/2017AJ....153...77C} {153, 77}

\bibitem[\protect\citeauthoryear{{Colton}, {Horch}, {Everett}, {Howell},
  {Davidson}, {Baptista}  \& {Casetti-Dinescu}}{{Colton}
  et~al.}{2021}]{colton2021}
{Colton} N.~M.,  {Horch} E.~P.,  {Everett} M.~E.,  {Howell} S.~B.,  {Davidson}
  James~W. J.,  {Baptista} B.~J.,   {Casetti-Dinescu} D.~I.,  2021, \mn@doi
  [\aj] {10.3847/1538-3881/abc9af}, \href
  {https://ui.adsabs.harvard.edu/abs/2021AJ....161...21C} {161, 21}

\bibitem[\protect\citeauthoryear{{Cutri} \& {et al.}}{{Cutri} \& {et
  al.}}{2012}]{Cutri2012}
{Cutri} R.~M.,  {et al.} 2012, VizieR Online Data Catalog, \href
  {https://ui.adsabs.harvard.edu/abs/2012yCat.2311....0C} {p. II/311}

\bibitem[\protect\citeauthoryear{{Cutri} et~al.,}{{Cutri}
  et~al.}{2003}]{Cutri:2003}
{Cutri} R.~M.,  et~al., 2003, VizieR Online Data Catalog, \href
  {http://cdsads.u-strasbg.fr/abs/2003yCat.2246....0C} {2246, 0}

\bibitem[\protect\citeauthoryear{{Dai}, {Masuda}  \& {Winn}}{{Dai}
  et~al.}{2018}]{Dai2018}
{Dai} F.,  {Masuda} K.,   {Winn} J.~N.,  2018, \mn@doi [\apjl]
  {10.3847/2041-8213/aadd4f}, \href
  {https://ui.adsabs.harvard.edu/abs/2018ApJ...864L..38D} {864, L38}

\bibitem[\protect\citeauthoryear{{Dai}, {Masuda}, {Winn}  \& {Zeng}}{{Dai}
  et~al.}{2019}]{Dai2019}
{Dai} F.,  {Masuda} K.,  {Winn} J.~N.,   {Zeng} L.,  2019, \mn@doi [\apj]
  {10.3847/1538-4357/ab3a3b}, \href
  {https://ui.adsabs.harvard.edu/abs/2019ApJ...883...79D} {883, 79}

\bibitem[\protect\citeauthoryear{{David}, {Petigura}, {Luger},
  {Foreman-Mackey}, {Livingston}, {Mamajek}  \& {Hillenbrand}}{{David}
  et~al.}{2019}]{David2019b}
{David} T.~J.,  {Petigura} E.~A.,  {Luger} R.,  {Foreman-Mackey} D.,
  {Livingston} J.~H.,  {Mamajek} E.~E.,   {Hillenbrand} L.~A.,  2019, \mn@doi
  [\apjl] {10.3847/2041-8213/ab4c99}, \href
  {https://ui.adsabs.harvard.edu/abs/2019ApJ...885L..12D} {885, L12}

\bibitem[\protect\citeauthoryear{{David} et~al.,}{{David}
  et~al.}{2020}]{David2020}
{David} T.~J.,  et~al., 2020, arXiv e-prints, \href
  {https://ui.adsabs.harvard.edu/abs/2020arXiv201109894D} {p. arXiv:2011.09894}

\bibitem[\protect\citeauthoryear{{Demory} et~al.,}{{Demory}
  et~al.}{2016}]{Demory2016}
{Demory} B.-O.,  et~al., 2016, \mn@doi [\nat] {10.1038/nature17169}, \href
  {https://ui.adsabs.harvard.edu/abs/2016Natur.532..207D} {532, 207}

\bibitem[\protect\citeauthoryear{{Dholakia}, {Dholakia}, {Mayo}  \&
  {Dressing}}{{Dholakia} et~al.}{2020}]{Dholakia2020}
{Dholakia} S.,  {Dholakia} S.,  {Mayo} A.~W.,   {Dressing} C.~D.,  2020,
  \mn@doi [\aj] {10.3847/1538-3881/ab594c}, \href
  {https://ui.adsabs.harvard.edu/abs/2020AJ....159...93D} {159, 93}

\bibitem[\protect\citeauthoryear{{Dotter}}{{Dotter}}{2016}]{Dotter:2016}
{Dotter} A.,  2016, \mn@doi [\apjs] {10.3847/0067-0049/222/1/8}, \href
  {https://ui.adsabs.harvard.edu/abs/2016ApJS..222....8D} {222, 8}

\bibitem[\protect\citeauthoryear{{Duncan} et~al.,}{{Duncan}
  et~al.}{1991}]{1991ApJS...76..383D}
{Duncan} D.~K.,  et~al., 1991, \mn@doi [\apjs] {10.1086/191572}, \href
  {https://ui.adsabs.harvard.edu/abs/1991ApJS...76..383D} {76, 383}

\bibitem[\protect\citeauthoryear{{Eastman}, {Gaudi}  \& {Agol}}{{Eastman}
  et~al.}{2013}]{Eastman:2013}
{Eastman} J.,  {Gaudi} B.~S.,   {Agol} E.,  2013, \mn@doi [\pasp]
  {10.1086/669497}, \href {http://adsabs.harvard.edu/abs/2013PASP..125...83E}
  {125, 83}

\bibitem[\protect\citeauthoryear{{Eastman} et~al.,}{{Eastman}
  et~al.}{2019}]{Eastman:2019}
{Eastman} J.~D.,  et~al., 2019, arXiv e-prints, \href
  {https://ui.adsabs.harvard.edu/abs/2019arXiv190709480E} {p. arXiv:1907.09480}

\bibitem[\protect\citeauthoryear{{Eisner}, {Hillenbrand}, {White}, {Akeson}  \&
  {Sargent}}{{Eisner} et~al.}{2005}]{Eisner2005}
{Eisner} J.~A.,  {Hillenbrand} L.~A.,  {White} R.~J.,  {Akeson} R.~L.,
  {Sargent} A.~I.,  2005, \mn@doi [\apj] {10.1086/428828}, \href
  {https://ui.adsabs.harvard.edu/abs/2005ApJ...623..952E} {623, 952}

\bibitem[\protect\citeauthoryear{{Fabrycky} et~al.,}{{Fabrycky}
  et~al.}{2014}]{Fabrycky2014}
{Fabrycky} D.~C.,  et~al., 2014, \mn@doi [\apj] {10.1088/0004-637X/790/2/146},
  \href {https://ui.adsabs.harvard.edu/abs/2014ApJ...790..146F} {790, 146}

\bibitem[\protect\citeauthoryear{{Fischer} \& {Valenti}}{{Fischer} \&
  {Valenti}}{2005}]{Fischer2005}
{Fischer} D.~A.,  {Valenti} J.,  2005, \mn@doi [\apj] {10.1086/428383}, \href
  {https://ui.adsabs.harvard.edu/abs/2005ApJ...622.1102F} {622, 1102}

\bibitem[\protect\citeauthoryear{{Fleming}, {Schmitt}  \& {Giampapa}}{{Fleming}
  et~al.}{1995}]{1995ApJ...450..401F}
{Fleming} T.~A.,  {Schmitt} J. H.~M.~M.,   {Giampapa} M.~S.,  1995, \mn@doi
  [\apj] {10.1086/176150}, \href
  {https://ui.adsabs.harvard.edu/abs/1995ApJ...450..401F} {450, 401}

\bibitem[\protect\citeauthoryear{Foreman-Mackey}{Foreman-Mackey}{2016}]{corner}
Foreman-Mackey D.,  2016, \mn@doi [The Journal of Open Source Software]
  {10.21105/joss.00024}, 1, 24

\bibitem[\protect\citeauthoryear{Foreman-Mackey, Luger, Czekala, Agol,
  Price-Whelan  \& Barclay}{Foreman-Mackey et~al.}{2020}]{exoplanet:exoplanet}
Foreman-Mackey D.,  Luger R.,  Czekala I.,  Agol E.,  Price-Whelan A.,
  Barclay T.,  2020, exoplanet-dev/exoplanet v0.3.2,
  \mn@doi{10.5281/zenodo.1998447}, \url
  {https://doi.org/10.5281/zenodo.1998447}

\bibitem[\protect\citeauthoryear{Fulton et~al.,}{Fulton et~al.}{2017}]{CKS}
Fulton B.~J.,  et~al., 2017, \mn@doi [The Astronomical Journal]
  {10.3847/1538-3881/aa80eb}, 154, 109

\bibitem[\protect\citeauthoryear{Furesz}{Furesz}{2008}]{phdthesis}
Furesz G.,  2008, PhD thesis, Univ. of Szeged, Hungary

\bibitem[\protect\citeauthoryear{{Gaia Collaboration} et~al.,}{{Gaia
  Collaboration} et~al.}{2018}]{Gaia:2018}
{Gaia Collaboration} et~al., 2018, \mn@doi [\aap]
  {10.1051/0004-6361/201833051}, \href
  {https://ui.adsabs.harvard.edu/abs/2018A&A...616A...1G} {616, A1}

\bibitem[\protect\citeauthoryear{{Giacalone} \& {Dressing}}{{Giacalone} \&
  {Dressing}}{2020}]{2020ascl.soft02004G}
{Giacalone} S.,  {Dressing} C.~D.,  2020, {triceratops: Candidate exoplanet
  rating tool} (\mn@eprint {ascl} {2002.004})

\bibitem[\protect\citeauthoryear{{Giacalone} et~al.,}{{Giacalone}
  et~al.}{2021}]{triceratops}
{Giacalone} S.,  et~al., 2021, \mn@doi [\aj] {10.3847/1538-3881/abc6af}, \href
  {https://ui.adsabs.harvard.edu/abs/2021AJ....161...24G} {161, 24}

\bibitem[\protect\citeauthoryear{{Gilliland} et~al.,}{{Gilliland}
  et~al.}{2011}]{gilland}
{Gilliland} R.~L.,  et~al., 2011, \mn@doi [\apjs] {10.1088/0067-0049/197/1/6},
  \href {https://ui.adsabs.harvard.edu/abs/2011ApJS..197....6G} {197, 6}

\bibitem[\protect\citeauthoryear{{Ginzburg}, {Schlichting}  \&
  {Sari}}{{Ginzburg} et~al.}{2016}]{Ginzburg2016}
{Ginzburg} S.,  {Schlichting} H.~E.,   {Sari} R.,  2016, \mn@doi [\apj]
  {10.3847/0004-637X/825/1/29}, \href
  {https://ui.adsabs.harvard.edu/abs/2016ApJ...825...29G} {825, 29}

\bibitem[\protect\citeauthoryear{{Ginzburg}, {Schlichting}  \&
  {Sari}}{{Ginzburg} et~al.}{2018}]{Ginzburg2018}
{Ginzburg} S.,  {Schlichting} H.~E.,   {Sari} R.,  2018, \mn@doi [\mnras]
  {10.1093/mnras/sty290}, \href
  {https://ui.adsabs.harvard.edu/abs/2018MNRAS.476..759G} {476, 759}

\bibitem[\protect\citeauthoryear{{Green}}{{Green}}{2018}]{2018JOSS....3..695M}
{Green} G.,  2018, \mn@doi [The Journal of Open Source Software]
  {10.21105/joss.00695}, \href
  {https://ui.adsabs.harvard.edu/abs/2018JOSS....3..695G} {3, 695}

\bibitem[\protect\citeauthoryear{{Green}, {Schlafly}, {Zucker}, {Speagle}  \&
  {Finkbeiner}}{{Green} et~al.}{2019}]{2019ApJ...887...93G}
{Green} G.~M.,  {Schlafly} E.,  {Zucker} C.,  {Speagle} J.~S.,   {Finkbeiner}
  D.,  2019, \mn@doi [\apj] {10.3847/1538-4357/ab5362}, \href
  {https://ui.adsabs.harvard.edu/abs/2019ApJ...887...93G} {887, 93}

\bibitem[\protect\citeauthoryear{Gupta \& Schlichting}{Gupta \&
  Schlichting}{2019}]{Gupta19}
Gupta A.,  Schlichting H.~E.,  2019, Monthly Notices of the Royal Astronomical
  Society, 487, 24

\bibitem[\protect\citeauthoryear{Gupta \& Schlichting}{Gupta \&
  Schlichting}{2020a}]{Gupta20}
Gupta A.,  Schlichting H.~E.,  2020a, Monthly Notices of the Royal Astronomical
  Society, 493, 792

\bibitem[\protect\citeauthoryear{{Gupta} \& {Schlichting}}{{Gupta} \&
  {Schlichting}}{2020b}]{Gupta2020}
{Gupta} A.,  {Schlichting} H.~E.,  2020b, \mn@doi [\mnras]
  {10.1093/mnras/staa315}, \href
  {https://ui.adsabs.harvard.edu/abs/2020MNRAS.493..792G} {493, 792}

\bibitem[\protect\citeauthoryear{{Hamer} \& {Schlaufman}}{{Hamer} \&
  {Schlaufman}}{2020}]{HamerSchlaufman2020}
{Hamer} J.~H.,  {Schlaufman} K.~C.,  2020, \mn@doi [\aj]
  {10.3847/1538-3881/aba74f}, \href
  {https://ui.adsabs.harvard.edu/abs/2020AJ....160..138H} {160, 138}

\bibitem[\protect\citeauthoryear{{H{\o}g} et~al.,}{{H{\o}g}
  et~al.}{2000}]{Hog:2000}
{H{\o}g} E.,  et~al., 2000, \aap, \href
  {https://ui.adsabs.harvard.edu/abs/2000A&A...355L..27H} {355, L27}

\bibitem[\protect\citeauthoryear{{Howard} et~al.,}{{Howard}
  et~al.}{2012}]{Howard2012}
{Howard} A.~W.,  et~al., 2012, \mn@doi [\apjs] {10.1088/0067-0049/201/2/15},
  \href {https://ui.adsabs.harvard.edu/abs/2012ApJS..201...15H} {201, 15}

\bibitem[\protect\citeauthoryear{{Howell}, {Everett}, {Sherry}, {Horch}  \&
  {Ciardi}}{{Howell} et~al.}{2011}]{Howell2011}
{Howell} S.~B.,  {Everett} M.~E.,  {Sherry} W.,  {Horch} E.,   {Ciardi} D.~R.,
  2011, \mn@doi [\aj] {10.1088/0004-6256/142/1/19}, \href
  {https://ui.adsabs.harvard.edu/abs/2011AJ....142...19H} {142, 19}

\bibitem[\protect\citeauthoryear{{Howell}, {Matson}, {Ciardi}, {Everett},
  {Livingston}, {Scott}, {Horch}  \& {Winn}}{{Howell}
  et~al.}{2021}]{howell2021}
{Howell} S.~B.,  {Matson} R.~A.,  {Ciardi} D.~R.,  {Everett} M.~E.,
  {Livingston} J.~H.,  {Scott} N.~J.,  {Horch} E.~P.,   {Winn} J.~N.,  2021,
  \mn@doi [\aj] {10.3847/1538-3881/abdec6}, \href
  {https://ui.adsabs.harvard.edu/abs/2021AJ....161..164H} {161, 164}

\bibitem[\protect\citeauthoryear{{Inamdar} \& {Schlichting}}{{Inamdar} \&
  {Schlichting}}{2016}]{Inamdar2016}
{Inamdar} N.~K.,  {Schlichting} H.~E.,  2016, \mn@doi [\apjl]
  {10.3847/2041-8205/817/2/L13}, \href
  {https://ui.adsabs.harvard.edu/abs/2016ApJ...817L..13I} {817, L13}

\bibitem[\protect\citeauthoryear{{Jackson}, {Davis}  \& {Wheatley}}{{Jackson}
  et~al.}{2012}]{Jackson2012}
{Jackson} A.~P.,  {Davis} T.~A.,   {Wheatley} P.~J.,  2012, \mn@doi [\mnras]
  {10.1111/j.1365-2966.2012.20657.x}, \href
  {https://ui.adsabs.harvard.edu/abs/2012MNRAS.422.2024J} {422, 2024}

\bibitem[\protect\citeauthoryear{{Jang-Condell}}{{Jang-Condell}}{2015}]{jang-condell2015}
{Jang-Condell} H.,  2015, \mn@doi [\apj] {10.1088/0004-637X/799/2/147}, \href
  {https://ui.adsabs.harvard.edu/abs/2015ApJ...799..147J} {799, 147}

\bibitem[\protect\citeauthoryear{{Jenkins}}{{Jenkins}}{2002}]{jenk1}
{Jenkins} J.~M.,  2002, \mn@doi [\apj] {10.1086/341136}, \href
  {http://adsabs.harvard.edu/abs/2002ApJ...575..493J} {575, 493}

\bibitem[\protect\citeauthoryear{{Jenkins} et~al.,}{{Jenkins}
  et~al.}{2010}]{jenk2}
{Jenkins} J.~M.,  et~al., 2010, in {Radziwill} N.~M.,  {Bridger} A.,  eds,
  Society of Photo-Optical Instrumentation Engineers (SPIE) Conference Series
  Vol. 7740, Software and Cyberinfrastructure for Astronomy. p. 77400D,
  \mn@doi{10.1117/12.856764}

\bibitem[\protect\citeauthoryear{{Jenkins} et~al.,}{{Jenkins}
  et~al.}{2016}]{tesspipeline}
{Jenkins} J.~M.,  et~al., 2016, in {Chiozzi} G.,  {Guzman} J.~C.,  eds,
  Society of Photo-Optical Instrumentation Engineers (SPIE) Conference Series
  Vol. 9913, Software and Cyberinfrastructure for Astronomy IV. p. 99133E,
  \mn@doi{10.1117/12.2233418}

\bibitem[\protect\citeauthoryear{{Jensen}}{{Jensen}}{2013}]{Jensen:2013}
{Jensen} E.,  2013, {Tapir: A web interface for transit/eclipse observability},
  Astrophysics Source Code Library (\mn@eprint {ascl} {1306.007})

\bibitem[\protect\citeauthoryear{{Jin}, {Mordasini}, {Parmentier}, {van
  Boekel}, {Henning}  \& {Ji}}{{Jin} et~al.}{2014}]{Jin2014}
{Jin} S.,  {Mordasini} C.,  {Parmentier} V.,  {van Boekel} R.,  {Henning} T.,
  {Ji} J.,  2014, \mn@doi [\apj] {10.1088/0004-637X/795/1/65}, \href
  {https://ui.adsabs.harvard.edu/abs/2014ApJ...795...65J} {795, 65}

\bibitem[\protect\citeauthoryear{{Kempton} et~al.,}{{Kempton}
  et~al.}{2018}]{TSM}
{Kempton} E. M.~R.,  et~al., 2018, \mn@doi [\pasp] {10.1088/1538-3873/aadf6f},
  \href {https://ui.adsabs.harvard.edu/abs/2018PASP..130k4401K} {130, 114401}

\bibitem[\protect\citeauthoryear{{Kipping}}{{Kipping}}{2013a}]{kipping_prior1}
{Kipping} D.~M.,  2013a, \mn@doi [\mnras] {10.1093/mnrasl/slt075}, \href
  {http://adsabs.harvard.edu/abs/2013MNRAS.434L..51K} {434, L51}

\bibitem[\protect\citeauthoryear{{Kipping}}{{Kipping}}{2013b}]{exoplanet:kipping13}
{Kipping} D.~M.,  2013b, \mn@doi [\mnras] {10.1093/mnras/stt1435}, \href
  {http://adsabs.harvard.edu/abs/2013MNRAS.435.2152K} {435, 2152}

\bibitem[\protect\citeauthoryear{{Kipping}}{{Kipping}}{2014}]{kipping_prior2}
{Kipping} D.~M.,  2014, \mn@doi [\mnras] {10.1093/mnras/stu1561}, \href
  {http://adsabs.harvard.edu/abs/2014MNRAS.444.2263K} {444, 2263}

\bibitem[\protect\citeauthoryear{{Kipping} \& {Sandford}}{{Kipping} \&
  {Sandford}}{2016}]{kipping16}
{Kipping} D.~M.,  {Sandford} E.,  2016, \mn@doi [\mnras]
  {10.1093/mnras/stw1926}, \href
  {https://ui.adsabs.harvard.edu/abs/2016MNRAS.463.1323K} {463, 1323}

\bibitem[\protect\citeauthoryear{{Klein} et~al.,}{{Klein}
  et~al.}{2021}]{Klein2021}
{Klein} B.,  et~al., 2021, \mn@doi [\mnras] {10.1093/mnras/staa3702}, \href
  {https://ui.adsabs.harvard.edu/abs/2021MNRAS.502..188K} {502, 188}

\bibitem[\protect\citeauthoryear{{Kreidberg} et~al.,}{{Kreidberg}
  et~al.}{2019}]{Kreidberg2019}
{Kreidberg} L.,  et~al., 2019, \mn@doi [\nat] {10.1038/s41586-019-1497-4},
  \href {https://ui.adsabs.harvard.edu/abs/2019Natur.573...87K} {573, 87}

\bibitem[\protect\citeauthoryear{{Kubyshkina} et~al.,}{{Kubyshkina}
  et~al.}{2019a}]{Kubyshkina2019b}
{Kubyshkina} D.,  et~al., 2019a, \mn@doi [\aap] {10.1051/0004-6361/201936581},
  \href {https://ui.adsabs.harvard.edu/abs/2019A&A...632A..65K} {632, A65}

\bibitem[\protect\citeauthoryear{{Kubyshkina} et~al.,}{{Kubyshkina}
  et~al.}{2019b}]{Kubyshkina2019a}
{Kubyshkina} D.,  et~al., 2019b, \mn@doi [\apj] {10.3847/1538-4357/ab1e42},
  \href {https://ui.adsabs.harvard.edu/abs/2019ApJ...879...26K} {879, 26}

\bibitem[\protect\citeauthoryear{{Kurucz}}{{Kurucz}}{1992}]{kurucz}
{Kurucz} R.~L.,  1992, in {Barbuy} B.,  {Renzini} A.,  eds, ~ Vol. 149, The
  Stellar Populations of Galaxies. p.~225

\bibitem[\protect\citeauthoryear{{Lacy} \& {Burrows}}{{Lacy} \&
  {Burrows}}{2020}]{lacy}
{Lacy} B.,  {Burrows} A.,  2020, \mn@doi [\apj] {10.3847/1538-4357/ab7017},
  \href {https://ui.adsabs.harvard.edu/abs/2020ApJ...892..151L} {892, 151}

\bibitem[\protect\citeauthoryear{{Lee} \& {Chiang}}{{Lee} \&
  {Chiang}}{2017}]{LeeChiang2017}
{Lee} E.~J.,  {Chiang} E.,  2017, \mn@doi [\apj] {10.3847/1538-4357/aa6fb3},
  \href {https://ui.adsabs.harvard.edu/abs/2017ApJ...842...40L} {842, 40}

\bibitem[\protect\citeauthoryear{{Lee} \& {Connors}}{{Lee} \&
  {Connors}}{2021}]{Lee2021}
{Lee} E.~J.,  {Connors} N.~J.,  2021, \mn@doi [\apj]
  {10.3847/1538-4357/abd6c7}, \href
  {https://ui.adsabs.harvard.edu/abs/2021ApJ...908...32L} {908, 32}

\bibitem[\protect\citeauthoryear{{Li}, {Tenenbaum}, {Twicken}, {Burke},
  {Jenkins}, {Quintana}, {Rowe}  \& {Seader}}{{Li}
  et~al.}{2019}]{Li:DVmodelFit2019}
{Li} J.,  {Tenenbaum} P.,  {Twicken} J.~D.,  {Burke} C.~J.,  {Jenkins} J.~M.,
  {Quintana} E.~V.,  {Rowe} J.~F.,   {Seader} S.~E.,  2019, \mn@doi [\pasp]
  {10.1088/1538-3873/aaf44d}, \href
  {https://ui.adsabs.harvard.edu/abs/2019PASP..131b4506L} {131, 024506}

\bibitem[\protect\citeauthoryear{{Lightkurve Collaboration}
  et~al.,}{{Lightkurve Collaboration} et~al.}{2018}]{Lightkurve}
{Lightkurve Collaboration} et~al., 2018, {Lightkurve: Kepler and TESS time
  series analysis in Python}, Astrophysics Source Code Library (\mn@eprint
  {ascl} {1812.013})

\bibitem[\protect\citeauthoryear{Lindegren}{Lindegren}{2018}]{ruwe}
Lindegren L.,  2018, {R}e-normalising the astrometric chi-square in {G}aia
  {D}{R}2, GAIA-C3-TN-LU-LL-124, \url
  {http://www.rssd.esa.int/doc_fetch.php?id=3757412}

\bibitem[\protect\citeauthoryear{{Lindegren} et~al.,}{{Lindegren}
  et~al.}{2018}]{Lindegren:2018}
{Lindegren} L.,  et~al., 2018, \mn@doi [\aap] {10.1051/0004-6361/201832727},
  \href {https://ui.adsabs.harvard.edu/abs/2018A&A...616A...2L} {616, A2}

\bibitem[\protect\citeauthoryear{{Linder}, {Mordasini}, {Molli{\`e}re},
  {Marleau}, {Malik}, {Quanz}  \& {Meyer}}{{Linder} et~al.}{2019}]{Linder2019}
{Linder} E.~F.,  {Mordasini} C.,  {Molli{\`e}re} P.,  {Marleau} G.-D.,  {Malik}
  M.,  {Quanz} S.~P.,   {Meyer} M.~R.,  2019, \mn@doi [\aap]
  {10.1051/0004-6361/201833873}, \href
  {https://ui.adsabs.harvard.edu/abs/2019A&A...623A..85L} {623, A85}

\bibitem[\protect\citeauthoryear{{Livingston} et~al.,}{{Livingston}
  et~al.}{2019}]{Livingston2019}
{Livingston} J.~H.,  et~al., 2019, \mn@doi [\mnras] {10.1093/mnras/sty3464},
  \href {https://ui.adsabs.harvard.edu/abs/2019MNRAS.484....8L} {484, 8}

\bibitem[\protect\citeauthoryear{{Lopez} \& {Fortney}}{{Lopez} \&
  {Fortney}}{2013}]{LopezFortney2013}
{Lopez} E.~D.,  {Fortney} J.~J.,  2013, \mn@doi [\apj]
  {10.1088/0004-637X/776/1/2}, \href
  {https://ui.adsabs.harvard.edu/abs/2013ApJ...776....2L} {776, 2}

\bibitem[\protect\citeauthoryear{{Lopez} \& {Rice}}{{Lopez} \&
  {Rice}}{2018}]{Lopez2018}
{Lopez} E.~D.,  {Rice} K.,  2018, \mn@doi [\mnras] {10.1093/mnras/sty1707},
  \href {https://ui.adsabs.harvard.edu/abs/2018MNRAS.479.5303L} {479, 5303}

\bibitem[\protect\citeauthoryear{{Luger}, {Agol}, {Foreman-Mackey}, {Fleming},
  {Lustig-Yaeger}  \& {Deitrick}}{{Luger} et~al.}{2019}]{exoplanet:luger18}
{Luger} R.,  {Agol} E.,  {Foreman-Mackey} D.,  {Fleming} D.~P.,
  {Lustig-Yaeger} J.,   {Deitrick} R.,  2019, \mn@doi [\aj]
  {10.3847/1538-3881/aae8e5}, \href
  {http://adsabs.harvard.edu/abs/2019AJ....157...64L} {157, 64}

\bibitem[\protect\citeauthoryear{{Luhn}, {Wright}, {Howard}  \&
  {Isaacson}}{{Luhn} et~al.}{2020}]{Luhn2020}
{Luhn} J.~K.,  {Wright} J.~T.,  {Howard} A.~W.,   {Isaacson} H.,  2020, \mn@doi
  [\aj] {10.3847/1538-3881/ab855a}, \href
  {https://ui.adsabs.harvard.edu/abs/2020AJ....159..235L} {159, 235}

\bibitem[\protect\citeauthoryear{{Malavolta} et~al.,}{{Malavolta}
  et~al.}{2018}]{Malavolta2018}
{Malavolta} L.,  et~al., 2018, \mn@doi [\aj] {10.3847/1538-3881/aaa5b5}, \href
  {https://ui.adsabs.harvard.edu/abs/2018AJ....155..107M} {155, 107}

\bibitem[\protect\citeauthoryear{{Mamajek} \& {Hillenbrand}}{{Mamajek} \&
  {Hillenbrand}}{2008}]{2008ApJ...687.1264M}
{Mamajek} E.~E.,  {Hillenbrand} L.~A.,  2008, \mn@doi [\apj] {10.1086/591785},
  \href {https://ui.adsabs.harvard.edu/abs/2008ApJ...687.1264M} {687, 1264}

\bibitem[\protect\citeauthoryear{{Mann} et~al.,}{{Mann}
  et~al.}{2017}]{Mann2017}
{Mann} A.~W.,  et~al., 2017, \mn@doi [\aj] {10.1088/1361-6528/aa5276}, \href
  {https://ui.adsabs.harvard.edu/abs/2017AJ....153...64M} {153, 64}

\bibitem[\protect\citeauthoryear{{Mann} et~al.,}{{Mann} et~al.}{2020}]{thyme2}
{Mann} A.~W.,  et~al., 2020, \mn@doi [\aj] {10.3847/1538-3881/abae64}, \href
  {https://ui.adsabs.harvard.edu/abs/2020AJ....160..179M} {160, 179}

\bibitem[\protect\citeauthoryear{{Millholland} \& {Spalding}}{{Millholland} \&
  {Spalding}}{2020}]{Millholland2020}
{Millholland} S.~C.,  {Spalding} C.,  2020, \mn@doi [\apj]
  {10.3847/1538-4357/abc4e5}, \href
  {https://ui.adsabs.harvard.edu/abs/2020ApJ...905...71M} {905, 71}

\bibitem[\protect\citeauthoryear{{Minkowski} \& {Abell}}{{Minkowski} \&
  {Abell}}{1963}]{poss1}
{Minkowski} R.~L.,  {Abell} G.~O.,  1963, {The National Geographic
  Society-Palomar Observatory Sky Survey}.
p.~481

\bibitem[\protect\citeauthoryear{{Montet} et~al.,}{{Montet}
  et~al.}{2020}]{Montet2020}
{Montet} B.~T.,  et~al., 2020, \mn@doi [\aj] {10.3847/1538-3881/ab6d6d}, \href
  {https://ui.adsabs.harvard.edu/abs/2020AJ....159..112M} {159, 112}

\bibitem[\protect\citeauthoryear{{Moraux} et~al.,}{{Moraux}
  et~al.}{2013}]{2013A&A...560A..13M}
{Moraux} E.,  et~al., 2013, \mn@doi [\aap] {10.1051/0004-6361/201321508}, \href
  {https://ui.adsabs.harvard.edu/abs/2013A&A...560A..13M} {560, A13}

\bibitem[\protect\citeauthoryear{{Mordasini}, {Alibert}  \& {Benz}}{{Mordasini}
  et~al.}{2009}]{Mordasini2009}
{Mordasini} C.,  {Alibert} Y.,   {Benz} W.,  2009, \mn@doi [\aap]
  {10.1051/0004-6361/200810301}, \href
  {https://ui.adsabs.harvard.edu/abs/2009A&A...501.1139M} {501, 1139}

\bibitem[\protect\citeauthoryear{{Morton}}{{Morton}}{2015}]{vespa}
{Morton} T.~D.,  2015, {VESPA: False positive probabilities calculator}
  (\mn@eprint {ascl} {1503.011})

\bibitem[\protect\citeauthoryear{{Murray} \& {Dermott}}{{Murray} \&
  {Dermott}}{1999}]{MD99}
{Murray} C.~D.,  {Dermott} S.~F.,  1999, {Solar system dynamics}

\bibitem[\protect\citeauthoryear{{Muzerolle}, {Calvet}, {Hartmann}  \&
  {D'Alessio}}{{Muzerolle} et~al.}{2003}]{Muzerolle2003}
{Muzerolle} J.,  {Calvet} N.,  {Hartmann} L.,   {D'Alessio} P.,  2003, \mn@doi
  [\apjl] {10.1086/379921}, \href
  {https://ui.adsabs.harvard.edu/abs/2003ApJ...597L.149M} {597, L149}

\bibitem[\protect\citeauthoryear{{Nardiello} et~al.,}{{Nardiello}
  et~al.}{2020}]{Nardiello2020}
{Nardiello} D.,  et~al., 2020, \mn@doi [\mnras] {10.1093/mnras/staa1465}, \href
  {https://ui.adsabs.harvard.edu/abs/2020MNRAS.495.4924N} {495, 4924}

\bibitem[\protect\citeauthoryear{{Narita} et~al.,}{{Narita}
  et~al.}{2019}]{Narita2019}
{Narita} N.,  et~al., 2019, \mn@doi [Journal of Astronomical Telescopes,
  Instruments, and Systems] {10.1117/1.JATIS.5.1.015001}, \href
  {https://ui.adsabs.harvard.edu/abs/2019JATIS...5a5001N} {5, 015001}

\bibitem[\protect\citeauthoryear{{Newton} et~al.,}{{Newton}
  et~al.}{2019}]{Newton2019}
{Newton} E.~R.,  et~al., 2019, \mn@doi [\apjl] {10.3847/2041-8213/ab2988},
  \href {https://ui.adsabs.harvard.edu/abs/2019ApJ...880L..17N} {880, L17}

\bibitem[\protect\citeauthoryear{{Newton} et~al.,}{{Newton}
  et~al.}{2021}]{thyme3}
{Newton} E.~R.,  et~al., 2021, \mn@doi [\aj] {10.3847/1538-3881/abccc6}, \href
  {https://ui.adsabs.harvard.edu/abs/2021AJ....161...65N} {161, 65}

\bibitem[\protect\citeauthoryear{Oh, Price-Whelan, Hogg, Morton  \& Spergel}{Oh
  et~al.}{2017}]{comovedisc}
Oh S.,  Price-Whelan A.~M.,  Hogg D.~W.,  Morton T.~D.,   Spergel D.~N.,  2017,
  The Astronomical Journal, 153, 257

\bibitem[\protect\citeauthoryear{{Owen}}{{Owen}}{2020}]{Owen2020}
{Owen} J.~E.,  2020, \mn@doi [\mnras] {10.1093/mnras/staa2784}, \href
  {https://ui.adsabs.harvard.edu/abs/2020MNRAS.498.5030O} {498, 5030}

\bibitem[\protect\citeauthoryear{{Owen} \& {Campos Estrada}}{{Owen} \& {Campos
  Estrada}}{2020}]{OwenCamposEstrada2020}
{Owen} J.~E.,  {Campos Estrada} B.,  2020, \mn@doi [\mnras]
  {10.1093/mnras/stz3435}, \href
  {https://ui.adsabs.harvard.edu/abs/2020MNRAS.491.5287O} {491, 5287}

\bibitem[\protect\citeauthoryear{{Owen} \& {Wu}}{{Owen} \&
  {Wu}}{2013}]{OwenWu2013}
{Owen} J.~E.,  {Wu} Y.,  2013, \mn@doi [\apj] {10.1088/0004-637X/775/2/105},
  \href {https://ui.adsabs.harvard.edu/abs/2013ApJ...775..105O} {775, 105}

\bibitem[\protect\citeauthoryear{{Parviainen} et~al.,}{{Parviainen}
  et~al.}{2019}]{Parviainen2019}
{Parviainen} H.,  et~al., 2019, \mn@doi [\aap] {10.1051/0004-6361/201935709},
  \href {https://ui.adsabs.harvard.edu/abs/2019A&A...630A..89P} {630, A89}

\bibitem[\protect\citeauthoryear{{Parviainen} et~al.,}{{Parviainen}
  et~al.}{2020}]{Parviainen2020}
{Parviainen} H.,  et~al., 2020, \mn@doi [\aap] {10.1051/0004-6361/201935958},
  \href {https://ui.adsabs.harvard.edu/abs/2020A&A...633A..28P} {633, A28}

\bibitem[\protect\citeauthoryear{{Pepper} et~al.,}{{Pepper}
  et~al.}{2007}]{kelt1}
{Pepper} J.,  et~al., 2007, \mn@doi [\pasp] {10.1086/521836}, \href
  {https://ui.adsabs.harvard.edu/abs/2007PASP..119..923P} {119, 923}

\bibitem[\protect\citeauthoryear{{Pepper}, {Kuhn}, {Siverd}, {James}  \&
  {Stassun}}{{Pepper} et~al.}{2012}]{kelt2}
{Pepper} J.,  {Kuhn} R.~B.,  {Siverd} R.,  {James} D.,   {Stassun} K.,  2012,
  \mn@doi [\pasp] {10.1086/665044}, \href
  {https://ui.adsabs.harvard.edu/abs/2012PASP..124..230P} {124, 230}

\bibitem[\protect\citeauthoryear{Pepper, Stassun  \& Gaudi}{Pepper
  et~al.}{2018}]{kelt3}
Pepper J.,  Stassun K.~G.,   Gaudi B.~S.,  2018, KELT: The Kilodegree Extremely
  Little Telescope, a Survey for Exoplanets Transiting Bright, Hot Stars.
Springer International Publishing, Cham, pp 969--980,
  \mn@doi{10.1007/978-3-319-55333-7_128}, \url
  {https://doi.org/10.1007/978-3-319-55333-7_128}

\bibitem[\protect\citeauthoryear{{Petigura} et~al.,}{{Petigura}
  et~al.}{2018}]{Petigura2018}
{Petigura} E.~A.,  et~al., 2018, \mn@doi [\aj] {10.3847/1538-3881/aaa54c},
  \href {https://ui.adsabs.harvard.edu/abs/2018AJ....155...89P} {155, 89}

\bibitem[\protect\citeauthoryear{{Petrovich}, {Deibert}  \& {Wu}}{{Petrovich}
  et~al.}{2019}]{Petrovich2019}
{Petrovich} C.,  {Deibert} E.,   {Wu} Y.,  2019, \mn@doi [\aj]
  {10.3847/1538-3881/ab0e0a}, \href
  {https://ui.adsabs.harvard.edu/abs/2019AJ....157..180P} {157, 180}

\bibitem[\protect\citeauthoryear{{Poppenhaeger}, {Ketzer}  \&
  {Mallonn}}{{Poppenhaeger} et~al.}{2021}]{Poppenhaeger2021}
{Poppenhaeger} K.,  {Ketzer} L.,   {Mallonn} M.,  2021, \mn@doi [\mnras]
  {10.1093/mnras/staa1462}, \href
  {https://ui.adsabs.harvard.edu/abs/2021MNRAS.500.4560P} {500, 4560}

\bibitem[\protect\citeauthoryear{Price-Whelan \& Brammer}{Price-Whelan \&
  Brammer}{2020}]{pyia}
Price-Whelan A.,  Brammer G.,  2020, adrn/pyia v1.2,
  \mn@doi{10.5281/zenodo.4300654}, \url
  {https://doi.org/10.5281/zenodo.4300654}

\bibitem[\protect\citeauthoryear{{Pu} \& {Lai}}{{Pu} \& {Lai}}{2019}]{Pu2019}
{Pu} B.,  {Lai} D.,  2019, \mn@doi [\mnras] {10.1093/mnras/stz1817}, \href
  {https://ui.adsabs.harvard.edu/abs/2019MNRAS.488.3568P} {488, 3568}

\bibitem[\protect\citeauthoryear{{Rebull} et~al.,}{{Rebull}
  et~al.}{2016}]{2016AJ....152..113R}
{Rebull} L.~M.,  et~al., 2016, \mn@doi [\aj] {10.3847/0004-6256/152/5/113},
  \href {https://ui.adsabs.harvard.edu/abs/2016AJ....152..113R} {152, 113}

\bibitem[\protect\citeauthoryear{{Rebull}, {Stauffer}, {Hillenbrand}, {Cody},
  {Bouvier}, {Soderblom}, {Pinsonneault}  \& {Hebb}}{{Rebull}
  et~al.}{2017}]{2017ApJ...839...92R}
{Rebull} L.~M.,  {Stauffer} J.~R.,  {Hillenbrand} L.~A.,  {Cody} A.~M.,
  {Bouvier} J.,  {Soderblom} D.~R.,  {Pinsonneault} M.,   {Hebb} L.,  2017,
  \mn@doi [\apj] {10.3847/1538-4357/aa6aa4}, \href
  {https://ui.adsabs.harvard.edu/abs/2017ApJ...839...92R} {839, 92}

\bibitem[\protect\citeauthoryear{{Reid} et~al.,}{{Reid} et~al.}{1991}]{poss2}
{Reid} I.~N.,  et~al., 1991, \mn@doi [\pasp] {10.1086/132866}, \href
  {https://ui.adsabs.harvard.edu/abs/1991PASP..103..661R} {103, 661}

\bibitem[\protect\citeauthoryear{{Ricker} et~al.,}{{Ricker}
  et~al.}{2015}]{TESS}
{Ricker} G.~R.,  et~al., 2015, \mn@doi [Journal of Astronomical Telescopes,
  Instruments, and Systems] {10.1117/1.JATIS.1.1.014003}, \href
  {https://ui.adsabs.harvard.edu/abs/2015JATIS...1a4003R} {1, 014003}

\bibitem[\protect\citeauthoryear{{Rizzuto} et~al.,}{{Rizzuto}
  et~al.}{2020}]{thyme1}
{Rizzuto} A.~C.,  et~al., 2020, \mn@doi [\aj] {10.3847/1538-3881/ab94b7}, \href
  {https://ui.adsabs.harvard.edu/abs/2020AJ....160...33R} {160, 33}

\bibitem[\protect\citeauthoryear{{Rogers} \& {Owen}}{{Rogers} \&
  {Owen}}{2021}]{Rogers2021}
{Rogers} J.~G.,  {Owen} J.~E.,  2021, \mn@doi [\mnras] {10.1093/mnras/stab529},
  \href {https://ui.adsabs.harvard.edu/abs/2021MNRAS.tmp..530R} {}

\bibitem[\protect\citeauthoryear{Salvatier, Wiecki  \& Fonnesbeck}{Salvatier
  et~al.}{2016}]{exoplanet:pymc3}
Salvatier J.,  Wiecki T.~V.,   Fonnesbeck C.,  2016, PeerJ Computer Science, 2,
  e55

\bibitem[\protect\citeauthoryear{{Sanchis-Ojeda}, {Rappaport}, {Winn},
  {Kotson}, {Levine}  \& {El Mellah}}{{Sanchis-Ojeda}
  et~al.}{2014}]{SanchisOjeda2014}
{Sanchis-Ojeda} R.,  {Rappaport} S.,  {Winn} J.~N.,  {Kotson} M.~C.,  {Levine}
  A.,   {El Mellah} I.,  2014, \mn@doi [\apj] {10.1088/0004-637X/787/1/47},
  \href {https://ui.adsabs.harvard.edu/abs/2014ApJ...787...47S} {787, 47}

\bibitem[\protect\citeauthoryear{{Sandoval}, {Contardo}  \& {David}}{{Sandoval}
  et~al.}{2020}]{Sandoval2020}
{Sandoval} A.,  {Contardo} G.,   {David} T.~J.,  2020, arXiv e-prints, \href
  {https://ui.adsabs.harvard.edu/abs/2020arXiv201209239S} {p. arXiv:2012.09239}

\bibitem[\protect\citeauthoryear{Savel, Dressing, Hirsch, Ciardi, Fleming,
  Giacalone, Mayo  \& Christiansen}{Savel et~al.}{2020}]{savel2020closer}
Savel A.~B.,  Dressing C.~D.,  Hirsch L.~A.,  Ciardi D.~R.,  Fleming J.~P.,
  Giacalone S.~A.,  Mayo A.~W.,   Christiansen J.~L.,  2020, The Astronomical
  Journal, 160, 287

\bibitem[\protect\citeauthoryear{{Schlafly} \& {Finkbeiner}}{{Schlafly} \&
  {Finkbeiner}}{2011}]{Schlafly:2011}
{Schlafly} E.~F.,  {Finkbeiner} D.~P.,  2011, \mn@doi [\apj]
  {10.1088/0004-637X/737/2/103}, \href
  {https://ui.adsabs.harvard.edu/abs/2011ApJ...737..103S} {737, 103}

\bibitem[\protect\citeauthoryear{{Schlaufman}, {Lin}  \& {Ida}}{{Schlaufman}
  et~al.}{2010}]{Schlaufman2010}
{Schlaufman} K.~C.,  {Lin} D.~N.~C.,   {Ida} S.,  2010, \mn@doi [\apjl]
  {10.1088/2041-8205/724/1/L53}, \href
  {https://ui.adsabs.harvard.edu/abs/2010ApJ...724L..53S} {724, L53}

\bibitem[\protect\citeauthoryear{{Schlegel}, {Finkbeiner}  \&
  {Davis}}{{Schlegel} et~al.}{1998}]{Schlegel:1998}
{Schlegel} D.~J.,  {Finkbeiner} D.~P.,   {Davis} M.,  1998, \mn@doi [\apj]
  {10.1086/305772}, \href {http://adsabs.harvard.edu/abs/1998ApJ...500..525S}
  {500, 525}

\bibitem[\protect\citeauthoryear{{Seager} \& {Mall{\'e}n-Ornelas}}{{Seager} \&
  {Mall{\'e}n-Ornelas}}{2003}]{Seager2003}
{Seager} S.,  {Mall{\'e}n-Ornelas} G.,  2003, \mn@doi [\apj] {10.1086/346105},
  \href {http://adsabs.harvard.edu/abs/2003ApJ...585.1038S} {585, 1038}

\bibitem[\protect\citeauthoryear{Stassun et~al.,}{Stassun et~al.}{2018}]{TIC}
Stassun K.~G.,  et~al., 2018, \mn@doi [The Astronomical Journal]
  {10.3847/1538-3881/aad050}, 156, 102

\bibitem[\protect\citeauthoryear{{Steffen} \& {Coughlin}}{{Steffen} \&
  {Coughlin}}{2016}]{Steffen2016}
{Steffen} J.~H.,  {Coughlin} J.~L.,  2016, \mn@doi [Proceedings of the National
  Academy of Science] {10.1073/pnas.1606658113}, \href
  {https://ui.adsabs.harvard.edu/abs/2016PNAS..11312023S} {113, 12023}

\bibitem[\protect\citeauthoryear{{Steffen} \& {Farr}}{{Steffen} \&
  {Farr}}{2013}]{Steffen2013}
{Steffen} J.~H.,  {Farr} W.~M.,  2013, \mn@doi [\apjl]
  {10.1088/2041-8205/774/1/L12}, \href
  {https://ui.adsabs.harvard.edu/abs/2013ApJ...774L..12S} {774, L12}

\bibitem[\protect\citeauthoryear{{Telting} et~al.,}{{Telting}
  et~al.}{2014}]{Telting2014}
{Telting} J.~H.,  et~al., 2014, \mn@doi [Astronomische Nachrichten]
  {10.1002/asna.201312007}, \href
  {https://ui.adsabs.harvard.edu/abs/2014AN....335...41T} {335, 41}

\bibitem[\protect\citeauthoryear{{Theano Development Team}}{{Theano Development
  Team}}{2016}]{exoplanet:theano}
{Theano Development Team} 2016, arXiv e-prints, abs/1605.02688

\bibitem[\protect\citeauthoryear{{Tofflemire} et~al.,}{{Tofflemire}
  et~al.}{2021}]{Tofflemire2021}
{Tofflemire} B.~M.,  et~al., 2021, \mn@doi [\aj] {10.3847/1538-3881/abdf53},
  \href {https://ui.adsabs.harvard.edu/abs/2021AJ....161..171T} {161, 171}

\bibitem[\protect\citeauthoryear{{Triaud}}{{Triaud}}{2018}]{Triaud2018}
{Triaud} A. H.~M.~J.,  2018, {The Rossiter-McLaughlin Effect in Exoplanet
  Research}.
p.~2, \mn@doi{10.1007/978-3-319-55333-7_2}

\bibitem[\protect\citeauthoryear{{Twicken} et~al.,}{{Twicken}
  et~al.}{2018}]{Twicken:DVdiagnostics2018}
{Twicken} J.~D.,  et~al., 2018, \mn@doi [\pasp] {10.1088/1538-3873/aab694},
  \href {http://adsabs.harvard.edu/abs/2018PASP..130f4502T} {130, 064502}

\bibitem[\protect\citeauthoryear{{Van Cleve} et~al.,}{{Van Cleve}
  et~al.}{2016}]{vancleve}
{Van Cleve} J.~E.,  et~al., 2016, \mn@doi [\pasp]
  {10.1088/1538-3873/128/965/075002}, \href
  {https://ui.adsabs.harvard.edu/abs/2016PASP..128g5002V} {128, 075002}

\bibitem[\protect\citeauthoryear{{Vanderburg} et~al.,}{{Vanderburg}
  et~al.}{2016}]{Vanderburg2016}
{Vanderburg} A.,  et~al., 2016, \mn@doi [\apjl] {10.3847/2041-8205/827/1/L10},
  \href {https://ui.adsabs.harvard.edu/abs/2016ApJ...827L..10V} {827, L10}

\bibitem[\protect\citeauthoryear{{Vanderburg} et~al.,}{{Vanderburg}
  et~al.}{2019}]{vanderburg2019}
{Vanderburg} A.,  et~al., 2019, \mn@doi [\apjl] {10.3847/2041-8213/ab322d},
  \href {https://ui.adsabs.harvard.edu/abs/2019ApJ...881L..19V} {881, L19}

\bibitem[\protect\citeauthoryear{{Vaughan}, {Preston}  \& {Wilson}}{{Vaughan}
  et~al.}{1978}]{1978PASP...90..267V}
{Vaughan} A.~H.,  {Preston} G.~W.,   {Wilson} O.~C.,  1978, \mn@doi [\pasp]
  {10.1086/130324}, \href
  {https://ui.adsabs.harvard.edu/abs/1978PASP...90..267V} {90, 267}

\bibitem[\protect\citeauthoryear{{Voges} et~al.,}{{Voges}
  et~al.}{2000}]{2000yCat.9029....0V}
{Voges} W.,  et~al., 2000, VizieR Online Data Catalog, \href
  {https://ui.adsabs.harvard.edu/abs/2000yCat.9029....0V} {p. IX/29}

\bibitem[\protect\citeauthoryear{{Wang}, {Xie}, {Barclay}  \& {Fischer}}{{Wang}
  et~al.}{2014a}]{Wang2014a}
{Wang} J.,  {Xie} J.-W.,  {Barclay} T.,   {Fischer} D.~A.,  2014a, \mn@doi
  [\apj] {10.1088/0004-637X/783/1/4}, \href
  {https://ui.adsabs.harvard.edu/abs/2014ApJ...783....4W} {783, 4}

\bibitem[\protect\citeauthoryear{{Wang}, {Fischer}, {Xie}  \& {Ciardi}}{{Wang}
  et~al.}{2014b}]{Wang2014b}
{Wang} J.,  {Fischer} D.~A.,  {Xie} J.-W.,   {Ciardi} D.~R.,  2014b, \mn@doi
  [\apj] {10.1088/0004-637X/791/2/111}, \href
  {https://ui.adsabs.harvard.edu/abs/2014ApJ...791..111W} {791, 111}

\bibitem[\protect\citeauthoryear{{Wilson}}{{Wilson}}{1978}]{1978ApJ...226..379W}
{Wilson} O.~C.,  1978, \mn@doi [\apj] {10.1086/156618}, \href
  {https://ui.adsabs.harvard.edu/abs/1978ApJ...226..379W} {226, 379}

\bibitem[\protect\citeauthoryear{{Winn} et~al.,}{{Winn}
  et~al.}{2017}]{Winn2017}
{Winn} J.~N.,  et~al., 2017, \mn@doi [\aj] {10.3847/1538-3881/aa7b7c}, \href
  {https://ui.adsabs.harvard.edu/abs/2017AJ....154...60W} {154, 60}

\bibitem[\protect\citeauthoryear{{Winn}, {Sanchis-Ojeda}  \&
  {Rappaport}}{{Winn} et~al.}{2018}]{Winn2018}
{Winn} J.~N.,  {Sanchis-Ojeda} R.,   {Rappaport} S.,  2018, \mn@doi [\nar]
  {10.1016/j.newar.2019.03.006}, \href
  {https://ui.adsabs.harvard.edu/abs/2018NewAR..83...37W} {83, 37}

\bibitem[\protect\citeauthoryear{{Wolfgang}, {Rogers}  \& {Ford}}{{Wolfgang}
  et~al.}{2016}]{angie}
{Wolfgang} A.,  {Rogers} L.~A.,   {Ford} E.~B.,  2016, \mn@doi [\apj]
  {10.3847/0004-637X/825/1/19}, \href
  {https://ui.adsabs.harvard.edu/abs/2016ApJ...825...19W} {825, 19}

\bibitem[\protect\citeauthoryear{{Wyatt}, {Kral}  \& {Sinclair}}{{Wyatt}
  et~al.}{2020}]{Wyatt2020}
{Wyatt} M.~C.,  {Kral} Q.,   {Sinclair} C.~A.,  2020, \mn@doi [\mnras]
  {10.1093/mnras/stz3052}, \href
  {https://ui.adsabs.harvard.edu/abs/2020MNRAS.491..782W} {491, 782}

\bibitem[\protect\citeauthoryear{{Yang}, {Xie}  \& {Zhou}}{{Yang}
  et~al.}{2020}]{Yang2020}
{Yang} J.-Y.,  {Xie} J.-W.,   {Zhou} J.-L.,  2020, \mn@doi [\aj]
  {10.3847/1538-3881/ab7373}, \href
  {https://ui.adsabs.harvard.edu/abs/2020AJ....159..164Y} {159, 164}

\bibitem[\protect\citeauthoryear{{Zacharias}, {Finch}  \&
  {Frouard}}{{Zacharias} et~al.}{2017}]{Zacharias:2017}
{Zacharias} N.,  {Finch} C.,   {Frouard} J.,  2017, VizieR Online Data Catalog,
  \href {http://cdsads.u-strasbg.fr/abs/2017yCat.1340....0Z} {1340}

\bibitem[\protect\citeauthoryear{{Zhou} et~al.,}{{Zhou}
  et~al.}{2020}]{Zhou2020}
{Zhou} G.,  et~al., 2020, \mn@doi [\apjl] {10.3847/2041-8213/ab7d3c}, \href
  {https://ui.adsabs.harvard.edu/abs/2020ApJ...892L..21Z} {892, L21}

\bibitem[\protect\citeauthoryear{{Zhou} et~al.,}{{Zhou}
  et~al.}{2021}]{Zhou2021}
{Zhou} G.,  et~al., 2021, \mn@doi [\aj] {10.3847/1538-3881/abba22}, \href
  {https://ui.adsabs.harvard.edu/abs/2021AJ....161....2Z} {161, 2}

\bibitem[\protect\citeauthoryear{{{\v{Z}}erjal} et~al.,}{{{\v{Z}}erjal}
  et~al.}{2017}]{2017ApJ...835...61Z}
{{\v{Z}}erjal} M.,  et~al., 2017, \mn@doi [\apj] {10.3847/1538-4357/835/1/61},
  \href {https://ui.adsabs.harvard.edu/abs/2017ApJ...835...61Z} {835, 61}

\makeatother
\end{thebibliography}




\end{document}